\documentclass[
superscriptaddress,
bibnotes,
amsmath,amssymb,
twocolumn,
]{revtex4}

\usepackage{graphicx}
\usepackage{dcolumn}
\usepackage{bm}
\usepackage{color}

\def\ga{~\mbox{\raisebox{-.6ex}{$\stackrel{>}{\sim}$}}~}

\begin{document}

\title{Role of electronic localization in the phosphorescence of iridium sensitizing dyes}

\author{Burak Himmetoglu}
\affiliation{Department of Chemical Engineering and Materials Science, 
University of Minnesota, Minneapolis, MN, 55455, USA}
\author{Alex Marchenko\footnote{present adress: Nuclear Engineering and Materials Science, 
Universitat Polit\`{e}cnica de Catalunya, Barcelona, Spain}}
\affiliation{Department of Chemical Engineering and Materials Science, 
University of Minnesota, Minneapolis, MN, 55455, USA}
\affiliation{Department of Chemistry, 
University of Minnesota, Minneapolis, MN, 55455, USA}
\author{Isma\"ila Dabo}
\affiliation{CERMICS, Universit\'{e} Paris-Est, Champus sur Marne 
77455 Marne la Vallee Cedex 2, France}
\author{Matteo Cococcioni}
\affiliation{Department of Chemical Engineering and Materials Science, University of Minnesota, Minneapolis, 
MN, 55455, USA}

\date{\today}

\begin{abstract}
In this work we present a systematic study of three representative iridium dyes, namely,
Ir(ppy)$_3$, FIrpic and PQIr, which are commonly used as sensitizers in organic optoelectronic devices.
We show that electronic correlations play a crucial role in determining the excited-state energies
in these systems, due to localization of electrons on Ir $d$ orbitals.
Electronic localization is captured by employing hybrid functionals within time-dependent density-functional 
theory (TDDFT) and with Hubbard-model corrections within the $\Delta$-SCF 
approach.
The performance of both methods are studied comparatively and shown to be 
in good agreement with experiment. The Hubbard-corrected functionals provide further insight
into the localization of electrons and on the charge-transfer character of excited-states.
The gained insight allows us to comment on envisioned functionalization strategies 
to improve the performance of these systems. Complementary discussions 
on the $\Delta$-SCF method are also presented in order to fill some of the gaps in the literature.
\end{abstract}

\maketitle
\section{Introduction}
Phosphorescent organometallic dyes have attracted considerable interest over the last few decades
as highly efficient sensitizers in organic light-emitting diodes (OLEDs)~\cite{oled-nature}.
These complexes are characterized by strong spin-orbit coupling due to heavy transition metals
in their core, such as Ir, Pt, and Os. The strong spin-orbit coupling leads to intersystem 
crossing from the singlet to triplet excited states and allows for the emission from the otherwise forbidden
triplet state (phosphorescence), in addition to emission from the singlet state (fluorescence). 
As a result, phosphorescent emitters can have an internal quantum efficiency of $100 \%$
~\cite{oled-nature}.
Moreover, the possibility to tune the colors of these emitters by modifications in the ligands
surrounding the transition-metal center has opened up the possibility to 
design color displays and efficient white light sources from organic materials 
(see Ref.~\onlinecite{xiao-review} for a review).

More recently, the strong intersystem-crossing mechanism that take place 
in these complexes have been exploited to design organic solar cells (OSCs) of improved power conversion
efficiency~\cite{holmes-irppy3,triplet-photovoltalic}. In fact, in
organic heterojunctions, sensitizing transition-metal complexes are
utilized as promoters in converting photogenerated singlet excitons into long-lived triplet 
excitons, thereby increasing the probability for an exciton to diffuse to a donor-acceptor interface 
and dissociate into collectible charges~\cite{holmes-irppy3,triplet-photovoltalic}. 
In a recent study~\cite{holmes-irppy3} it was reported that incorporating Ir(ppy)$_3$ (in
concentrations as low as few mass percents) into electron-donating polymers nearly doubled 
the photovoltaic efficiency of the considered organic heterojunctions.  

These remarkable experimental achievements motivate further theoretical effort to understand
the optical properties of transition-metal complexes and ultimately
optimize their sensitizing functions taking advantage of their
almost unlimited chemical versatility, i.e., modifying the transition-metal
center and attached chromophores. To assist in this endeavor,
computational methods of increasing predictive ability are now
available. Among successful electronic-structure techniques, the
Green's function many-body perturbation theory (GW) with
Bethe-Salpeter post-treatment has been accurate in predicting
excited-state energies~\cite{hedin,GW}, yet requiring considerable effort to achieve
full self-consistent convergence. Computationally less demanding
predictions are based upon time-dependent density-functional theory
(TDDFT) and linear-response theory~\cite{runge-gross-thm,tddft-gross-1,tddft-gross-2}. 
Recently, alternative $\Delta$-SCF~\cite{d-scf} techniques
that consist in evaluating excited-state energies from moderately
demanding constrained density-functional theory (DFT) calculations
have also seen a revival of interest~\cite{dscf-better-1,dscf-better-2,dscf-better-3}.

In spite of their limited cost, the accuracy of TDDFT and $\Delta$-SCF
approximations depends crucially on the ability of the underlying DFT
functional to properly capture orbital localization. In this regard,
local and semilocal DFT functionals are known to insufficiently
localize electronic states. 
Moreover, conventional TDDFT calculations rely on the
adiabatic approximation whereby the frequency dependence of the TDDFT
kernel is neglected, representing another important source of error in
capturing excited states. In practical terms, TDDFT approximations do
not properly describe charge-transfer states that consist of an
electron weakly coupled to a hole~\cite{ct-error-1,tddft-ct-1,tddft-ct-2,tddft-ct-3,tddft-ct-4,
tddft-ct-5,tddft-ct-6,tddft-ct-7}.

In this study, we examine the performance of TDDFT and $\Delta$-SCF
approximations in predicting singlet and triplet excitonic energies
in three representative Ir complexes, namely, Ir(ppy)$_3$, FIrpic and PQIr,
which emit in the green, the blue and the red~\cite{erickson-all}, respectively.
Specifically, we compare TDDFT with $\Delta$-SCF that employ Hubbard 
corrections, enabling us to gain needed insight into the effects of electronic
localization and charge-transfer for excited states and comment on envisioned 
functionalization strategies. Such insight will be useful to
study the influence of ligand modification on
singlet and triplet energies with the ultimate goal of guiding experiments
in designing more efficient OLEDs and OSCs.

The paper is organized as follows: We provide some
technical details about the computational approach used in this work in Section~\ref{sec:method} and summarize the
methods employed to compute excited-state energies within TDDFT
and $\Delta$-SCF in Section~\ref{sec:overview}. 
In Section~\ref{sec:results}, we present the main results and provide 
a detailed discussion on the performance of the 
methods used and some comments on strategies to tune excited-state energies. 
Finally, we provide some concluding remarks. We also 
include an Appendix, which provides a brief analysis of the calculation of the
singlet excited-state energy, within the $\Delta$-SCF approach.

\section{\label{sec:method}Computational methods}
%
%
 
Structural optimizations and $\Delta$-SCF calculations presented in this paper
were performed using the plane-waves pseudopotential 
implementation of DFT contained in the PWSCF code of the {\it Quantum ESPRESSO} package~\cite{espresso}.
The TDDFT calculations were performed using  
the {\it Gaussian 09} package~\cite{g09}. For plane-wave calculations, the exchange-correlation
energy was approximated using the generalized-gradient approximation (GGA) with the
Perdew-Burke-Ernzherof (PBE) parametrization~\cite{pbe}. The Ir, N, C, and H atoms were all represented by
ultrasoft pseudopotentials~\cite{vanderbilt}. 
The electronic wavefunctions and charge density were expanded up to kinetic energy cutoffs of
$30\, {\rm Ry}$ and $360\, {\rm Ry}$, respectively. In performing TDDFT calculations, 
the SDD basis set of Refs.
~\onlinecite{g09-impl-1,g09-impl-2,g09-impl-3,g09-impl-4,g09-impl-5,g09-impl-6,g09-impl-7}
was used.
The semilocal PBE~\cite{pbe} and hybrid functionals B3LYP~\cite{b3lyp-1,b3lyp-2} 
and M06~\cite{m06} were used in the TDDFT calculations.  

In our plane wave calculations, we employed the Hubbard-model corrected DFT+U
~\cite{anisimov-1991,anisimov-1993,mazin-1997,Ucalc} method, in order to 
capture the effects of electronic localization accurately.
An improved version of the DFT+U formalism, which includes inter-site interactions (DFT+U+V)~\cite{HubV}, 
was also used in order to obtain  
a better description of interactions between localized electrons on Ir $d$ states and the surrounding
organic ligands. The values of the Coulomb interaction parameters $U$ and $V$ 
were computed using the linear-response method introduced in Ref.~\onlinecite{Ucalc}. 
The molecular structures and charge densities 
presented in this paper were generated using XCrysden~\cite{xcrysden}.
\section{\label{sec:overview} Theoretical Overview}
In this section, we provide a brief overview of the computational approaches used in this paper. We refer the reader
to the original literature for a complete discussion of these methods.
\subsection{\label{subsec:tddft} TDDFT}
The theorem of Runge and Gross~\cite{runge-gross-thm} states that there is a one-to-one
correspondence between a time-dependent external potential and the electronic density of a 
many-body system. A particular application of the Runge-Gross theorem was used to 
compute the energies of excited states of many-body systems, using linear-response functions
~\cite{tddft-gross-1,tddft-gross-2}. Within this approach, 
the poles of the response function provide the excited-state
energies of the system, which can be determined through the solution of the following 
eigenvalue problem
~\cite{tddft-gross-2,g09-impl-1,g09-impl-2,g09-impl-3,g09-impl-4,g09-impl-5,g09-impl-6,g09-impl-7}:
\begin{equation}
\left(
 \begin{array}{cc}
  {\bf L} & {\bf M} \\
  {\bf M}^* & {\bf L}^*
 \end{array}
\right)\, 
\left(
 \begin{array}{c}
  {\bf X}(\omega) \\
  {\bf Y}(\omega)
 \end{array}
\right) = \omega\,
\left(
 \begin{array}{cc}
  -{\bf I} & 0 \\
   0 & {\bf I}
 \end{array}
\right)\, 
\left( 
 \begin{array}{c}
  {\bf X}(\omega) \\
  {\bf Y}(\omega)
 \end{array}
\right) \label{casida-eqn}
\end{equation}
where the solutions $\omega$ represent excited-state energies. The matrices ${\bf L}$ and ${\bf M}$ are
given by
\begin{eqnarray}
&& L_{i\, a\, \sigma ; j\, b\, \tau} = \delta_{\sigma\, \tau}\, \delta_{i\, j}\, \delta_{a\, b}\, 
   \left( \epsilon_a^{\sigma} - \epsilon_i^{\sigma} \right) + M_{i\, a\, \sigma ; b\, j\, \tau}  \nonumber\\
&& M_{i\, a\, \sigma ; j\, b\, \tau} = \int d^3{\bf r}\, \int d^3{\bf r'}\, \psi_i^{\sigma *}({\bf r})\, 
   \psi_a^{\sigma}({\bf r}) \nonumber\\
&& \qquad\qquad\qquad \times K^{\sigma\, \tau}({\bf r}, {\bf r'})\, 
                      \psi_j^{\tau *}({\bf r'})\, \psi_b^{\tau}({\bf r}), \nonumber\\ 
&& K^{\sigma\, \tau}({\bf r}, {\bf r'}) =
   \frac{1}{\vert {\bf r} - {\bf r'} \vert} +
   \frac{\delta^2 E_{\rm xc}}{\delta \rho_{\sigma}({\bf r})\, \delta \rho_{\tau}({\bf r'})} \label{LandM}
\end{eqnarray}
where $\psi_i^{\sigma}$ are solutions to the Kohn-Sham (KS) equations with eigenvalues $\epsilon_i^{\sigma}$,
$E_{\rm xc}$ is the exchange-correlation functional, and $\rho_{\sigma}$ denotes the single-particle 
density constructed from occupied Kohn-Sham states. In addition, we adopt the
convention that the indices $i, \, j$ run over occupied states, 
while $a, \, b$ run over unoccupied states, and the
frequency-dependent vector coefficients $X_{i\, a\, \sigma} = Y_{a\, i\, \sigma}$ are
related to the response of the electronic density to a time-dependent perturbation through
\begin{equation}
\delta \rho_{\sigma}({\bf r}, t) = \int d\omega\, \sum_{i, \, a} \left[ X_{i\, a\, \sigma}(\omega)\, 
        \psi_a^{\sigma *}({\bf r})\, \psi_i^{\sigma}({\bf r})\, e^{-i \omega t} + (a \leftrightarrow i) \right] 
\label{XandY}
\end{equation}
Very often, the frequency dependence of the approximate exchange-correlation 
functional $E_{\rm xc}$ is ignored (the adiabatic approximation), in obtaining 
solutions to Eq. (\ref{casida-eqn}).
It is well known that the approach briefly summarized
above usually fails in reproducing experimental excited-state energies for Rydberg and charge-transfer-type
excitations
~\cite{ct-error-1,tddft-ct-1,tddft-ct-2,tddft-ct-3,tddft-ct-4,tddft-ct-5,tddft-ct-6,tddft-ct-7}.
In charge-transfer excitations, the occupied $\psi_i^{\sigma}$ and 
unoccupied $\psi_a^{\sigma}$ orbitals are spatially separated. Therefore, the matrix ${\bf M}$, 
representing the off-diagonal block in Eq. (\ref{casida-eqn}) becomes negligible for local or semilocal
exchange-correlation functionals (such as PBE), while ${\bf L}$ reduces to a diagonal matrix 
composed of energy differences of occupied and unoccupied KS states. As a result, 
the excitation energies are simply given by KS energy differences, which underestimate the 
experimental values~\cite{ct-error-1}. This inaccuracy can be corrected to some extent with the use
of nonlocal, hybrid functionals (such as B3LYP and M06). 
The inaccuracy related to Rydberg and charge-transfer type excitations stems from the
incorrect asymptotic behavior of most approximate exchange-correlation functionals, and can be
mitigated by improving their long-range behavior
~\cite{tddft-ct-5,tddft-ct-7,ct-correct-1,ct-correct-2,ct-correct-3,ct-correct-4,ct-correct-5,ct-correct-6,
ct-correct-7,ct-correct-8}.
The modifications to the long-range interactions
are generally parameterized and determined empirically. Alternatively, 
they can be determined from first-principles in a system-dependent fashion
~\cite{kronik-1,kronik-2,kronik-3}.
Several works also considered the $\Delta$-SCF method (discussed
in the next subsection) to be more precise for these types of excitations compared to TDDFT 
~\cite{tddft-ct-4,dscf-better-1,dscf-better-2,dscf-better-3}.
A mixture of the two approaches has also been shown to be effective in treating 
charge-transfer excitations~\cite{tddft-dscf-mix-1,tddft-dscf-mix-2}. 
\subsection{\label{subsec:d-scf} $\Delta$-SCF}
\begin{figure}[!ht]
\includegraphics[width=0.45\textwidth]{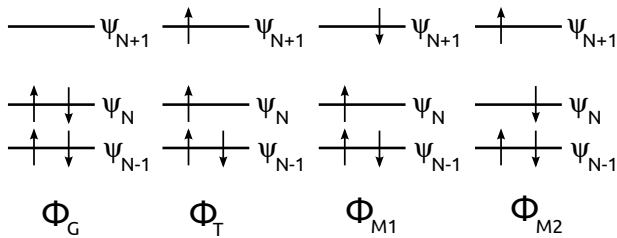}
\caption{\label{fig:levels} Schematic representation of the single determinant configurations 
used in $\Delta$-SCF calculations}
\end{figure}

$\Delta$-SCF method is based on the construction of excited-state electronic densities, using a non-Aufbau
scheme for orbital occupations~\cite{d-scf}. For a system of $N$ electrons, the electronic density 
is constructed by filling the lowest $N-1$ orbitals and the $(N+1)^{\rm th}$ orbital at each iteration
of the electronic-structure optimization until
self-consistency is reached. More precisely, the electronic density of an excited-state 
of spin $\sigma$ is given by
\begin{equation}
\rho_{\sigma}^{\rm ex}({\bf r}) = \sum_{i=1}^{N-1} \vert \psi_i^{\sigma}({\bf r}) \vert^2 
 + \vert \psi_{N+1}^{\sigma}({\bf r}) \vert^2 \label{rho-ex}
\end{equation}
where the orbitals $\psi_i^{\sigma}$ are self-consistently determined from the minimization of the
energy functional depending on the density in Eq.(\ref{rho-ex}).
Examples of such configurations are schematically represented in Fig.~\ref{fig:levels}, where 
the ground-state is assumed to be in a closed shell configuration. The constructed density can be
reproduced from a single excited Slater determinant of specific space and spin symmetry
(for some cases). 
In Fig.~\ref{fig:levels}, $\Phi_{\rm G}$ refers to the ground-state wavefunction, 
$\Phi_{\rm T}$ is a triplet excited-state,
while $\Phi_{\rm M1}$ and $\Phi_{\rm M2}$ are excited-states with mixed spin symmetry.
For the moment, orbital relaxation due to KS self-consistency is ignored. 
The spin symmetries of these states can be explicitly determined through the action of the spin operators
${\vec S}^2$ and $S_z$, which yields (ignoring orbital relaxations),
\begin{eqnarray}
&& {\vec S}^2\, \vert \Phi_{\rm G} \rangle = 0, \quad S_z\, \vert \Phi_{\rm G} \rangle = 0 \nonumber\\
&& {\vec S}^2\, \vert \Phi_{\rm T} \rangle = 2\, \vert \Phi_{\rm T} \rangle, \quad 
    S_z\, \vert \Phi_{\rm T} \rangle = 1\, \vert \Phi_{\rm T} \rangle \nonumber\\
&& {\vec S}^2\, \vert \Phi_{\rm M1} \rangle = \vert \Phi_{\rm M1} \rangle - \vert \Phi_{\rm M2} \rangle, 
   \quad S_z\, \vert \Phi_{\rm M1} \rangle = 0 \nonumber\\
&& {\vec S}^2\, \vert \Phi_{\rm M2} \rangle = \vert \Phi_{\rm M2} \rangle - \vert \Phi_{\rm M1} \rangle ,
   \quad S_z\, \vert \Phi_{\rm M2} \rangle = 0 \label{S2onPhi}
\end{eqnarray}
The above equations show that $\Phi_{\rm M1,2}$ are not eigenvectors of the total spin operator, 
but rather a mixture of singlet and $m_z=0$ triplet-states with 
$\langle \Phi_{\rm M1,2} \vert {\vec S}^2 \vert \Phi_{\rm M1,2} \rangle = 1$. Although
not being eigenfunctions of the actual many-body Hamiltonian, such mixed-states 
appear as variational extrema in $\Delta$-SCF calculations since approximate
exchange-correlation functionals depend on $m_z$ rather than on ${\vec S}^2$.
Therefore, determining the energy of the singlet excited state requires spin-purification
~\cite{spin-pur-1,spin-pur-2,spin-pur-3}, whereby the energy of the singlet state is 
approximated through
\begin{equation}
E_{\rm S} = 2\, E_{\rm M} - E_{\rm T} .\label{pur}
\end{equation}
In this equation, 
$E_{\rm S}$ is the energy of the singlet state, $E_{\rm T}$ is the energy of the triplet state calculated from the 
density corresponding to $\Phi_{\rm T}$ and $E_{\rm M}$ is the energy corresponding to either of 
$\Phi_{\rm M1}$ or $\Phi_{\rm M2}$ (note that they have the same energy, since the external potential is spin-independent). 
A brief derivation of Eq. (\ref{pur}) is provided in the Appendix.

A commonly used approach for the calculation of the singlet excited state consists in performing the 
$\Delta$-SCF calculation with the spin unpolarized (NSP) exchange-correlation functional
~\cite{nsp-o2-1,nsp-o2-2,nsp-o2-3,nsp-new}. This type of calculation
has shown to yield good agreement with experimental 
excitation energies. In the next section, we use and compare both the spin-purified approximation (SPA) and the 
NSP method for singlet energies. (In the Appendix, we provide a brief discussion of the
validity of both approaches.) 

As a consequence of the variational principle, the $\Delta$-SCF 
calculation yields the lowest excited-state energy compatible with the imposed
symmetry of the Slater determinant that was constructed from the one-electron KS states. 
Although it is recognized that the $\Delta$-SCF approach lacks a formal justification 
(at variance with TDDFT that is based on the Runge-Gross
theorem), the $\Delta$-SCF technique is widely used in the literature and has been quite successful
~\cite{dscf-better-1,dscf-better-2,dscf-better-3}. 
One of the formal drawbacks of the $\Delta$-SCF method is that the exchange-correlation
functional is the same for both ground and excited states (i.e., the functional
form is the same with the sole difference that the excited-state density is used in place
of the ground-state density).
More elaborate schemes with formal justification
requires the excited-state calculation to be carried out by different functionals
~\cite{nagy,nagy-2}, by DFT methods with specific many-body wavefunction dependence
~\cite{dft-wvfc-1,dft-wvfc-2}
or by construction of ensembles with mixed ground and excited states~\cite{edft-1,edft-2,edft-3}. 
In either of these cases, the needed exchange-correlation functionals are unknown, and certain approximations are
necessary. For constructing ensembles, one generally relies on statistical arguments extrapolated from the
interacting electron gas~\cite{kohn-edft}. The success of $\Delta$-SCF for lowest energy excitations
is plausibly related to the ground-state exchange-correlation functional being 
sufficiently accurate to describe excited-states in some systems. 
\subsection{\label{subsec:hubbard}Hubbard model based functionals}
Typically, conventional approximate exchange-correlation functionals, such as PBE, yield a poor 
description of localized $d$ states for transition metals.
The DFT+U approach based on an orbital-dependent correction inspired from the Hubbard model
has been widely used to obtain an accurate description of electronic localization
in these cases~\cite{anisimov-1991,anisimov-1993,mazin-1997,Ucalc}.
More recently, the extension of DFT+U, to 
include inter-site interactions between atomic sites (DFT+U+V), has been quite successful, not only
in systems with strong electronic localization (e.g. transition-metal oxides), but also in 
cases where electrons are  delocalized (e.g., band insulators)~\cite{HubV}. The $V$ 
correction has also proved to be crucial in obtaining accurate structural properties 
of transition-metal dioxide molecules~\cite{hubv-molecule}, and 
in describing the dimerization of V atoms in VO$_2$ accross the high-temperature to low-temperature
phase transition through a DMFT approach~\cite{vo2}. 
In Ir complexes, while the on-site interaction $U$ helps capturing the localization of
$d$ electrons on the central Ir atom, the inter-site $V$ is expected to improve the description of 
the interactions with the surrounding organic ligands.

In explicit form, the corrective functional in the DFT+U+V method is given by~\cite{HubV}
\begin{eqnarray}
E_{UV} &=& \sum_{I, \sigma}\, \frac{U^I}{2}\, {\rm Tr}\left[ {\bf n}^{I\, I\, \sigma}\, 
 \left( {\bf 1} - {\bf n}^{I\, I\, \sigma} \right) \right] \nonumber\\
&& \quad - \sum_{I, J, \sigma}\, \frac{V^{I\, J}}{2}\, 
         {\rm Tr}\left[ {\bf n}^{I\, J\, \sigma}\, {\bf n}^{J\, I\, \sigma} \right]
\label{euv}
\end{eqnarray}
where $U^I$ and $V^{I\, J}$ are on-site and inter-site interaction parameters respectively and indices
$I$ and $J$ denote atomic sites. In Eq. (\ref{euv}), the occupation matrices 
${\bf n}^{I\, J\, \sigma}$ are computed as
\begin{equation}
n_{m\, m'}^{I\, J\, \sigma} = \sum_{i}\, f_i^{\sigma}\, 
 \langle \phi_m^I \vert \psi_i^{\sigma} \rangle\, \langle \psi_i^{\sigma} \vert \phi_{m'}^J \rangle
\label{nsIJ}
\end{equation}
where $\psi_i^{\sigma}$ are KS states, $f_i^{\sigma}$ are their occupations, and $\phi_m^{I}$ are 
atomic orbitals centered on site $I$. The first term in Eq. (\ref{euv}) that is proportional to $U$
favors electronic localization on atomic sites, while the second term proportional to $V$ leads to
hybridization between orbitals on different sites, the ground-state configuration 
being eventually governed by the competition between the two opposite tendencies. 
\section{\label{sec:results}Results and Discussions}

The structures of the three molecules [Ir(ppy)$_3$, FIrpic, and PQIr]
have been optimized using the PBE functional, and kept
fixed in all TDDFT and $\Delta$-SCF calculations.
The TDDFT calculations are performed using PBE, B3LYP, and M06 functionals.

The $\Delta$-SCF calculations were performed using the GGA, GGA+U, and GGA+U+V functionals
based upon PBE using 
the $U$ and $V$ parameters obtained from linear response~\cite{Ucalc}. The 
details of the linear-response calculations for each of the molecules are summarized in the following
subsections. In some cases, re-optimizing the structure with forces due the $U$ and $V$ corrections
was needed.
\subsection{\label{subsec:res-irppy}Ir(ppy)$_3$} 
\begin{figure}[!ht]
\begin{minipage}[b]{0.45\linewidth}
\centering
\includegraphics[width=1.0\textwidth]{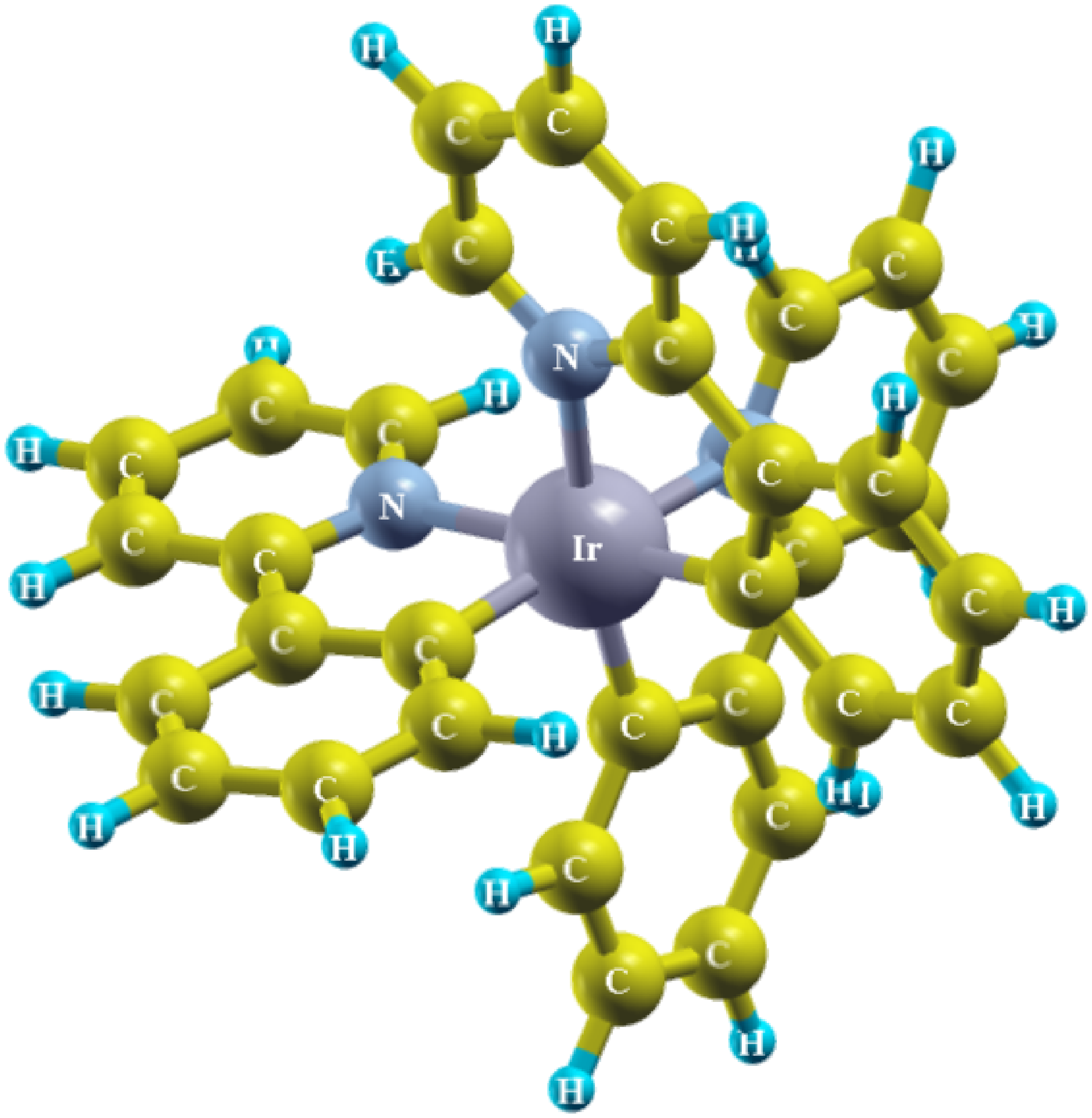}
\end{minipage}
\hspace{0.3cm}
\begin{minipage}[b]{0.45\linewidth}
\centering
\includegraphics[width=0.5\textwidth]{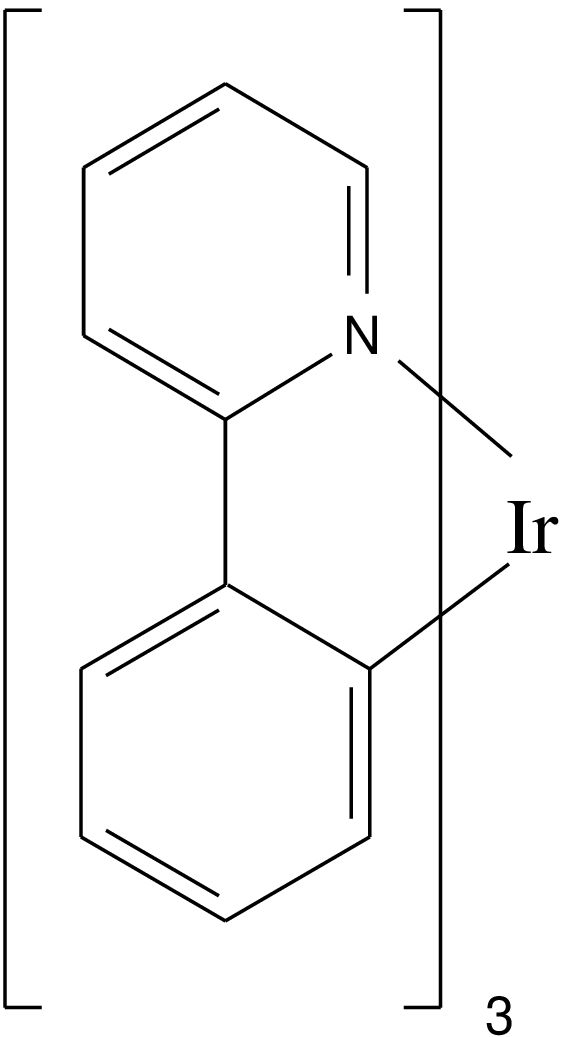}
\end{minipage}
\caption{\label{fig:molecule} 
(Color online) Molecular structure of Ir(ppy)$_3$ as a 3d model and a chemical formula.}
\end{figure}

The molecular structure of Ir(ppy)$_3$ is shown in Fig.~\ref{fig:molecule}.
The lowest triplet and singlet energies obtained within TDDFT using the different
semilocal and hybrid functionals are reported and compared with experiments
in Table.~\ref{tab:tddft}.
\begin{table}[!ht]
\caption{\label{tab:tddft} Lowest TDDFT triplet and singlet excited state energies
compared with experiment. All values are in eV and given relative to the ground-state
energy.}
\begin{ruledtabular}
\begin{tabular}{ccccc}
& Exp & GGA & B3LYP & M06 \\
\hline
T  & 2.4\footnotemark[1] & 2.09 & 2.55 & 2.55 \\
S  & 2.6\footnotemark[2]-2.7\footnotemark[3] & 2.16 & 2.75 & 2.79 \\
\end{tabular}
\end{ruledtabular}
\footnotetext[1]{From Refs.~\onlinecite{holmes-irppy3,baldo-irppy3,goushi-irppy3,eom-irppy3,
adachi-irppy3,tsuboi-irppy3,lamansky-irppy3,hofbeck-irppy3,colombo-irppy3,tsuboyama-irppy3,holzer-irppy3},
from the highest peak in the phosphorescence spectra.}
\footnotetext[2]{From Ref.~\onlinecite{holmes-irppy3}.}
\footnotetext[3]{From Refs.~\onlinecite{tsuboi-irppy3,lamansky-irppy3,holzer-irppy3} from the second
peak/shoulder in the absorption spectrum (the first absorption feature is assumed to be due to the 
triplet state).}
\end{table}
Due to strong spin-orbit coupling present in these systems, it is in principle not possible to distinguish between 
singlet and triplet states since the actual eigenstates are mixtures of them.  
However, the triplet energy can be measured quite accurately in experiment from the phosphorescence spectra.
The largest peak in the phosphorescence spectrum determines the energy of the triplet state, and can 
be determined with high accuracy~\cite{holmes-irppy3,baldo-irppy3,goushi-irppy3,eom-irppy3,
adachi-irppy3,tsuboi-irppy3,lamansky-irppy3,hofbeck-irppy3,colombo-irppy3,tsuboyama-irppy3,holzer-irppy3}.
Nevertheless, in contrast to the triplet state, 
the experimental determination of the singlet energy is very difficult. One could
in principle measure the onset of optical absorption of singlet excitons for this purpose. However, due
to strong spin-orbit coupling, the optical absorption starts at much lower energies than the actual 
singlet energy, leaving a long tail that extends deeply into the high-wavelength region
~\cite{tsuboi-irppy3,lamansky-irppy3,hofbeck-irppy3,colombo-irppy3,tsuboyama-irppy3,holzer-irppy3,ichimura-irppy3}.
Another possibility is to determine the strongest peaks in the absorption spectra, which would yield 
much higher singlet-state energies than the lowest one. Alternatively, the LUMO-HOMO gap
when the LUMO is measured by optical absorption (the so-called optical LUMO)
has also been used to determine the singlet state energy~\cite{baldo-optical-lumo,djurovich-optical-lumo}.
It is important to note that the
optical LUMO contains the binding energy of the exciton  
unlike an inverse photo-emission measurement where the exciton binding energy is
eliminated~\cite{ipes-1,ipes-2}. However, the strong spin-orbit coupling also affects such measurements since
they also rely on optical absorption. Therefore, comparison of the calculated lowest singlet 
energy with the experimental data is not always valid, and one needs in principle to compute the absorption spectrum including 
spin-orbit coupling effects, which can be computationally very expensive. Relativistic effects have only recently been
studied for similar dyes and was shown to result in better agreement with experiment
~\cite{ircomplex-relativistic}.
Another uncertainty that might affect the analysis of the experimental data is 
whether the measurements are performed in the gas phase, in solution or for the molecule absorbed
on the surface of a thin film. However, the correspondence between
different types of measurements are well-known in the literature~\cite{solution}.

In spite of recognized difficulties in making comparison with optical measurements,
we report the calculated lowest singlet excited-state energies  
(and trends in splittings from the triplet excited states) because they are critical 
for the purpose of guiding the design of OSCs with Ir dyes used as sensitizers~\cite{holmes-irppy3}.

As can be seen from Table~\ref{tab:tddft}, the triplet excited-state energy is 
predicted to be within $0.1\, {\rm eV}$ from the experimental value
by the hybrid functionals (B3LYP and M06), whereas it is notably underestimated 
by the semilocal functional (PBE). Although as already mentioned, 
it is very difficult to determine the lowest singlet energy experimentally,
the hybrid functionals yield predictions within an error 
of $0.2\, {\rm eV}$ relative to the reported experimental values, whereas PBE again underestimates it. 
The excited-state energies reported in Table~\ref{tab:tddft} are also in agreement with the previous results
in the literature
~\cite{hay-irppy3,ir-complex-calc-1,ir-complex-calc-2,ir-complex-calc-3}.

\begin{figure}[!ht]
\begin{minipage}[b]{0.49\linewidth}
\centering
\includegraphics[width=1.1\textwidth]{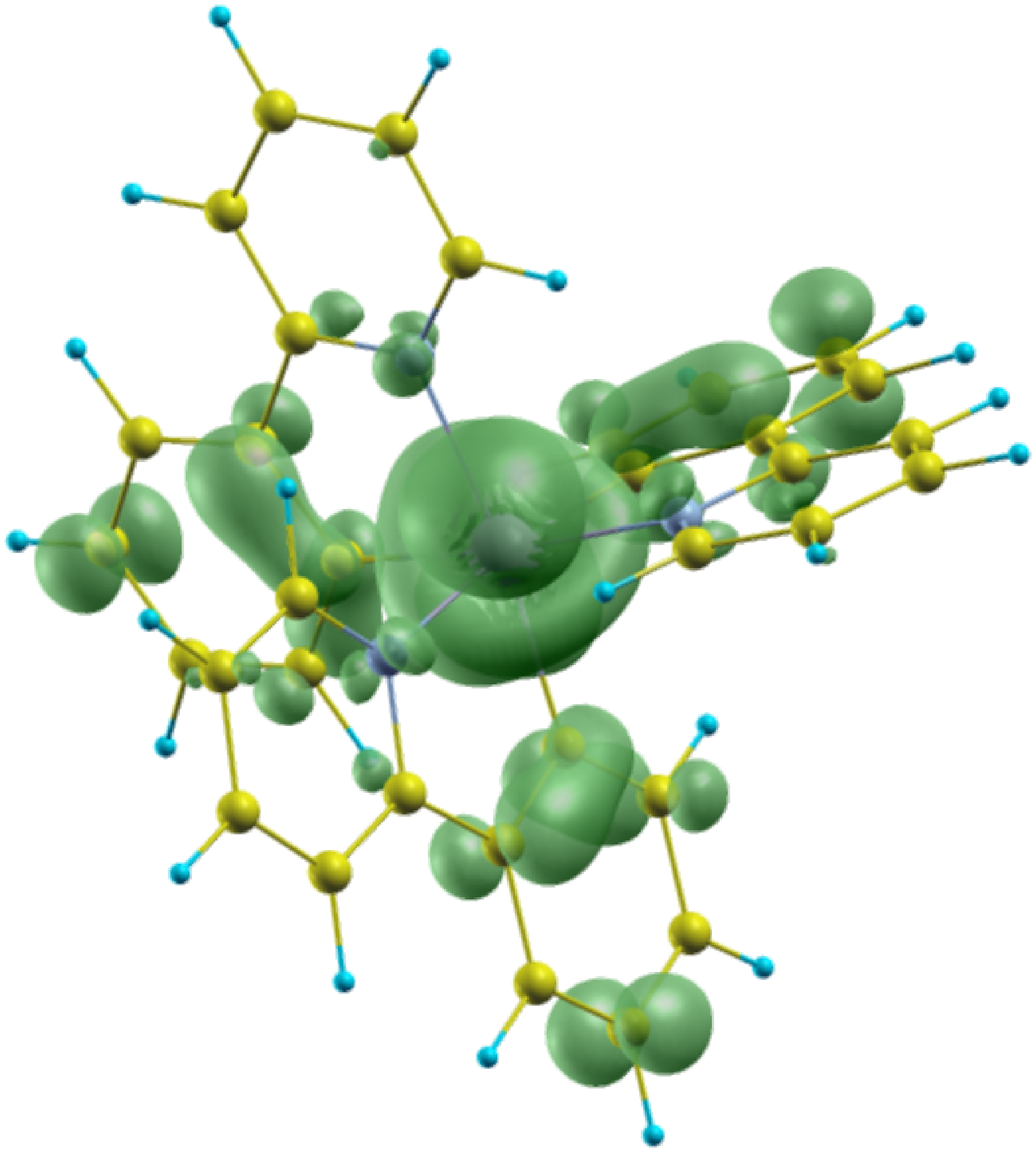}
\end{minipage}
\begin{minipage}[b]{0.49\linewidth}
\centering
\includegraphics[width=1.1\textwidth]{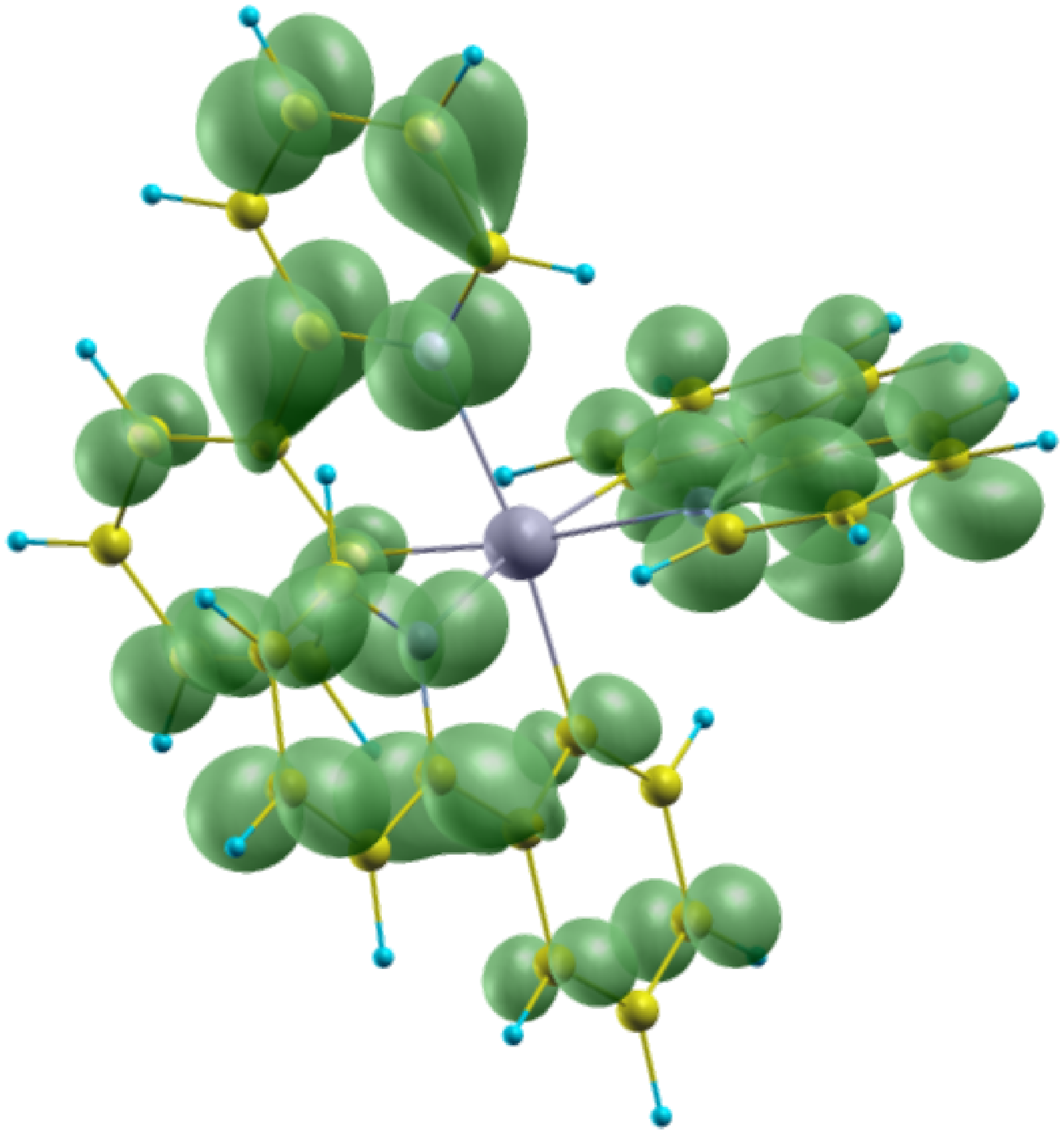}
\end{minipage}
\caption{\label{fig:homo-lumo}
(Color online) Ir(ppy)$_3$ HOMO(left) and LUMO(right) calculated using the ground-state electronic density.}
\end{figure}

In order to gain further insight into computational predictions, we depict the probability density of the HOMO 
and the LUMO for the ground state within PBE in Fig.~\ref{fig:homo-lumo}.
As can be seen, the HOMO is predominantly a metal-centered state with some contribution from the
ligands while the LUMO is almost entirely ligand-centered, 
indicating that optical excitations correlate with metal-to-ligand charge-transfer (MLCT)
processes, as was previously recognized in the literature
~\cite{mlct-irppy3-1,mlct-irppy3-2,hay-irppy3,ir-complex-calc-1,ir-complex-calc-2,ir-complex-calc-3}.
This charge-transfer property of the excited states also explains why the local and
semilocal adiabatic approximation underestimate their TDDFT energies.

\begin{figure}[!ht]
\includegraphics[width=0.45\textwidth]{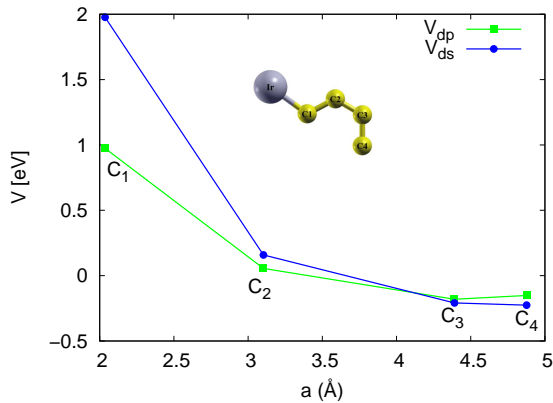}
\caption{\label{fig:chain.C} Inter-site $V$ parameters calculated for a Ir-C-C-C-C chain, as 
a function of distance. $V_{dp}$ denotes the interaction parameter between Ir $d$ and C $p$ states,
while $V_{ds}$ denotes that between Ir $d$ and C $s$ states.}
\end{figure}

\begin{figure}[!ht]
\includegraphics[width=0.45\textwidth]{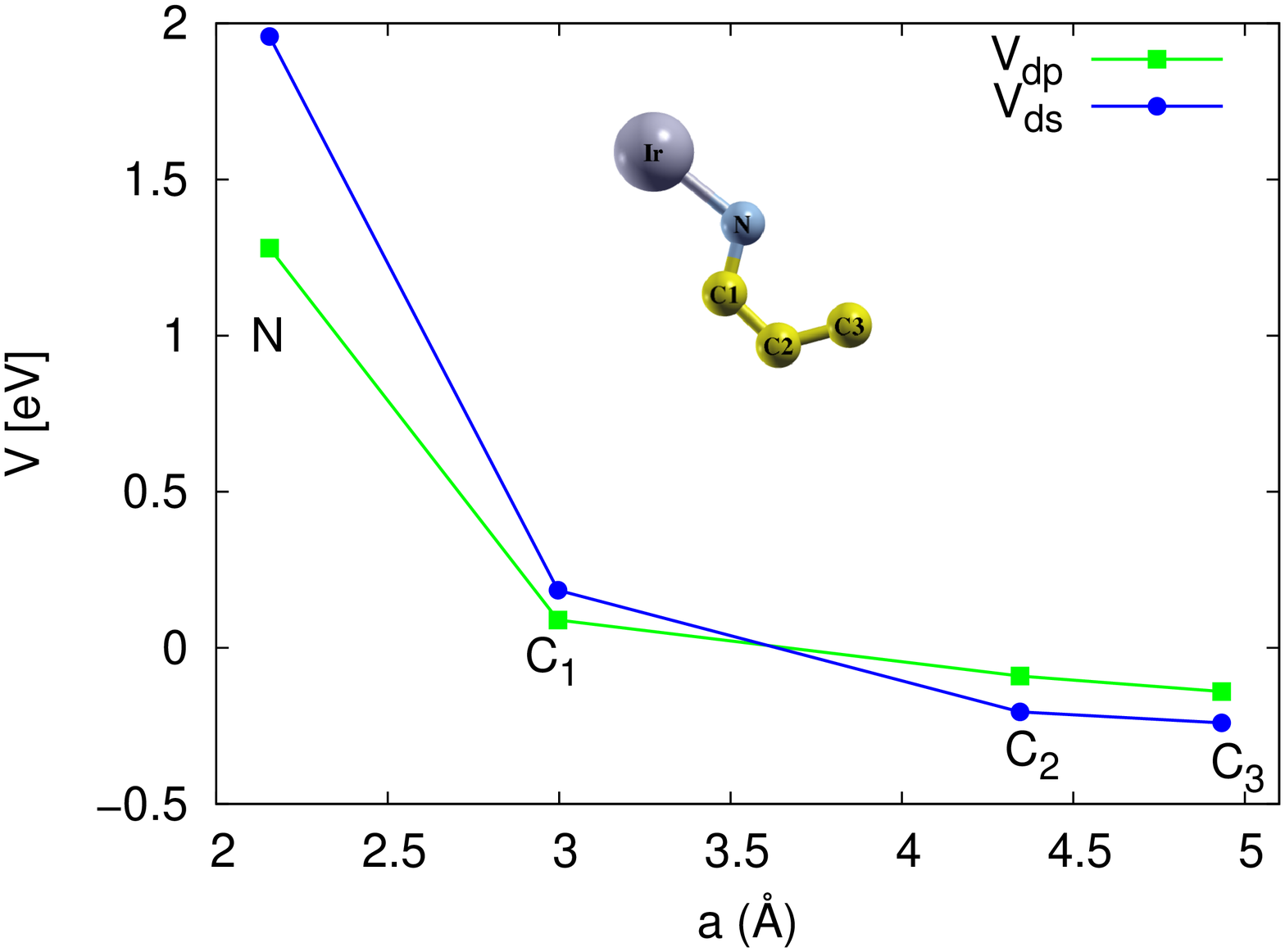}
\caption{\label{fig:chain.N} Inter-site $V$ parameters calculated for a Ir-N-C-C-C chain, as 
a function of distance. $V_{dp}$ denote interaction parameter between Ir $d$ and C,N $p$ states,
while $V_{ds}$ denote those between Ir $d$ and C,N $s$ states.}
\end{figure}

The calculation of $U$ and $V$ parameters requires perturbing each atomic site separately for the 
construction of the response matrices.
Instead of performing such a computationally demanding calculation including the response of
all inequivalent sites, we concentrate only on the central Ir atom and its nearest neighbor C and N atoms. 
This approach is justified by our expectation
that the electronic correlations are only important for electrons localized on the Ir atom.
The atoms further away from Ir 
in the organic ligands (i.e., the distant C and H atoms) are taken into account
in an average manner as a charge reservoir centered at the Ir site, 
as explained in Ref.~\onlinecite{hwc}. The linear response calculation yields the on-site $U$ 
parameter for Ir-$d$ ($U_d$) and Ir-$s$ ($U_s$) states, 
the $V$ parameter between Ir-$d$,$s$ and neighboring
$C$-$p$,$s$ and $N$-$p$,$s$ states [$V({\rm Ir}_{d,s}\, , N_{p,s})$ and $V({\rm Ir}_{d,s}\, , C_{p,s})$]
and the on-site interaction parameter between Ir $d$ and $s$ states 
($V_{\rm on}$). In Table.~\ref{tab:irppy3-uandv} we report only a subset of these values, which 
affects the electronic structure most significantly. Further calculations, not reported here,
have proven that other interaction parameters have no effect on the excited-state energies.
\begin{table}[!ht]
\caption{\label{tab:irppy3-uandv} Linear-response values of $U$ and $V$ parameters.
We use the notation $V({\rm Ir}_d, N_{p,s}) \equiv V_{d-p,s}^N$ and 
$V({\rm Ir}_d, C_{p,s}) \equiv V_{d-p,s}^C$. All values are in $eV$.}
\begin{ruledtabular}
\begin{tabular}{ccccc}
 $U_d$  &  $V_{d-p}^N$  &  $V_{d-p}^C$ 
        &  $V_{d-s}^N$  &  $V_{d-s}^C$ \\
\hline
 7.36 &  1.62 &  1.32 & 2.39 & 2.65 \\
\end{tabular}
\end{ruledtabular}
\end{table}
The reported values of $V({\rm Ir}_d, \, N_{p,s})$ and $V({\rm Ir}_d, \, C_{p,s})$ are average values, since the 
distances between the three N atoms and the three C atoms to Ir atom differ slightly from one ligand
to another. 
However, the differences are within the numerical precision of the linear-response calculations.
We have also computed the inter-site interaction parameters between the Ir-$d$ states and C-$p,s$ 
states which are not nearest neighbors to the Ir atom. This calculation is performed by isolating a chain
of atoms which starts from the Ir atom and ends at a C atom in one of the ligands. The rest of the atoms
in the molecule is treated as a charge reservoir, which is placed at the Ir site.
As can be seen from the results shown in Figs.~\ref{fig:chain.C}, and~\ref{fig:chain.N}, 
the inter-site $V$ parameters decay as a function of the distance from the Ir atom.
This result provides a quantitative justification for the use of the nearest neighbor $V$
parameters between Ir-N and Ir-C pairs only. 

The results of the $\Delta$-SCF calculations using both the SPA singlet-state energy in 
Eq. (\ref{pur}), and its NSP counterpart are reported in Table~\ref{tab:d-scf}.
\begin{table}[!ht]
\caption{\label{tab:d-scf} $\Delta$-SCF calculation of the lowest triplet and singlet states for 
Ir(ppy)$_3$. All values are in eV and measured from the ground-state energy.}
\begin{ruledtabular}
\begin{tabular}{cccccc}
 &  GGA  &  U  &  U+V & (U+V)$_{\rm rel}$ & Exp \\
\hline
T &  2.32  &  2.27  &  2.41 & 2.44 & 2.4 \\
M &  2.34  &  2.29  &  2.43 & 2.46 & -- \\
S [NSP] &  2.50  &  2.90  &  2.78 & 2.73 & 2.6-2.7 \\
$\Delta\, E_{\rm TS}$ [SPA] &  0.05  &  0.04  &  0.02 & 0.04 & -- \\
$\Delta\, E_{\rm TS}$ [NSP] & 0.18 & 0.62 & 0.37 & 0.29 & 0.2-0.3 \\
\end{tabular}
\end{ruledtabular}
\end{table}
Compared with the TDDFT calculations reported in Table~\ref{tab:tddft}, 
the triplet energy resulting from the GGA+U and GGA+U+V calculations are slightly smaller
than the triplet state energy calculated with hybrid functionals. 
Notice that the GGA functional within the $\Delta$-SCF
approach yields results closer to experiments than TDDFT(PBE). This is 
due to the ability of $\Delta$-SCF to partially rectify charge-transfer errors.  
Although the energy of the triplet-state was obtained accurately, 
when the singlet energy is evaluated through the spin-purification formula of Eq. (\ref{pur}),
the $\Delta$-SCF singlet-triplet splitting is almost vanishing.
On the other hand, the NSP calculation of the singlet state (S) yields results in 
better agreement with experiment and TDDFT. In quantitative terms within NSP,
GGA+U overestimates the singlet energy by $0.3\, {\rm eV}$, whereas GGA+U+V improves the 
agreement to an accuracy of approximately $0.2\, {\rm eV}$. This improvement results from 
the inclusion of inter-site interactions through $V$, which corrects
the overlocalization resulting from the straight use of the on-site $U_d$.
In addition, when the molecular structure is re-optimized
with the GGA+U+V functional [i.e. with additional forces coming from the corrective terms in Eq. (\ref{euv})]
, the triplet-state almost matches with the experimental
value, while the singlet energy is overestimated by only $0.1\, {\rm eV}$. 
Although both TDDFT and the Hubbard-corrected functionals yield results that are close to
experiments, the experimental values themselves have an accuracy close to
$0.2\, {\rm eV}$~\cite{djurovich-optical-lumo}. Errors in the singlet energies are in
principle much larger due to problems inherent to their experimental determination, as discussed 
above.
\subsection{\label{subsec:res-firpic}FIrpic}
\begin{figure}[!ht]
\begin{minipage}[b]{0.49\linewidth}
\centering
\includegraphics[width=1.0\textwidth]{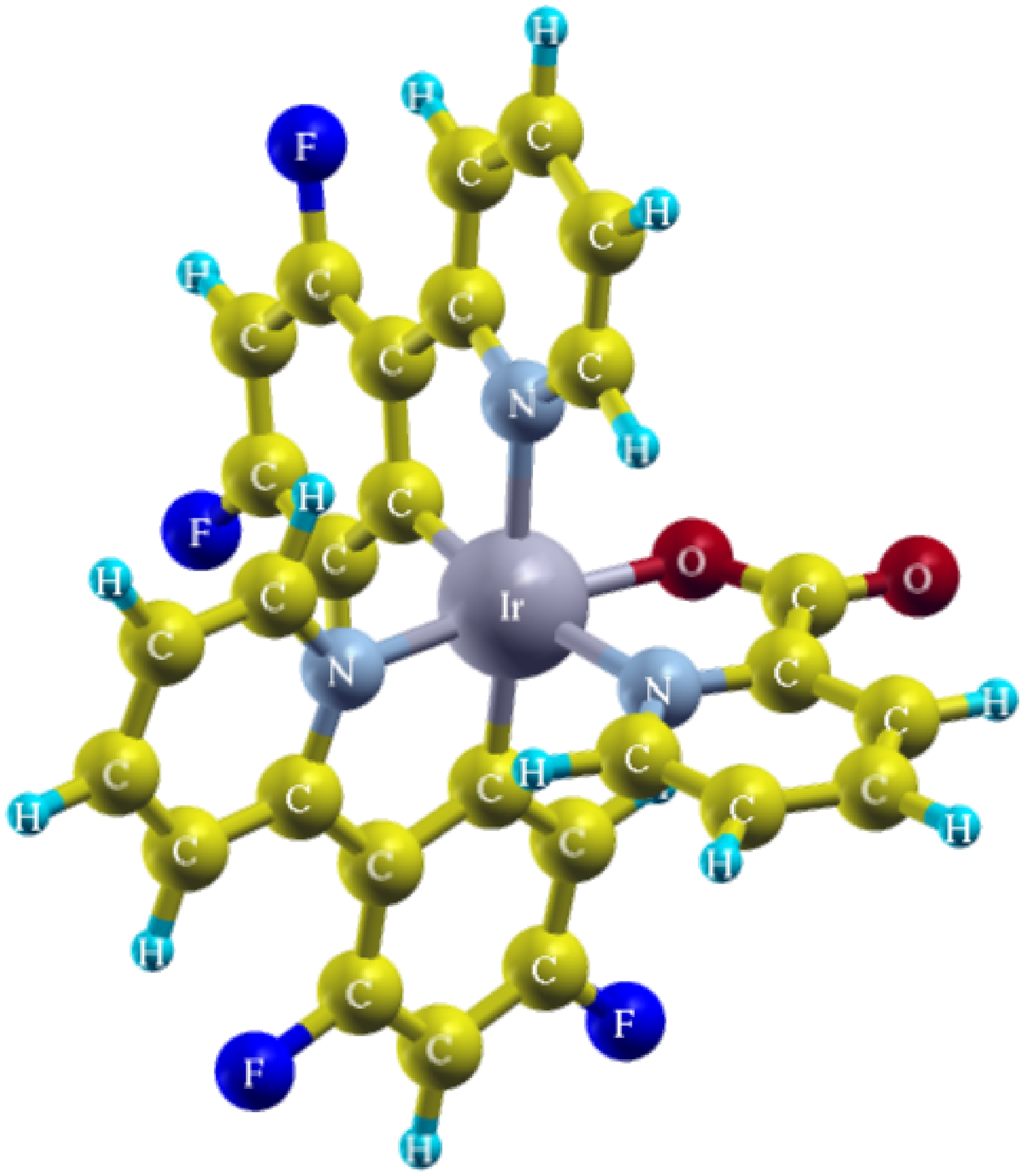}
\end{minipage}
\begin{minipage}[b]{0.49\linewidth}
\centering
\includegraphics[width=1.0\textwidth]{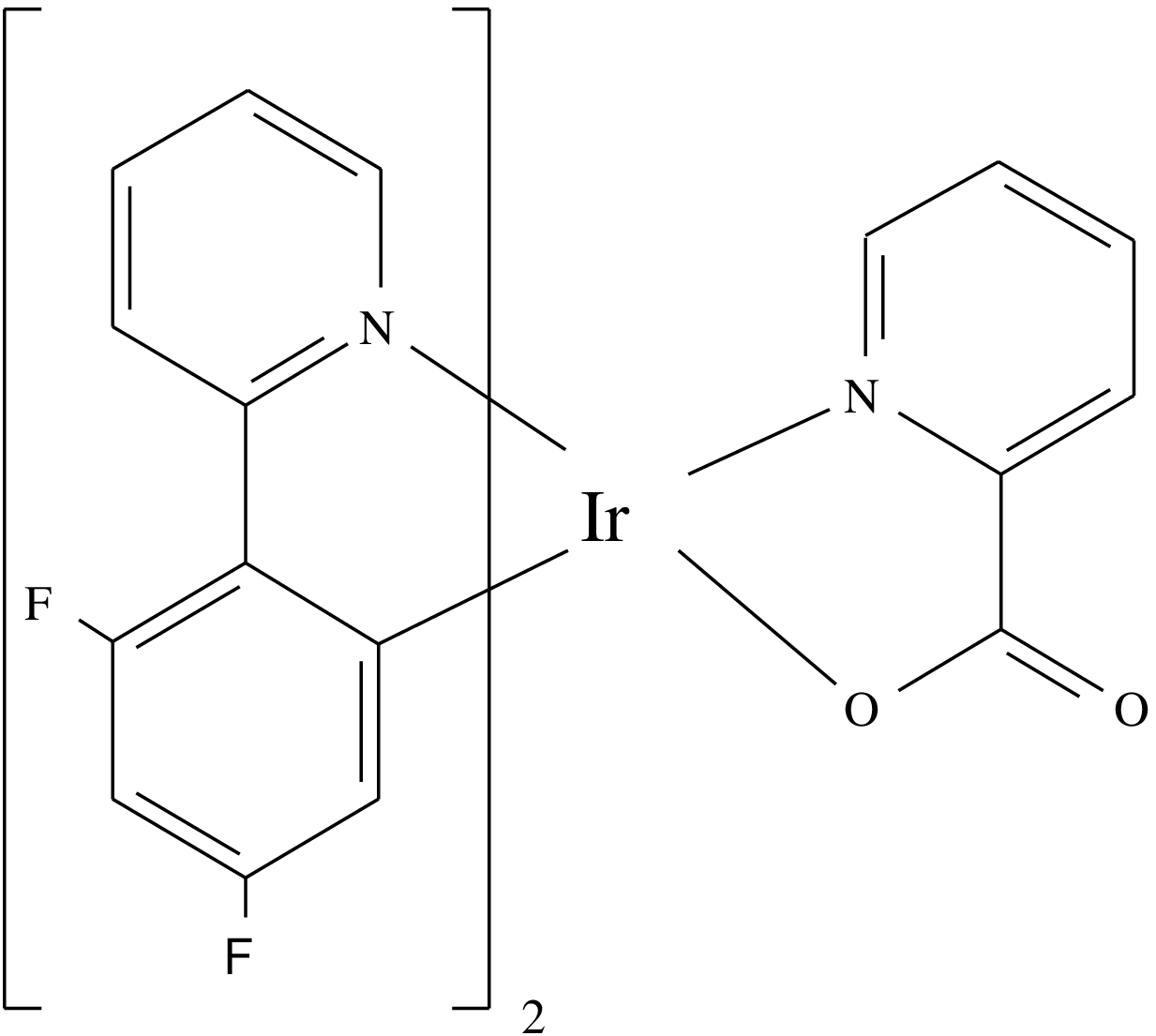}
\end{minipage}
\caption{\label{fig:molecule.2}
(Color online) Molecular structure of FIrpic as a 3d model and a chemical formula.}
\end{figure}

The molecular structure of FIrpic is shown in Fig.~\ref{fig:molecule.2}.
The lowest triplet and singlet excited-state energies calculated with TDDFT using PBE, B3LYP,
and M06 functionals are also compared to experimental data in Table~\ref{tab:tddft-2}.
\begin{table}[!ht]
\caption{\label{tab:tddft-2} Lowest triplet and singlet excited-state energies obtained with TDDFT
and from experiments for FIrpic. All values are in $eV$ and measured from the ground-state energy.}
\begin{ruledtabular}
\begin{tabular}{ccccc}
& Exp & GGA & B3LYP & M06 \\
\hline
T & 2.6\footnotemark[1] & 2.14 & 2.67 & 2.66 \\
S & 3.3\footnotemark[2]-2.9\footnotemark[3]  & 2.24 & 2.94 & 3.00 \\
\end{tabular}
\end{ruledtabular}
\footnotetext[1]{From Refs.~\onlinecite{tsuboi-firpic,adachi-firpic,holmes-firpic,tokito-firpic,
you-firpic,lee-firpic,xiao-review} from the highest peak in the phosphorescence spectrum.}
\footnotetext[2]{From Ref.~\onlinecite{tsuboi-firpic}, from the second peak in the absorption
spectrum.}
\footnotetext[2]{From Ref.~\onlinecite{lee-firpic}, from the onset of optical absorption.}
\end{table}
A salient feature in this comparison is the fact that the PBE functional 
underestimates the energies of both excited states similarly 
to Ir(ppy)$_3$ (cf. Table~\ref{tab:tddft}). Instead, hybrid B3LYP and
M06 show a significant improvement, and yield a triplet energy very close to experimental value.
Despite the difficulties discussed in the previous section, 
the hybrid functionals are always within the experimental uncertainty. Therefore,
we argue that they are fairly accurate.  

\begin{figure}[!ht]
\begin{minipage}[b]{0.49\linewidth}
\centering
\includegraphics[width=1.1\textwidth]{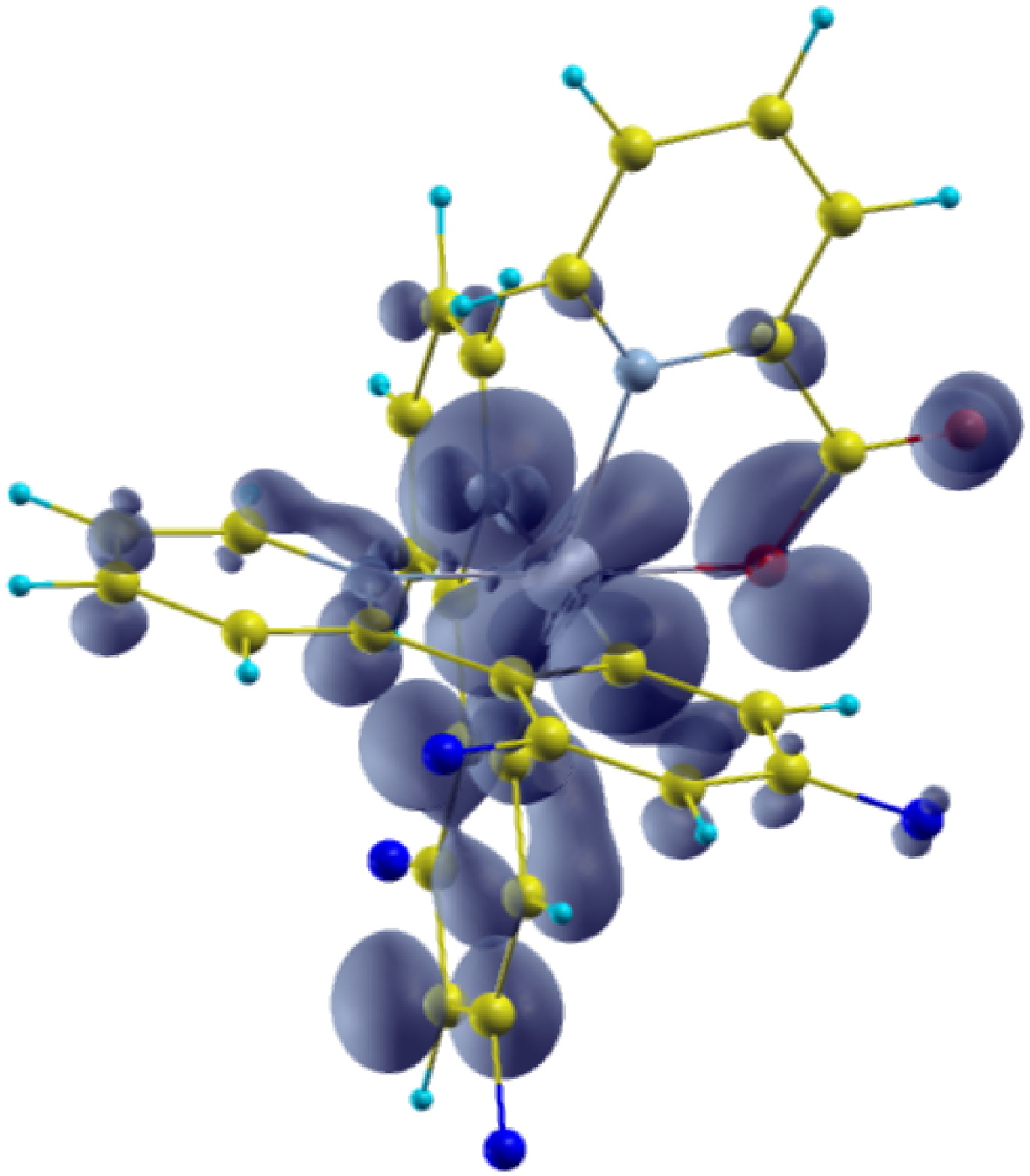}
\end{minipage}
\begin{minipage}[b]{0.49\linewidth}
\centering
\includegraphics[width=1.1\textwidth]{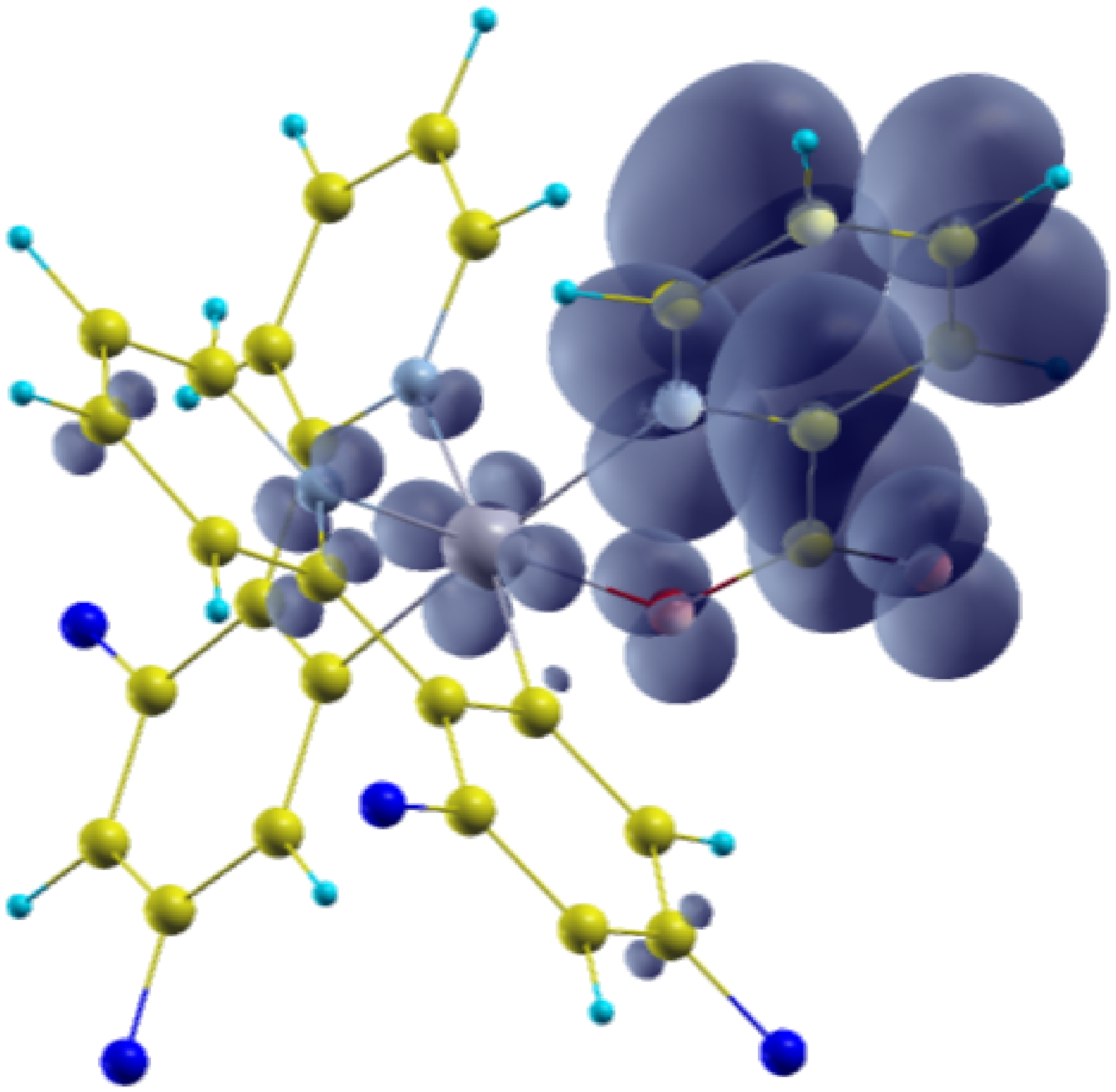}
\end{minipage}
\caption{\label{fig:homo-lumo-firpic}
(Color online) FIrpic HOMO(left) and LUMO(right) calculated using the ground-state electronic density.}
\end{figure}

To further explore the properties of the excited states, 
the HOMO and the LUMO densities in the ground state of FIrpic,
are plotted in Fig.~\ref{fig:homo-lumo-firpic} within PBE.
Similarly to Ir(ppy)$_3$, the HOMO is mostly located on Ir with considerable contribution
from the ligands, whereas the LUMO is almost entirely ligand-centered on the
attached chromophores [predominantly the picolinate (pic) ligand], 
with some contribution coming from the metal center. 

The $\Delta$-SCF calculations were performed using the GGA, GGA+U, and GGA+U+V functionals.
The $U$ and $V$ interaction parameters were calculated using the approach described in 
subsection~\ref{subsec:res-irppy}. 
We have also checked the validity of this 
approach by isolating several chains containing Ir, C, N and O atoms and computing $V$ between pairs that
are not nearest neighbors. We verified that the interaction parameters between the Ir and its
nearest neighbor atoms are important, while the interaction parameters between Ir and
more distant shells of neighbors
vanish similarly as in Figs.~\ref{fig:chain.C} and~\ref{fig:chain.N}. Due to this strong
similarity of the results with Ir(ppy)$_3$, we do not report these calculations here. 

The calculated $U$ and $V$ interaction parameters for FIrpic are reported in Table.~\ref{tab:firpic-uandv}.
\begin{table}[!ht]
\caption{\label{tab:firpic-uandv} Linear-response values of $U$ and $V$ parameters. 
All values are in $eV$.}
\begin{ruledtabular}
\begin{tabular}{ccccccc}
 $U_d$  &  $V_{d-p}^N$  &  $V_{d-p}^C$  &  $V_{d-p}^O$ & 
        $V_{d-s}^N$  &  $V_{d-s}^C$  &  $V_{d-s}^O$ \\   
\hline
  7.17 & 1.69 &  1.29 & 2.0 & 3.06 &  2.56 & 5.77  \\
\end{tabular}
\end{ruledtabular}
\end{table}
We report only the parameters that affect the electronic structure of the molecule most significantly, and
disregard the ones that do not contribute at all. Notice that the parameters $U_d$, $V({\rm Ir}_d, \, C_p)$,
and $V({\rm Ir}_d, \, N_p)$ are very close to the ones which are calculated for Ir(ppy)$_3$
reported in Table.~\ref{tab:irppy3-uandv}. 

The results of the $\Delta$-SCF calculations using both the SPA and NSP calculations for the singlet state 
are reported in Table~\ref{tab:d-scf-2}.
\begin{table}[!ht]
\caption{\label{tab:d-scf-2} $\Delta$-SCF calculation of the lowest triplet and singlet states.
All values are in eV and given relative to the ground-state.}
\begin{ruledtabular}
\begin{tabular}{cccccc}
 &  GGA  &  U  & U+V & (U+V)$_{\rm rel}$ & Exp \\
\hline
T &  2.46  &  2.79  &  2.56 & 2.52 & 2.6 \\
M &  2.50  &  N.A.\footnotemark[1]  &  2.60 & 2.56 & -- \\
S [NSP] &  2.65  &  3.00  &  2.88 & 2.81 & 3.3-2.9 \\
$\Delta\, E_{\rm TS}$ [SPA] & 0.08   &  N.A\footnotemark[1]  &  0.08 & 0.08 & -- \\
$\Delta\, E_{\rm TS}$ [NSP] & 0.18 & 0.20 & 0.31 & 0.29 & 0.3-0.7 \\
\end{tabular}
\end{ruledtabular}
\footnotetext[1]{No convergence is achieved for these calculations.}
\end{table}
As for Ir(ppy)$_3$, GGA within $\Delta$-SCF yields excited-state energies
in better agreement with experimental data compared to GGA within TDDFT.
While GGA+U overestimates the triplet
energy by approximately $0.2\, {\rm eV}$, GGA+U+V lowers it by $0.2\, {\rm eV}$, that is, 
within $0.1\, {\rm eV}$ of experiments. Additionally, the singlet state energies predicted 
by GGA+U and GGA+U+V fall within the experimental range, as was the 
case for TDDFT with hybrid functionals. 

Yet, compared to Ir(ppy)$_3$, the excited-state energies are higher in FIrpic, explaining the
blue phosphorescence of FIrpic. We thus confirm a well-known experimental result
that the attachment of F atoms
in the ligands stabilize the HOMO level and increase the LUMO-HOMO energy gap
~\cite{func-1,func-2}. Since F is highly electronegative, it decreases the amount of charge 
localized on the metal center, thereby reducing the total Coulomb repulsion felt by these electrons
and stabilizing the HOMO level. 
This comparison with Ir(ppy)$_3$ suggests that
further adjustments on the triplet emission energy can be obtained
by functionalizing the (pic) ligand, since the LUMO is mainly localized on it, as can be 
seen from Fig.~\ref{fig:homo-lumo-firpic}.
\subsection{\label{subsec:res-pqir}PQIr}
\begin{figure}[!ht]
\begin{minipage}[b]{0.45\linewidth}
\centering
\includegraphics[width=1.0\textwidth]{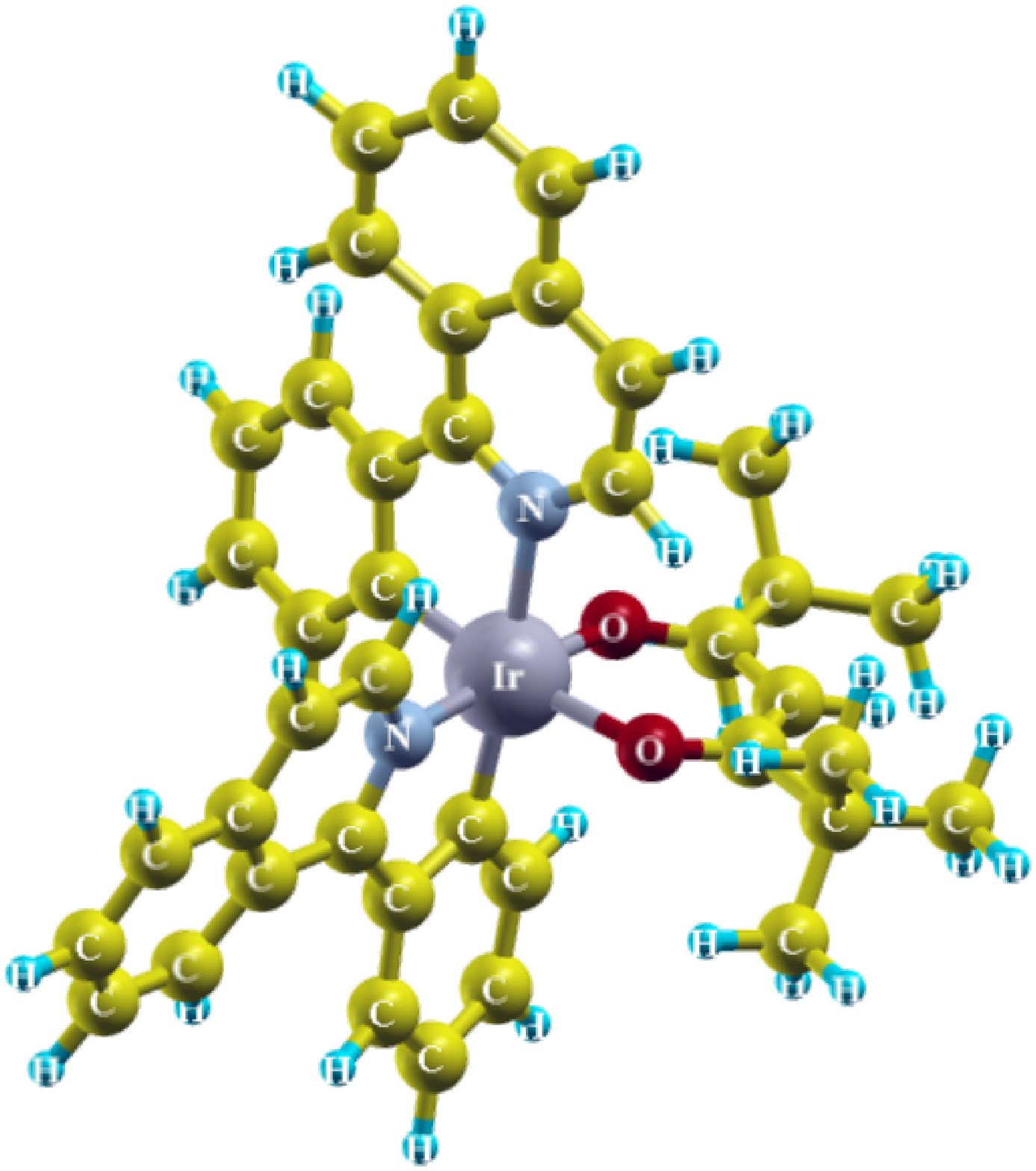}
\end{minipage}
\hspace{0.3cm}
\begin{minipage}[b]{0.45\linewidth}
\centering
\includegraphics[width=1.0\textwidth]{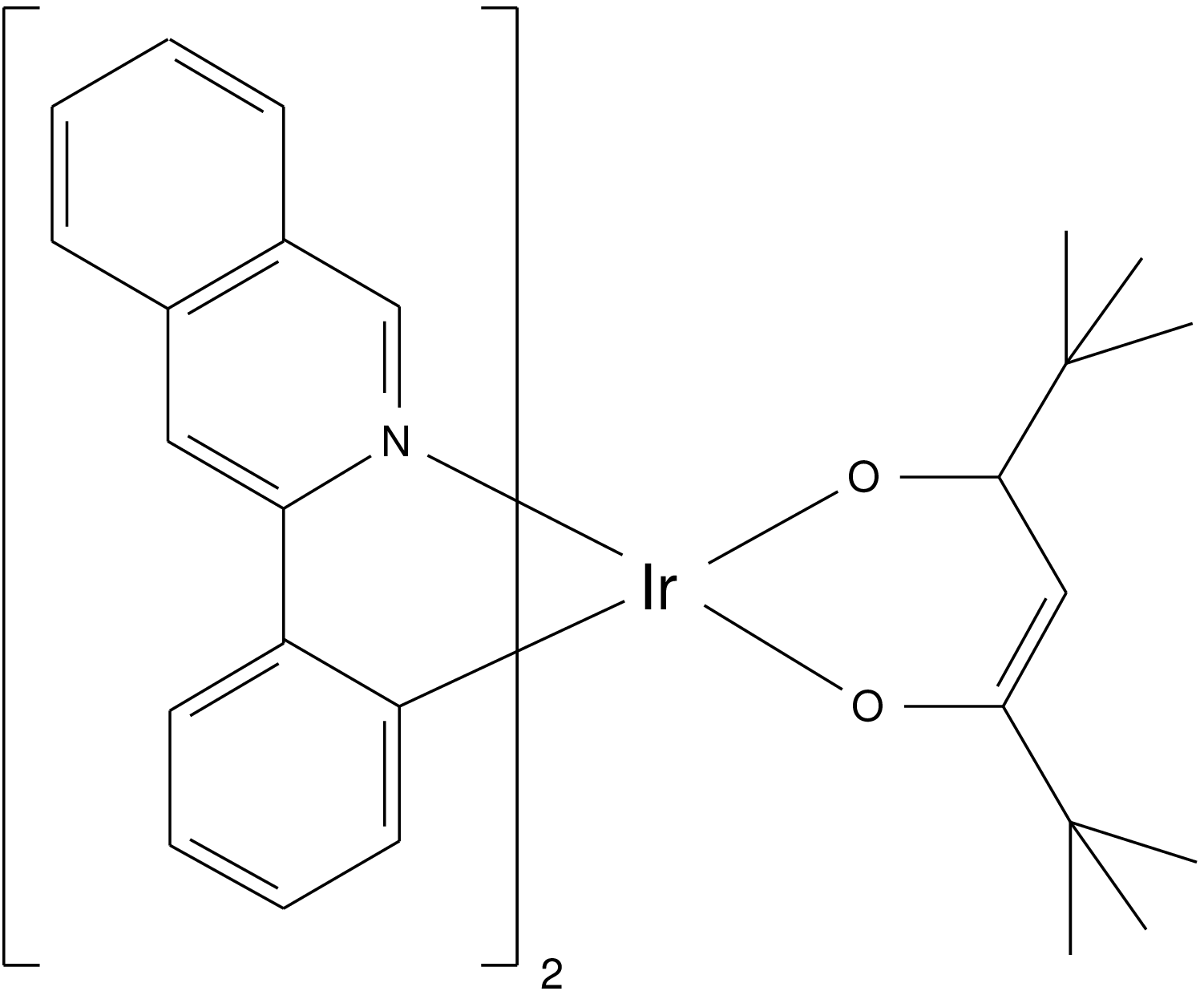}
\end{minipage}
\caption{\label{fig:molecule.3}
(Color online) Molecular structure of PQIr as a 3d model and a chemical formula.}
\end{figure}

The molecular structure of PQIr is shown in Fig.~\ref{fig:molecule.3}.
The lowest triplet and singlet excited-state energies within TDDFT using GGA (PBE), B3LYP,
and M06 are reported in Table.~\ref{tab:tddft-3}.
\begin{table}[!ht]
\caption{\label{tab:tddft-3} Lowest triplet and singlet excited-state energies obtained with TDDFT
and from experiments for PQIr. All values are in eV and measured from the ground state.}
\begin{ruledtabular}
\begin{tabular}{ccccc}
& Exp & GGA & B3LYP & M06 \\
\hline
T & 2.1\footnotemark[1] & 1.63 & 2.02 & 2.01 \\
S & 2.3\footnotemark[2] & 1.74 & 2.35 & 2.37 \\
\end{tabular}
\end{ruledtabular}
\footnotetext[1]{From Refs.~\onlinecite{eom-irppy3,xiao-review,su-pqir} 
from the highest peak in the phosphorescence spectrum.}
\footnotetext[2]{From Ref.~\onlinecite{d2004-pqir,d2004-pqir2,su-pqir}, from the LUMO-HOMO
energy difference, where the LUMO is treated as an ``optical LUMO'' which also contains the 
exciton binding energy~\cite{baldo-optical-lumo,djurovich-optical-lumo} (which is a singlet state).}
\end{table}
As for the other molecules studied in this paper, TDDFT(PBE)
underestimates both the triplet and the singlet energies while the hybrid functionals yield 
results within an error of $0.1\, {\rm eV}$ relative to experiment.
 
\begin{figure}[!ht]
\begin{minipage}[b]{0.49\linewidth}
\centering
\includegraphics[width=1.1\textwidth]{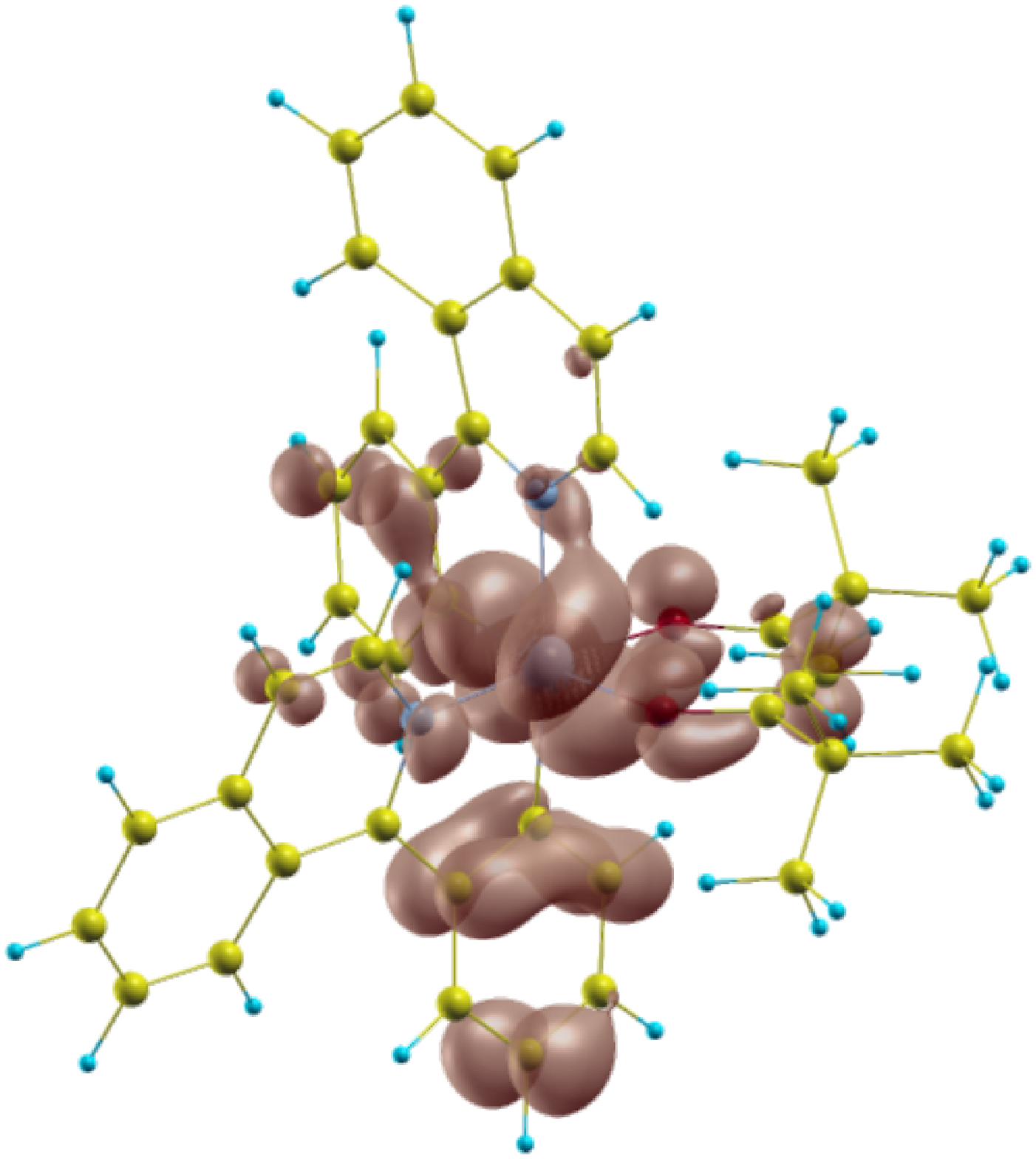}
\end{minipage}
\begin{minipage}[b]{0.49\linewidth}
\centering
\includegraphics[width=1.1\textwidth]{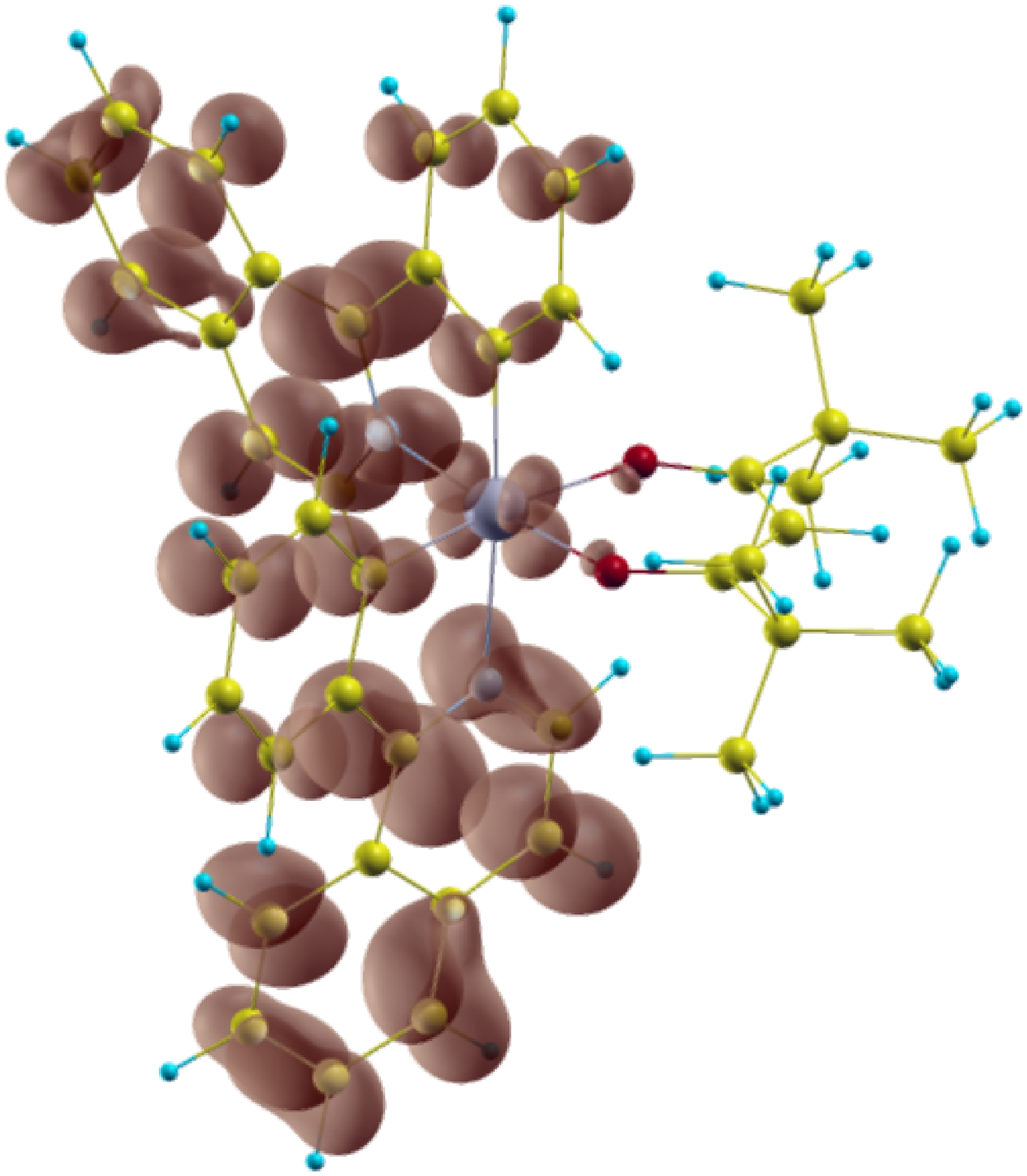}
\end{minipage}
\caption{\label{fig:homo-lumo-pqir}
(Color online) PQIr HOMO(left) and LUMO(right) calculated using the ground-state electronic density.}
\end{figure}

The HOMO and the LUMO charge distributions are depicted in Fig.~\ref{fig:homo-lumo-pqir}
for the ground state within PBE. 
Here, in accordance with the other Ir dyes, 
the HOMO is predominantly centered on the metal, with some contributions
from the ligands. Instead, the LUMO is almost entirely ligand-centered. Notice that the 
tetramethyl heptanedionate (tmd) ligand 
only contributes (marginally) to the HOMO, and the LUMO is concentrated on the larger 
phenyl-quinoline (pq) ligand.  

The calculated $U$ and $V$ parameters are reported in Table~\ref{tab:pqir-uandv}.
\begin{table}[!ht]
\caption{\label{tab:pqir-uandv} Linear-response values of $U$ and $V$ parameters. All values are in eV.}
\begin{ruledtabular}
\begin{tabular}{ccccccc}
 $U_d$  &  $V_{d-p}^N$  &  $V_{d-p}^C$  &  $V_{d-p}^O$
        &  $V_{d-s}^N$  &  $V_{d-s}^C$  &  $V_{d-s}^O$ \\
\hline
  7.26 & 1.71 &  1.39 & 1.79 & 2.66 &  4.43 & 5.40  \\
\end{tabular}
\end{ruledtabular}
\end{table}
As before, we only report interaction parameters that affect the electronic structure most significantly.
The validity of restricting our calculations to Ir and its nearest neighbors was also verified 
for PQIr, by computing $U$ and $V$ parameters between the atoms along several ``chains``
from the central Ir towards the external ligands
and showing that only nearest neighbor interactions
are important. The calculated $U_d$ and $V$ for Ir-N, Ir-C, and Ir-O pairs between $d$ and $p$ 
states are also very close to the ones computed previously for the other two molecules. This finding
suggests that these parameters could be considered to be universal and used for future studies without 
recomputing them for each molecule (provided that the same pseudo-potential and exchange-correlation 
functional are used).

The results of the $\Delta$-SCF calculations
using the GGA, GGA+U, and GGA+U+V functionals are reported in Table~\ref{tab:d-scf-3}. 
\begin{table}[!ht]
\caption{\label{tab:d-scf-3} $\Delta$-SCF calculation of the lowest triplet and singlet states 
for PQIr. All values are in eV and given relative to the ground-state.}
\begin{ruledtabular}
\begin{tabular}{cccccc}
 &  GGA  &  U  &  U+V & (U+V)$_{\rm rel}$ & Exp \\
\hline
T &  1.81  &  2.11  &  1.93 & 1.90 & 2.1\\
M &  1.79  &  2.18  &  1.98 & 1.95 & -- \\
S [NSP] &  2.01  &  2.34  &  2.23 & 2.18 & 2.3 \\
$\Delta\, E_{\rm TS}$ [SPA] & 0.06   &  0.14  &  0.10 & 0.10 & -- \\
$\Delta\, E_{\rm TS}$ [NSP] & 0.19 & 0.23 & 0.30 & 0.28 & 0.2 \\
\end{tabular}
\end{ruledtabular}
\end{table}
Again, GGA yields results in better accordance with experiment within $\Delta$-SCF
than within TDDFT. 
The best agreement with the reported experimental values in Table~\ref{tab:tddft-3}
are obtained with GGA+U, using the NSP calculation of the singlet state. This 
might seem to contradict the results for Ir(ppy)$_3$ and FIrpic, where the 
best agreement was found with the inclusion of $V$. However, it should be 
noted that the experimental results have an accuracy of approximately 
$0.2\, {\rm eV}$~\cite{djurovich-optical-lumo}. Thus, GGA+U+V results are within this
range of this accuracy. 

The excited-state energies of PQIr are smaller than those of Ir(ppy)$_3$ and FIrpic, which accounts
for its red phosphorescent emission. 
This is due to the stabilization of the LUMO level resulting from the larger size of the (pq)
ligand where it is mostly centered on (see Fig.~\ref{fig:homo-lumo-pqir}). In fact the (pq)
ligand has an additional benzene ring compared to Ir(ppy)$_3$. A larger ligand 
structure can be expected to lower the electron-electron interactions and reduce the electronic
kinetic energy through a more pronounced delocalization, hence the stabilization of the 
electronic states it hosts (the LUMO level in this case)~\cite{oled-book}.
\subsection{\label{subsec:discussion-1}Comparison of TDDFT and $\Delta$-SCF}
As discussed briefly in subsection~\ref{subsec:tddft}, TDDFT yields inaccurate energies for 
charge-transfer excitations, specifically if used with local and semilocal functionals, like GGA(PBE). 
As we have shown in 
Figs.~\ref{fig:homo-lumo},~\ref{fig:homo-lumo-firpic} and~\ref{fig:homo-lumo-pqir}, 
the excited states of the Ir dyes studied in this work are of MLCT type. 
The MLCT character of the excited-states explains why GGA underestimates the excited-state
energies, whereas the
hybrid functionals provide better agreement with the experimental measurements, which are reported in 
Tables~\ref{tab:tddft},~\ref{tab:tddft-2}, and~\ref{tab:tddft-3}. 
Instead, PBE within the $\Delta$-SCF approach using the NSP representation of 
the singlet state proves to be more accurate than
TDDFT, as can be seen from Tables~\ref{tab:d-scf},~\ref{tab:d-scf-2}, and~\ref{tab:d-scf-3}. 
Indeed, the fact that $\Delta$-SCF could perform better than TDDFT for charge-transfer and 
Rydberg excitations is already known from the literature
~\cite{tddft-dscf-mix-1,tddft-dscf-mix-2,dscf-better-1,dscf-better-2,dscf-better-3}.
TDDFT relies on KS energies determined
from the ground-state configuration, and in the case of charge-transfer excitations, the effect of 
unoccupied orbitals are not accurately taken into account by semilocal functionals since 
the off-diagonal block matrices (${\bf M}$) nearly vanishes in Eq. (\ref{casida-eqn}). This is 
due to the fact that the exchange-correlation potential decays exponentially with distance $r$,
rather than as ${\rm O}(r^{-1})$. 
Instead, $\Delta$-SCF takes into account the orbital
relaxation of ground-state KS states, since those are reconstructed at each iteration 
from an excited-state density [Eq. (\ref{rho-ex})]. 
If expanded using the basis of unrelaxed orbitals, the relaxed orbitals include components
from the unoccupied manifolds that are not properly taken into account by TDDFT used with 
(semi)local functionals. As a result, the effect of the 
unoccupied manifold (with respect to ground-state KS orbitals) is partially included in $\Delta$-SCF.
Along the same lines, Ref.~\onlinecite{tddft-dscf-mix-1} has shown 
that the long-range behavior of the $\Delta$-SCF energies can be more accurate than TDDFT.

Comparing Tables~\ref{tab:tddft},~\ref{tab:tddft-2},~\ref{tab:tddft-3} 
with Tables~\ref{tab:d-scf},~\ref{tab:d-scf-2},~\ref{tab:d-scf-3} one can observe that hybrid functionals
within TDDFT performs similarly to $\Delta$-SCF calculations with Hubbard corrections.
This shows that both methods can be used interchangeably. In explicit terms, one can use $\Delta$-SCF
with hybrid functionals or TDDFT with Hubbard U and V corrections, where the latter approach 
would allow for significantly less computational effort for large systems than 
hybrid functionals.
One other notable advantage of the Hubbard corrected functionals is that the $U$ and $V$ parameters directly control
the amount of electronic localization on the Ir atom and the hybridization between Ir and the surrounding ligands.
This knowledge could provide a functionalization strategy for tuning the excited-state energy levels,
as discussed briefly in the next subsection.

\subsection{\label{subsec:discussion-2}Effects of $U$ and $V$}
\begin{figure}[!ht]
\includegraphics[width=0.45\textwidth]{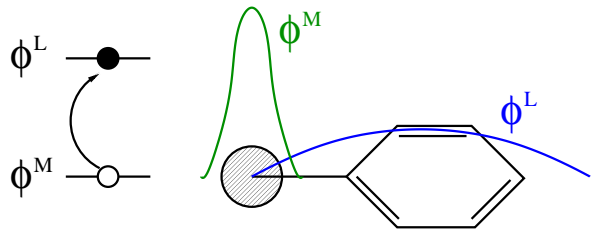}
\caption{\label{fig:MLCT} A simple description of the MLCT character of the excited states
in the Ir complexes.}
\end{figure}

\begin{figure}[!ht]
\includegraphics[width=0.4\textwidth]{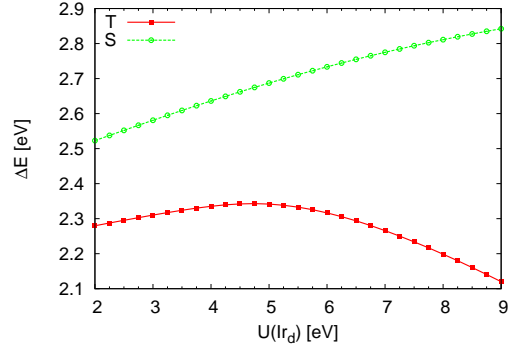}
\caption{\label{fig:d_vs_u} Triplet and singlet energies computed as a function of on-site $U$ 
on Ir $d$ states. No inter-site $V$ is included.}
\end{figure}

In order to investigate the effects of the Hubbard corrections more systematically, we provide a simple model
of the system, based on two atomic states, one centered on the metal and the other centered on the
ligand, as shown schematically in Fig.~\ref{fig:MLCT}.
In this simplified picture, we assume that the HOMO levels $\psi_N^{\sigma}$ have large overlap
with the metal orbital $\phi^M$, and a small overlap with the ligand orbital $\phi^L$. Instead,
the LUMO levels have large overlap with the ligand orbital $\phi^L$ but a small overlap with
the metal orbital $\phi^M$. The contribution of the Hubbard corrective energy in Eq. (\ref{euv}) in
the case of two atomic orbitals $\phi^M$ and $\phi^L$ (which are assumed non-degenerate) becomes
\begin{equation}
E_{\rm UV} = \frac{U_M}{2}\, \sum_{\sigma}\, n^{M\sigma}\, \left( 1 - n^{M\sigma} \right) 
              - V^{ML}\, \sum_{\sigma}\, n^{ML\sigma}\, n^{LM\sigma} \label{euv-2}
\end{equation}
where
\begin{eqnarray}
n^{M\sigma} &=& \sum_i\, f_i^{\sigma}\, \vert \langle \phi^M \vert \psi_i^{\sigma} \rangle \vert^2 \nonumber\\
n^{ML\sigma} &=& \sum_i\, f_i^{\sigma}\, \langle \phi^M \vert \psi_i^{\sigma} \rangle
                \langle \psi_i^{\sigma} \vert \phi^L \rangle \label{nML}
\end{eqnarray}
In the above equations, the occupation numbers $f_i^{\sigma}$ are either 1 or 0. We also assume 
that the occupation matrices are symmetric, i.e., $n^{ML\sigma}=n^{LM\sigma}$.
We can construct the occupation matrices $n^M, \, n^L, \, n^{ML}$ by ignoring the orbital 
relaxation and using the levels schematically represented in Fig.~\ref{fig:levels}.
For example, in the triplet state, the occupation matrices are given by
\begin{eqnarray}
n^{M\uparrow}_{\rm T} = n^{M\, \uparrow}_{\rm G} + \Delta\, n_{N+1}^{\uparrow} \,\,\, , \,\,\,
n^{M\downarrow}_{\rm T} = n^{M\downarrow}_{\rm G} - \Delta\, n_{N}^{\downarrow} \nonumber\\
n^{ML \uparrow}_{\rm T} = n^{ML \uparrow}_{\rm G} + \Delta\, n_{N+1}^{ML\uparrow} \,\,\, , \,\,\,
n^{ML \downarrow}_{\rm T} = n^{ML \downarrow}_{\rm G} - \Delta\, n_{N}^{ML\downarrow} \nonumber\\
\label{nML-2}
\end{eqnarray}
where
\begin{eqnarray}
\Delta n_{N+1}^{\uparrow} &\equiv& \vert \langle \phi^M \vert \psi_{N+1}^{\uparrow} \rangle \vert^2 \,\,\, , \,\,\,
\Delta n_N^{\downarrow} \equiv \vert \langle \phi^M \vert \psi_N^{\downarrow} \rangle \vert^2 \nonumber\\
\Delta n_{N+1}^{ML\uparrow} &\equiv& \langle \phi^M \vert \psi_{N+1}^{\uparrow} \rangle\, 
                                   \langle \psi_{N+1}^{\uparrow} \vert \phi^L \rangle \nonumber\\
\Delta n_{N}^{ML\downarrow} &\equiv& \langle \phi^M \vert \psi_{N}^{\downarrow} \rangle\, 
                                   \langle \psi_{N}^{\downarrow} \vert \phi^L \rangle \label{dnML}
\end{eqnarray}
Since $\psi_N^{\sigma}$ is dominantly metal-centered, while $\psi_{N+1}^{\sigma}$ is ligand centered,
$\Delta n_{N+1}^{\uparrow} \ll 1$ is satisfied, whereas $\Delta n_{N+1}^{ML\uparrow}$ and 
$\Delta n_{N}^{ML\downarrow}$ can in principle be larger, but still considerably smaller than 1. Instead, 
$\Delta n_N^{\downarrow}$ is of the order of 1. 
In Eq.(\ref{nML-2}), the subscripts G and T refers to occupation matrices evaluated from the ground-state 
and triplet-state, respectively.
We can calculate the contribution to the excited-state energies from 
Hubbard terms as $E_{UV}^{\rm T} - E_{UV}^{\rm G}$ by using 
Eqs. (\ref{euv-2})-(\ref{dnML}). The contribution coming from the on-site interactions
[i.e., the first term in Eq. (\ref{euv-2})]  is given by
\begin{eqnarray}
E_{U}^{\rm T} - E_{U}^{\rm G} \equiv \omega^U_{\rm T} &=& -\frac{U_M}{2}\, \Delta n_N^{\downarrow}\, 
\left( 1+\Delta n_N^{\downarrow} - n^{M}_{\rm G} \right) \nonumber\\
&+& \frac{U_M}{2}\, \Delta n_{N+1}^{\uparrow}\, \left( 1 - n_{\rm G}^M \right) 
+ {\rm O}\left( \Delta n_{N+1}^2 \right) \nonumber\\ \label{omega-U}
\end{eqnarray}
where we have ignored quadratic terms in $\Delta n_{N+1}^{\uparrow}$ and used the fact that
$n_{\rm G}^{\uparrow} = n_{\rm G}^{\downarrow} = \frac{1}{2}\, n_{\rm G}$. The first term in Eq. (\ref{omega-U})
is negative, since $1+\Delta_N^{\downarrow} > 1$, while $n_{\rm G}^{M} < 1$. Instead, the second term 
is positive. Therefore, Eq. (\ref{omega-U}) shows that the excited-state energy decreases with increasing
$U_M$, which is expected since larger $U_M$ destabilizes the HOMO level by adding larger penalty when it is 
doubly occupied in the ground-state. However, for cases where there is considerable metal and ligand overlap,
the second term in Eq. (\ref{omega-U}) can reverse this effect through a larger $\Delta n_{N+1}^{\uparrow}$.
To test this prediction, 
we have plotted in Fig.~\ref{fig:d_vs_u} the energies of the triplet and 
singlet state (obtained from a NSP calculation) as a function of $U_d$ for Ir(ppy)$_3$. 
As can be seen from Fig.~\ref{fig:d_vs_u}, the triplet energy increases for $U_M \le 5\, {\rm eV}$, but 
decreases for larger values of $U_M$. This decreasing behavior is correctly predicted by Eq. (\ref{omega-U}).
Instead, the increase in the triplet energy for $U_M \le 5\, {\rm eV}$ is probably a result of important 
modifications in the KS states due to orbital relaxation which is implicitly ignored in deriving Eq. (\ref{omega-U}).
For the case of
the molecules studied in this work, $U_d \ga 7\, {\rm eV}$. Therefore an increasing $U_d$ would
decrease the triplet energy, mainly due to the destabilization of the HOMO level.   
The singlet state is a result of a NSP calculation, 
and its behavior as a function of $U_M$ is thus difficult to assess since in principle it 
corresponds to an ensemble which contains an arbitrary combination of possible 
different Slater determinants, as discussed in the Appendix.

\begin{figure}[!ht]
\includegraphics[width=0.4\textwidth]{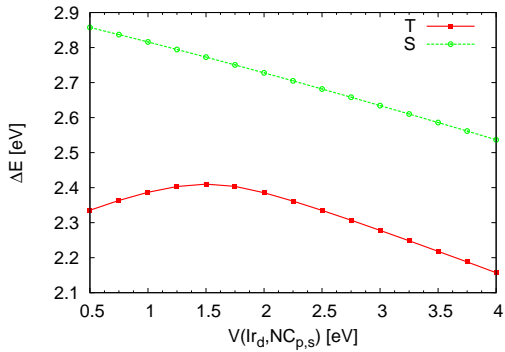}
\caption{\label{fig:d_vs_v} Triplet and singlet energies computed as a function of inter-site $V$ 
between Ir $d$ and C,N $s,p$ states. $U=7.36\, {\rm eV}$ fixed at the computed value.}
\end{figure}

The contribution from the inter-site interaction terms in Eq. (\ref{euv-2}) to the excited-state energy
can be calculated similarly. Doing so, we obtain
\begin{eqnarray}
E_{V}^{\rm T} - E_{V}^{\rm G} \equiv \omega_{\rm T}^V &=& -V^{ML}\, n_{\rm G}^{ML}\, 
\left( \Delta n^{LM\uparrow}_{N+1} - \Delta n^{ML\downarrow}_{N} \right) \nonumber\\
&& \qquad\qquad + {\rm O}\left( \Delta^2 \right)
\label{omega-V}
\end{eqnarray}
where we have ignored the quadratic terms in $\Delta n_N$ and $\Delta n_{N+1}$. 
Note that the term given in Eq. (\ref{omega-V})
could be either positive or negative depending on the relative magnitude of the projections
$\Delta n^{LM\uparrow}_{N+1}$ and $\Delta n^{ML\downarrow}_{N}$. For example, in the case of 
a strong metal-LUMO overlap but a smaller ligand-HOMO overlap, $V$ decreases the triplet energy.
Instead, in the opposite case where the ligand-HOMO overlap is larger than the metal-LUMO overlap,
$V$ increases the triplet energy. In order to validate these predictions, we have plotted the triplet
and singlet (from NSP calculation) state energies as a function of the $V$ between Ir $d$ and
C and N $p$ and $s$ states in Fig.~\ref{fig:d_vs_v}. 
In these calculations, we have fixed the value of $U_d$ to the calculated value of $7.36\, {\rm eV}$
and used the same $V$ between all the Ir-C and Ir-N pair interactions.
As can be seen from Fig.~\ref{fig:d_vs_v}, the triplet energy increases for $V \le 1.5\, {\rm eV}$
and decreases for $V \ge 1.5\, {\rm eV}$, reflecting both type of behaviors shown in Eq. (\ref{omega-V}).
In this case, the orbital relaxation effects, which are ignored in obtaining Eq. (\ref{omega-V}) are more
critical since they affect the overlap of the HOMO and LUMO levels ($\psi_N^{\sigma}$, $\psi_{N+1}^{\sigma}$)
with the metal-and-ligand-centered states. This contribution can very easily invert the behavior 
of the triplet energy as a function 
of $V$. In summary, Figs.~\ref{fig:d_vs_u} and~\ref{fig:d_vs_v} show that the excited-state
energies critically depend on the $U$ and $V$ values, which highlights the sensitivity of these
energies to small
chemical changes. 
In fact, as discussed in previous sections, a change in the chemical composition of the ligand 
groups can result in modifications to the electronic structure of these molecules analogous 
to those obtained by varying $U$ and $V$.
This conclusion was also obtained in a different way by using model Hamiltonians in
Refs.~\onlinecite{model-1} and~\onlinecite{model-2}. 
In fact, the results presented in Tables~\ref{tab:irppy3-uandv},~\ref{tab:firpic-uandv}
and~\ref{tab:pqir-uandv} can be used as effective parameters in a model-Hamiltonian study for 
these systems.

Summarizing, for the three molecules that we have studied in this work, two distinct behaviors are
observed. For Ir(ppy)$_3$, the effect of the $U$ and $V$ corrections is mainly to tune the HOMO level. 
The addition of $U$ decreases the excited-state energies 
(see Table~\ref{tab:d-scf})  due to the destabilization of the HOMO level, which is dominantly metal-centered. 
This destabilization arises from an increased Coulomb repulsion between the two electrons
occupying the HOMO level. Instead, with the addition of $V$, the HOMO level is stabilized, increasing
the excitation energies. This is due to the increased hybridization between Ir and neighboring atoms,
which depletes the Ir $d$ states, lowering the Coulomb repulsion on them.
In contrast, for FIrpic and PQIr, the addition of $U$ and $V$ affects mainly the LUMO level. Including $U$ increases
the excited-state energies (see Tables~\ref{tab:d-scf-2} and~\ref{tab:d-scf-3}) due to the 
destabilization of the LUMO level. This trend can be ascribed to the fact that the LUMO 
level contains a non-negligible contribution 
from the metal, as compared to Ir(ppy)$_3$ (compare Fig.~\ref{fig:homo-lumo} with 
Figs.~\ref{fig:homo-lumo-firpic} and~\ref{fig:homo-lumo-pqir}). Instead, the addition of $V$ stabilizes the
LUMO by lowering the fraction of electrons localized on the metal-center. 
In a nutshell, this study 
suggests that the substitution of atoms close to the Ir center can be as effective in the functionalization of 
the complex as the well known addition of F or other electronegative species to the external part of the 
ligands or modifications of their structure.
%

\section{\label{sec:conclusion}Conclusion}
In this work, we have studied the electronic structure of the lowest triplet and singlet excited-states of three
representative iridium dyes. The calculation of the excited-state energies were performed 
using TDDFT with hybrid functionals and $\Delta$-SCF with Hubbard corrections namely,
GGA+U and GGA+U+V. The results obtained in both approaches are in good agreement with experiment.
The Hubbard corrections $U$ and $V$ were computed from ab-initio calculations and provide
a measure for localization and hybridization of Ir $d$ states with neighboring organic ligands in the ground and 
excited-states. This knowledge is used to infer possible strategies for tuning the excited-state 
energies of the studied molecules. The gained insight underscores the interest of Hubbard corrections in 
unveiling the electronic origin of dye phosphorescence.
In addition, we have also investigated the validity of 
the spin purification (SPA) and the nonspin-polarized (NSP) calculations
for the computation of the singlet excited-state energy.
We found that the SPA approach (based on the construction of a state with mixed spin symmetry) clearly
underestimates the experimental singlet energies. Instead, the NSP calculation yields much more
accurate results. While the failure of the spin purification formula can be understood from the 
inadequacies of the conventional exchange-correlation functionals, the remarkable success of the NSP approach in
capturing the singlet remains a relevant open question. 

The present work can be considered as the starting point of two main research directions 
involving methodological developments on one side and the design and optimization of better 
molecular complexes, on the other.
From a methodological point of view this study highlights the necessity to 
improve functionals that can distinguish between different $\langle \vec{S}^2 \rangle$
configurations (with the inclusion of spin-orbit interactions).  
On the application side this work establishes DFT+U+V as a valid computational tool 
(thanks to its ability to capture the charge-transfer character of excited states) 
to efficiently screen useful modification to the ligand structure of the considered 
molecules that could improve their performance. Furthermore, the dependence of the 
energy splittings on the values of U and, more importantly, of V, suggests that the 
choice of the chemical species directly bonded with the Ir center and of the structure 
of the ligands in its coordination shell may represent another valuable route to 
functionalization that can be explored in addition to the modification of their most external part.

\begin{acknowledgments}
We thank Minnesota Supercomputing Institute (MSI) for providing the computational
resources used in this work. This work is supported by the Seed grant provided by 
Materials Research Science and Engineering Center (MRSEC) of the University of Minnesota.
M.C also acknowledges partial support from NSF EAR 0810272 and from the NSF CAREER award 
DMR 1151738.
\end{acknowledgments}

\appendix
\section{\label{sec:app} Comparison of spin-purification and NSP calculations}
In section~\ref{sec:results}, we have shown that the $\Delta$-SCF energy for the mixed spin states $\Phi_{\rm M1,2}$ 
used in the SPA formula Eq. (\ref{pur}) yields much lower singlet state energies
than the experimental results. Instead, the NSP $\Delta$-SCF calculation yields results
within the experimental accuracy of $0.2\, {\rm eV}$. In this Appendix,
we provide some insight into this finding. 

Approximate exchange-correlation functionals depend only on the spin component $m_z$, 
and cannot distinguish between $m_z=0$ triplet and singlet states. 
Moreover, with approximate exchange-correlation functionals
and integer occupations, only configurations that can be represented by a single Slater determinant are
obtained as stationary densities in $\Delta$-SCF calculations. 
Instead, the $m_z=0$ triplet and singlet states correspond to multi-Slater-determinant configurations, 
represented by the following density matrices (excluding orbital relaxation effects)
\begin{eqnarray}
&& {\hat D}_{1,0} = \frac{1}{2}\, \Big[ \vert \Phi_{\rm M1} \rangle \langle \Phi_{\rm M1} \vert - 
   \vert \Phi_{\rm M1} \rangle \langle \Phi_{\rm M2} \vert  \nonumber\\ 
&& \qquad\qquad\quad -\vert \Phi_{\rm M2} \rangle \langle \Phi_{\rm M1} \vert +
          \vert \Phi_{\rm M2} \rangle \langle \Phi_{\rm M2} \vert \Big] \label{TR} \\
&& {\hat D}_{0,0} = \frac{1}{2}\, \Big[ \vert \Phi_{\rm M1} \rangle \langle \Phi_{\rm M1} \vert +
   \vert \Phi_{\rm M1} \rangle \langle \Phi_{\rm M2} \vert \nonumber\\ 
&& \qquad\qquad\quad + \vert \Phi_{\rm M2} \rangle \langle \Phi_{\rm M1} \vert +
          \vert \Phi_{\rm M2} \rangle \langle \Phi_{\rm M2} \vert \Big] \label{SG}
\end{eqnarray}
where ${\hat D}_{1,0}$ corresponds to the density matrix of the $m_z=0$ triplet, while
${\hat D}_{0,0}$ corresponds to the density matrix of the singlet state.
Eqs. (\ref{TR}) and (\ref{SG}) can easily be verified by evaluating the expectation value of the 
total spin square operator, and making use of Eq. (\ref{S2onPhi}) as
\begin{eqnarray}
&& \langle {\vec S}^2 \rangle_{1,0} = {\rm Tr}\left[ {\hat D}_{1,0}\, {\vec S}^2 \right] = 2 \nonumber\\
&& \langle {\vec S}^2 \rangle_{0,0} = {\rm Tr}\left[ {\hat D}_{0,0}\, {\vec S}^2 \right] = 0. 
\label{S2onD}
\end{eqnarray}
Instead, the single-particle density is the same for both states, and is given by
\begin{eqnarray}
&& n({\bf r}) = {\rm Tr}\left[ {\hat D} {\hat n}({\bf r}) \right] \nonumber\\ 
&& \qquad = \frac{1}{2}\, \langle \Phi_{\rm M1} \vert {\hat n}({\bf r}) \vert \Phi_{\rm M1} \rangle + 
   \frac{1}{2}\, \langle \Phi_{\rm M2} \vert {\hat n}({\bf r}) \vert \Phi_{\rm M2} \rangle . \nonumber\\
\label{nD}
\end{eqnarray}
Notice that the off-diagonal terms $\langle \Phi_{\rm M1} \vert {\hat n}({\bf r}) \vert \Phi_{\rm M2} \rangle$
vanish when orbital relaxation effects are ignored (i.e. the states $\Phi_{\rm M1,2}$ are fixed).
With the inclusion of orbital relaxation effects, the densities of $m_z=0$ triplet and singlet states
would be different. However, realization of this difference requires an ensemble
dependent functional~\cite{edft-1,edft-2,edft-3,kohn-edft}. More specifically, the ensemble 
exchange-correlation functional should depend on the weights of the Slater determinants appearing
in the density matrices Eq. (\ref{TR}) and (\ref{SG}). In fact, any density matrix of the form 
$q\, {\hat D}_{1,0} + (1-q)\, {\hat D}_{0,0}$ with $q\le 1$, results precisely in the same
one particle density Eq. (\ref{nD}), with $\langle {\vec S}^2 \rangle = 2\, q$. Without 
a functional that depends on the ensemble weights (i.e., $q$ in this case) or a functional
that can distinguish between states with different ${\vec S}^2$, it is not possible to know
which ensemble $n({\bf r})$ corresponds to. In the absence of such a functional, 
one relies on the calculation of the expectation value of the many-body Hamiltonian operator
using DFT with standard exchange-correlation functionals, which are given by
\begin{eqnarray}
&& E_{1,0} = {\rm Tr}\left[ {\hat D}_{1,0}\, {\cal H} \right] \nonumber\\
&& \qquad = \frac{1}{2}\, \langle \Phi_{\rm M1} \vert {\cal H} \vert \Phi_{\rm M1} \rangle +
           \frac{1}{2}\, \langle \Phi_{\rm M2} \vert {\cal H} \vert \Phi_{\rm M2} \rangle \nonumber\\
&& \qquad\quad - \frac{1}{2}\, \left[ \langle \Phi_{\rm M1} \vert {\cal H} \vert \Phi_{\rm M2} \rangle + {\rm c.c.} \right]
\nonumber\\
&& E_{0,0} = {\rm Tr}\left[ {\hat D}_{0,0}\, {\cal H} \right] \nonumber\\
&& \qquad = \frac{1}{2}\, \langle \Phi_{\rm M1} \vert {\cal H} \vert \Phi_{\rm M1} \rangle +
           \frac{1}{2}\, \langle \Phi_{\rm M2} \vert {\cal H} \vert \Phi_{\rm M2} \rangle \nonumber\\
&& \qquad\quad + \frac{1}{2}\, \left[ \langle \Phi_{\rm M1} \vert {\cal H} \vert \Phi_{\rm M2} \rangle + {\rm c.c.} \right]
\nonumber\\ \label{DH}
\end{eqnarray}
Assuming that the expectation value of the many-body Hamiltonian can approximately be given 
by $\Delta$-SCF calculations, one can identify the mixed-state energy as the variational extrema given by 	
\begin{equation}
\langle \Phi_{\rm M1,2} \vert {\cal H} \vert \Phi_{\rm M1,2} \rangle = E_{\rm M}. 
\label{EM}
\end{equation}
Since the Hamiltonian does not contain any external magnetic fields, $m_z=1$ and
$m_z=0$ triplet states should have the same energy, which requires
\begin{equation}
\langle \Phi_{\rm T} \vert {\cal H} \vert \Phi_{\rm T} \rangle = E_{\rm T} = E_{1,0} \label{ET}
\end{equation}
Then, adding the two equations in Eq. (\ref{DH}) to cancel the off-diagonal terms, 
and using Eqs. (\ref{ET}) and (\ref{EM}), one obtains the spin purification formula given in 
Eq. (\ref{pur}): $E_{\rm T} + E_{\rm S} = 2 E_{\rm M}$ 
(with $E_{\rm S} = E_{0,0}$)~\cite{spin-pur-1,spin-pur-2,spin-pur-3}. 

The reason why the $\Delta$-SCF energy of the mixed-state ($E_{\rm M}$), used in spin-purification formula 
Eq. (\ref{pur}), underestimates the experimental singlet energies can be understood, in part, by 
carefully examining the assumptions that were made in the discussion above. First, the assumption
in Eq. (\ref{ET}) is not justified. Approximate ground-state exchange-correlation functionals 
can distinguish between states with different $m_z$, but not with different $\langle {\vec S}^2 \rangle$,
as discussed previously. 
Therefore, Eq. (\ref{ET}) assumes the equality of energies of 
a single determinant state $\Phi_{\rm T}$ and an ensemble ${\hat D}_{1,0}$, which have different $m_z$. 
In other words, $\langle \Phi_{\rm T} \vert {\cal H} \vert \Phi_{\rm T} \rangle$ is not a good estimate
of $E_{\rm T}$.
Such an identification is clearly beyond the capabilities of approximate 
exchange-correlation functionals. Eq. (\ref{ET}) can be justified only if an appropriate 
ensemble exchange-correlation functional is used. Second, the 
spin purification formula is plagued by the negligence of orbital relaxation effects. 
The ensemble density matrices of the singlet and $m_z=0$ triplet states of Eqs. (\ref{TR}) and (\ref{SG})
suggests that they are equal-weight combinations of the mixed-states $\Phi_{\rm M1}$ and $\Phi_{\rm M2}$.
The spin purification formula critically depends on this equivalence of the weights.
However, when orbital relaxation effects are included, KS orbitals corresponding to $\Phi_{\rm M1}$ are $\Phi_{\rm M2}$
are different and, in general, the transformation between them requires an infinite expansion, 
including infinite set of states in the unoccupied manifolds. Thus, an actual ensemble 
representation of the singlet and $m_z=0$ triplet-states should contain contribution from such an 
infinite expansion (e.g. on unrelaxed KS orbitals). 
Namely, the ensembles must include a linear combination of 
many excited-state determinants with orbitals $\psi_a^{\sigma}$ occupied where $a > N+1$. Such 
an approach is not viable, since constructing a functional that depends on the weights of each
excited Slater determinant is not easy. 
At this point we would like to stress that ensemble 
DFT of Refs.~\onlinecite{edft-1,edft-2,edft-3,kohn-edft}, does not suffer from such problems, provided
that an ensemble dependent exchange-correlation functional exists. In this case,
the energy of each state in the ensemble is uniquely determined by the Hohenberg-Kohn
theorem for ensembles~\cite{edft-1}. 

\begin{figure}[!ht]
\begin{minipage}[b]{0.4\linewidth}
\centering
\includegraphics[width=0.8\textwidth]{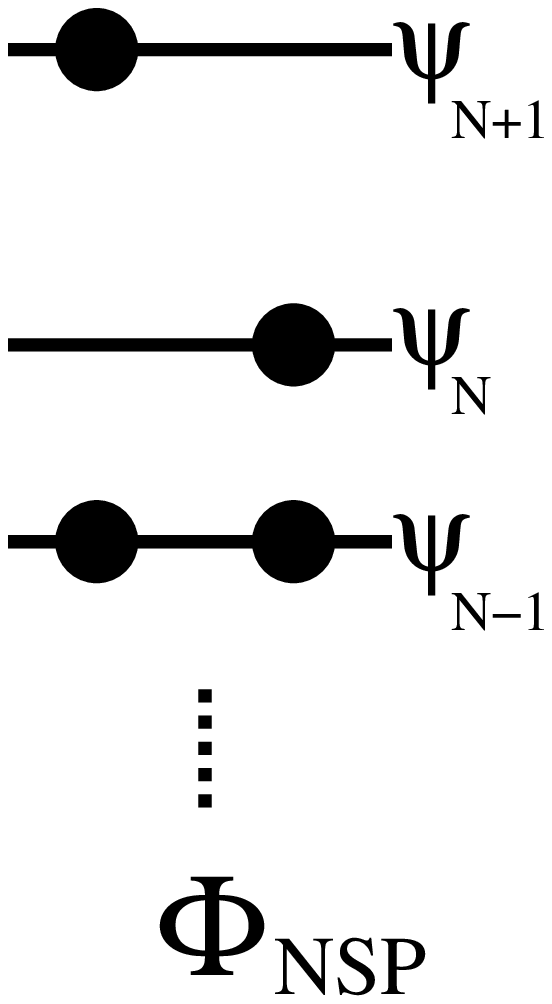}
\end{minipage}
\hspace{0.2cm}
\begin{minipage}[b]{0.4\linewidth}
\centering
\includegraphics[width=0.8\textwidth]{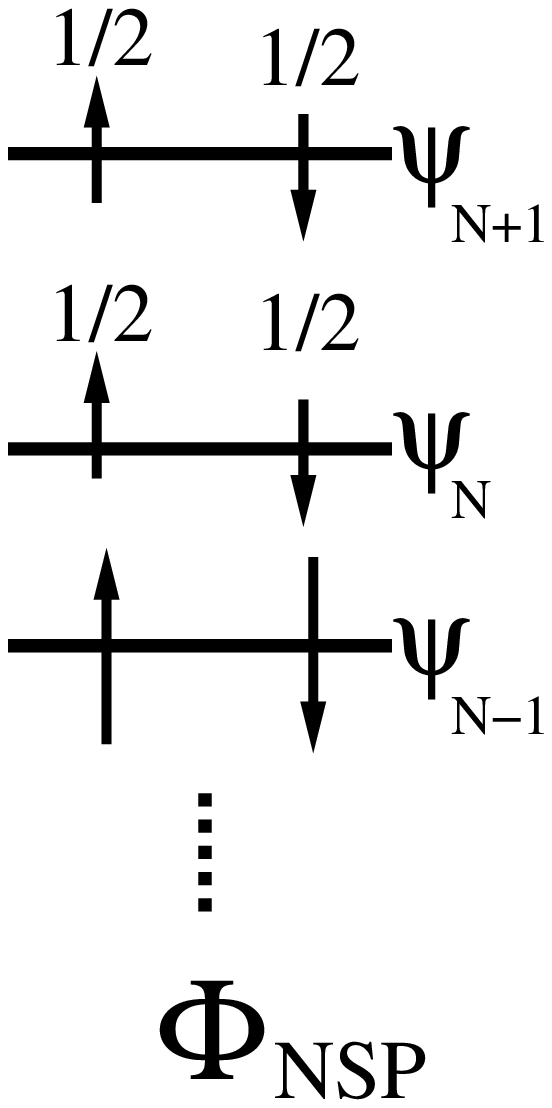}
\end{minipage}
\caption{\label{fig:NSP}
$\Phi_{\rm NSP}$ schematically represented both by spin-unpolarized orbitals and an 
equivalent transition state.}
\end{figure}

\begin{figure}[!ht]
\includegraphics[width=0.15\textwidth]{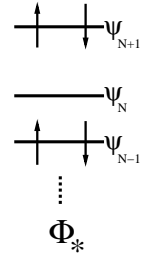}
\caption{\label{fig:Phist} Doubly excited-state that can participate in $n_{\rm NSP}({\bf r})$.}
\end{figure}

The success of the NSP $\Delta$-SCF calculation is more difficult to understand. 
The NSP state used for the $\Delta$-SCF calculation, $\Phi_{\rm NSP}$, 
exactly corresponds to a transition state with orbitals $\psi_N$ and $\psi_{N+1}$
occupied by half spin-up and half-spin down electrons, as shown schematically in 
Fig.~\ref{fig:NSP}.
It is well-known that the single particle density of a 
transition state~\cite{slater} is equivalent to an ensemble density
~\cite{gross-book,edft-2}. Indeed, the corresponding single particle density obtained from
$\Phi_{\rm NSP}$ is given by
\begin{eqnarray}
&& n_{\rm NSP}({\bf r}) = \sum_{i=1, \sigma}^{N-1}\, \vert \psi_i^{\sigma}({\bf r})\vert^2 
+ \frac{1}{2}\, \Big[ \vert \psi_N^{\uparrow}({\bf r}) \vert^2 \nonumber\\
&& \qquad\qquad\qquad         + \vert \psi_N^{\downarrow}({\bf r}) \vert^2 +
                                \vert \psi_{N+1}^{\uparrow}({\bf r}) \vert^2 +
                                \vert \psi_{N+1}^{\downarrow}({\bf r}) \vert^2 \Big] 
\nonumber\\ \label{nNSP}
\end{eqnarray}
When orbital relaxation effects are taken into account, Eq. (\ref{nNSP}) still holds,
since spin-up and spin-down orbitals are always equivalent due to magnetization density
being identically zero (so that the exchange-correlation functional is spin non-polarized).
Moreover, Eq. (\ref{nNSP}) is identical to Eq. (\ref{nD}), i.e.,  the densities of the 
$m_z=0$ triplet and singlet states, when orbital relaxation effects are ignored. 
Since $n_{\rm NSP}({\bf r})$ already provides a direct calculation of an ensemble
density, unlike the spin-purification formula which relies on a spurious 
mixed-state $\Phi_{\rm M1,2}$, it can be expected that $n_{\rm NSP}({\bf r})$ could provide
more accurate results. However, $n_{\rm NSP}({\bf r})$ could in principle be a linear 
combination of other types of excited single determinant states. For instance, 
consider a doubly excited state $\Phi_{*}$, which is schematically represented in 
Fig.~\ref{fig:Phist}.
The single-particle density $n_{\rm NSP}({\bf r})$ can be obtained from the following 
ensemble density matrix
\begin{equation}
{\hat D}_{*} = \frac{1}{2}\, \vert \Phi_{\rm G} \rangle \langle \Phi_{\rm G} \vert 
  + \frac{1}{2}\, \vert \Phi_{*} \rangle \langle \Phi_{*} \vert \label{Dst}
\end{equation}
More generally, $n_{\rm NSP}({\bf r})$ can be obtained from an ensemble of the form
\begin{equation}
{\hat D} = \frac{1-q}{2}\, {\hat D}_{1,0} + \frac{q}{2}\, {\hat D}_{0,0}
 + \frac{1}{4}\, \vert \Phi_{\rm G} \rangle \langle \Phi_{\rm G} \vert 
 + \frac{1}{4}\, \vert \Phi_{*} \rangle \langle \Phi_{*} \vert \label{Dst2}
\end{equation}
where $q \le 1$.
The reason why $n_{\rm NSP}({\bf r})$ results in better singlet energies
could be related to the fact that it corresponds to an ensemble Eq. (\ref{Dst2}),
rather than a single determinant description of mixed spin states.
Moreover, it is the only variational minimum that one can obtain using approximate
ground-state exchange-correlation functionals which has this property.
Due to the variational principle, one could expect that contributions
from $\Phi_{*}$ to be suppressed, since it is not the lowest energy configuration
with vanishing spin polarization. 
However, it is not possible to have control over the ensemble weights
appearing in Eq. (\ref{Dst2}) with approximate ground-state exchange-correlation 
functionals. Indeed, hints of this problem can be seen in Fig.~\ref{fig:d_vs_u}
where the singlet energy was found to increase with $U_d$ almost linearly.
One expects that larger $U_d$ to destabilize the HOMO level, leading to a
decreasing singlet energy. Due to the same reason, the ground-state energy should increase with $U_d$. Thus,
an increasing singlet energy in Fig.~\ref{fig:d_vs_u} is an indication that the NSP state contains 
ground-state contributions.
The arbitrariness in the ensembles that represent 
$n_{\rm NSP}({\bf r})$, limits the understanding of the NSP $\Delta$-SCF calculation.
Therefore, the use of $n_{\rm NSP}({\bf r})$ is only partially justified, and 
mostly motivated by its empirical success presented in section~\ref{sec:results} and previous
studies in the literature~\cite{nsp-o2-1,nsp-o2-2,nsp-o2-3}. 

\providecommand{\noopsort}[1]{}\providecommand{\singleletter}[1]{#1}%


\begin{thebibliography}{112}
\expandafter\ifx\csname natexlab\endcsname\relax\def\natexlab#1{#1}\fi
\expandafter\ifx\csname bibnamefont\endcsname\relax
  \def\bibnamefont#1{#1}\fi
\expandafter\ifx\csname bibfnamefont\endcsname\relax
  \def\bibfnamefont#1{#1}\fi
\expandafter\ifx\csname citenamefont\endcsname\relax
  \def\citenamefont#1{#1}\fi
\expandafter\ifx\csname url\endcsname\relax
  \def\url#1{\texttt{#1}}\fi
\expandafter\ifx\csname urlprefix\endcsname\relax\def\urlprefix{URL }\fi
\providecommand{\bibinfo}[2]{#2}
\providecommand{\eprint}[2][]{\url{#2}}

\bibitem[{\citenamefont{Baldo et~al.}(1998)\citenamefont{Baldo, O'brien, You,
  Shoustikov, Sibley, Thompson, and Forrest}}]{oled-nature}
\bibinfo{author}{\bibfnamefont{M.}~\bibnamefont{Baldo}},
  \bibinfo{author}{\bibfnamefont{D.}~\bibnamefont{O'brien}},
  \bibinfo{author}{\bibfnamefont{Y.}~\bibnamefont{You}},
  \bibinfo{author}{\bibfnamefont{A.}~\bibnamefont{Shoustikov}},
  \bibinfo{author}{\bibfnamefont{S.}~\bibnamefont{Sibley}},
  \bibinfo{author}{\bibfnamefont{M.}~\bibnamefont{Thompson}}, \bibnamefont{and}
  \bibinfo{author}{\bibfnamefont{S.}~\bibnamefont{Forrest}},
  \bibinfo{journal}{Nature} \textbf{\bibinfo{volume}{395}},
  \bibinfo{pages}{151} (\bibinfo{year}{1998}).

\bibitem[{\citenamefont{Xiao et~al.}(2011)\citenamefont{Xiao, Chen, Qu, Luo,
  Kong, Gong, and Kido}}]{xiao-review}
\bibinfo{author}{\bibfnamefont{L.}~\bibnamefont{Xiao}},
  \bibinfo{author}{\bibfnamefont{Z.}~\bibnamefont{Chen}},
  \bibinfo{author}{\bibfnamefont{B.}~\bibnamefont{Qu}},
  \bibinfo{author}{\bibfnamefont{J.}~\bibnamefont{Luo}},
  \bibinfo{author}{\bibfnamefont{S.}~\bibnamefont{Kong}},
  \bibinfo{author}{\bibfnamefont{Q.}~\bibnamefont{Gong}}, \bibnamefont{and}
  \bibinfo{author}{\bibfnamefont{J.}~\bibnamefont{Kido}},
  \bibinfo{journal}{Adv. Mater.} \textbf{\bibinfo{volume}{23}},
  \bibinfo{pages}{926} (\bibinfo{year}{2011}).

\bibitem[{\citenamefont{Luhman and Holmes}(2009)}]{holmes-irppy3}
\bibinfo{author}{\bibfnamefont{W.~A.} \bibnamefont{Luhman}} \bibnamefont{and}
  \bibinfo{author}{\bibfnamefont{R.~J.} \bibnamefont{Holmes}},
  \bibinfo{journal}{Appl. Phys. Lett} \textbf{\bibinfo{volume}{94}},
  \bibinfo{pages}{3} (\bibinfo{year}{2009}).

\bibitem[{\citenamefont{Shao and Yang}(2005)}]{triplet-photovoltalic}
\bibinfo{author}{\bibfnamefont{Y.}~\bibnamefont{Shao}} \bibnamefont{and}
  \bibinfo{author}{\bibfnamefont{Y.}~\bibnamefont{Yang}},
  \bibinfo{journal}{Adv. Mater.} \textbf{\bibinfo{volume}{17}},
  \bibinfo{pages}{2841} (\bibinfo{year}{2005}).

\bibitem[{\citenamefont{Hedin}(1965)}]{hedin}
\bibinfo{author}{\bibfnamefont{L.}~\bibnamefont{Hedin}},
  \bibinfo{journal}{Phys. Rev} \textbf{\bibinfo{volume}{139}},
  \bibinfo{pages}{A796} (\bibinfo{year}{1965}).

\bibitem[{\citenamefont{Onida et~al.}(2002)\citenamefont{Onida, Reining, and
  Rubio}}]{GW}
\bibinfo{author}{\bibfnamefont{G.}~\bibnamefont{Onida}},
  \bibinfo{author}{\bibfnamefont{L.}~\bibnamefont{Reining}}, \bibnamefont{and}
  \bibinfo{author}{\bibfnamefont{A.}~\bibnamefont{Rubio}},
  \bibinfo{journal}{Rev. Mod. Phys.} \textbf{\bibinfo{volume}{74}},
  \bibinfo{pages}{601} (\bibinfo{year}{2002}).

\bibitem[{\citenamefont{Runge and Gross}(1984)}]{runge-gross-thm}
\bibinfo{author}{\bibfnamefont{E.}~\bibnamefont{Runge}} \bibnamefont{and}
  \bibinfo{author}{\bibfnamefont{E.}~\bibnamefont{Gross}},
  \bibinfo{journal}{Phys. Rev. Lett} \textbf{\bibinfo{volume}{52}},
  \bibinfo{pages}{997} (\bibinfo{year}{1984}).

\bibitem[{\citenamefont{Petersilka et~al.}(1996)\citenamefont{Petersilka,
  Gossmann, and Gross}}]{tddft-gross-1}
\bibinfo{author}{\bibfnamefont{M.}~\bibnamefont{Petersilka}},
  \bibinfo{author}{\bibfnamefont{U.}~\bibnamefont{Gossmann}}, \bibnamefont{and}
  \bibinfo{author}{\bibfnamefont{E.}~\bibnamefont{Gross}},
  \bibinfo{journal}{Phys. Rev. Lett} \textbf{\bibinfo{volume}{76}},
  \bibinfo{pages}{1212} (\bibinfo{year}{1996}).

\bibitem[{\citenamefont{Grabo et~al.}(2000)\citenamefont{Grabo, Petersilka, and
  Gross}}]{tddft-gross-2}
\bibinfo{author}{\bibfnamefont{T.}~\bibnamefont{Grabo}},
  \bibinfo{author}{\bibfnamefont{M.}~\bibnamefont{Petersilka}},
  \bibnamefont{and} \bibinfo{author}{\bibfnamefont{E.}~\bibnamefont{Gross}},
  \bibinfo{journal}{J. Mol. Struct. THEOCHEM} \textbf{\bibinfo{volume}{501}},
  \bibinfo{pages}{353} (\bibinfo{year}{2000}).

\bibitem[{\citenamefont{Ziegler
  et~al.}(1977{\natexlab{a}})\citenamefont{Ziegler, Rauk, and
  Baerends}}]{d-scf}
\bibinfo{author}{\bibfnamefont{T.}~\bibnamefont{Ziegler}},
  \bibinfo{author}{\bibfnamefont{A.}~\bibnamefont{Rauk}}, \bibnamefont{and}
  \bibinfo{author}{\bibfnamefont{E.}~\bibnamefont{Baerends}},
  \bibinfo{journal}{Theor. Chem. Acc.} \textbf{\bibinfo{volume}{43}},
  \bibinfo{pages}{261} (\bibinfo{year}{1977}{\natexlab{a}}).

\bibitem[{\citenamefont{Casida et~al.}(2000)\citenamefont{Casida, Gutierrez,
  Guan, Gadea, Salahub, and Daudey}}]{dscf-better-1}
\bibinfo{author}{\bibfnamefont{M.}~\bibnamefont{Casida}},
  \bibinfo{author}{\bibfnamefont{F.}~\bibnamefont{Gutierrez}},
  \bibinfo{author}{\bibfnamefont{J.}~\bibnamefont{Guan}},
  \bibinfo{author}{\bibfnamefont{F.}~\bibnamefont{Gadea}},
  \bibinfo{author}{\bibfnamefont{D.}~\bibnamefont{Salahub}}, \bibnamefont{and}
  \bibinfo{author}{\bibfnamefont{J.}~\bibnamefont{Daudey}},
  \bibinfo{journal}{J. Chem. Phys.} \textbf{\bibinfo{volume}{113}},
  \bibinfo{pages}{7062} (\bibinfo{year}{2000}).

\bibitem[{\citenamefont{Cheng et~al.}(2008)\citenamefont{Cheng, Wu, and
  Van~Voorhis}}]{dscf-better-2}
\bibinfo{author}{\bibfnamefont{C.}~\bibnamefont{Cheng}},
  \bibinfo{author}{\bibfnamefont{Q.}~\bibnamefont{Wu}}, \bibnamefont{and}
  \bibinfo{author}{\bibfnamefont{T.}~\bibnamefont{Van~Voorhis}},
  \bibinfo{journal}{J. Chem. Phys.} \textbf{\bibinfo{volume}{129}},
  \bibinfo{pages}{124112} (\bibinfo{year}{2008}).

\bibitem[{\citenamefont{Kowalczyk et~al.}(2011)\citenamefont{Kowalczyk, Yost,
  and Voorhis}}]{dscf-better-3}
\bibinfo{author}{\bibfnamefont{T.}~\bibnamefont{Kowalczyk}},
  \bibinfo{author}{\bibfnamefont{S.}~\bibnamefont{Yost}}, \bibnamefont{and}
  \bibinfo{author}{\bibfnamefont{T.}~\bibnamefont{Voorhis}},
  \bibinfo{journal}{J. Chem. Phys.} \textbf{\bibinfo{volume}{134}},
  \bibinfo{pages}{054128} (\bibinfo{year}{2011}).

\bibitem[{\citenamefont{Tozer}(2003)}]{ct-error-1}
\bibinfo{author}{\bibfnamefont{D.}~\bibnamefont{Tozer}}, \bibinfo{journal}{J.
  Chem. Phys} \textbf{\bibinfo{volume}{119}}, \bibinfo{pages}{12697}
  (\bibinfo{year}{2003}).

\bibitem[{\citenamefont{Dreuw et~al.}(2003)\citenamefont{Dreuw, Weisman, and
  Head-Gordon}}]{tddft-ct-1}
\bibinfo{author}{\bibfnamefont{A.}~\bibnamefont{Dreuw}},
  \bibinfo{author}{\bibfnamefont{J.}~\bibnamefont{Weisman}}, \bibnamefont{and}
  \bibinfo{author}{\bibfnamefont{M.}~\bibnamefont{Head-Gordon}},
  \bibinfo{journal}{J. Chem. Phys.} \textbf{\bibinfo{volume}{119}},
  \bibinfo{pages}{2943} (\bibinfo{year}{2003}).

\bibitem[{\citenamefont{Casida et~al.}(1998{\natexlab{a}})\citenamefont{Casida,
  Jamorski, Casida, and Salahub}}]{tddft-ct-2}
\bibinfo{author}{\bibfnamefont{M.}~\bibnamefont{Casida}},
  \bibinfo{author}{\bibfnamefont{C.}~\bibnamefont{Jamorski}},
  \bibinfo{author}{\bibfnamefont{K.}~\bibnamefont{Casida}}, \bibnamefont{and}
  \bibinfo{author}{\bibfnamefont{D.}~\bibnamefont{Salahub}},
  \bibinfo{journal}{J. Chem. Phys.} \textbf{\bibinfo{volume}{108}},
  \bibinfo{pages}{4439} (\bibinfo{year}{1998}{\natexlab{a}}).

\bibitem[{\citenamefont{Dreuw and Head-Gordon}(2004)}]{tddft-ct-3}
\bibinfo{author}{\bibfnamefont{A.}~\bibnamefont{Dreuw}} \bibnamefont{and}
  \bibinfo{author}{\bibfnamefont{M.}~\bibnamefont{Head-Gordon}},
  \bibinfo{journal}{J. Am. Chem. Soc.} \textbf{\bibinfo{volume}{126}},
  \bibinfo{pages}{4007} (\bibinfo{year}{2004}).

\bibitem[{\citenamefont{Tozer and Handy}(2000)}]{tddft-ct-4}
\bibinfo{author}{\bibfnamefont{D.}~\bibnamefont{Tozer}} \bibnamefont{and}
  \bibinfo{author}{\bibfnamefont{N.}~\bibnamefont{Handy}},
  \bibinfo{journal}{Phys. Chem. Chem. Phys.} \textbf{\bibinfo{volume}{2}},
  \bibinfo{pages}{2117} (\bibinfo{year}{2000}).

\bibitem[{\citenamefont{Peach et~al.}(2008)\citenamefont{Peach, Benfield,
  Helgaker, and Tozer}}]{tddft-ct-5}
\bibinfo{author}{\bibfnamefont{M.}~\bibnamefont{Peach}},
  \bibinfo{author}{\bibfnamefont{P.}~\bibnamefont{Benfield}},
  \bibinfo{author}{\bibfnamefont{T.}~\bibnamefont{Helgaker}}, \bibnamefont{and}
  \bibinfo{author}{\bibfnamefont{D.}~\bibnamefont{Tozer}}, \bibinfo{journal}{J.
  Chem. Phys.} \textbf{\bibinfo{volume}{128}}, \bibinfo{pages}{044118}
  (\bibinfo{year}{2008}).

\bibitem[{\citenamefont{Maitra}(2005)}]{tddft-ct-6}
\bibinfo{author}{\bibfnamefont{N.}~\bibnamefont{Maitra}}, \bibinfo{journal}{J.
  Chem. Phys.} \textbf{\bibinfo{volume}{122}}, \bibinfo{pages}{234104}
  (\bibinfo{year}{2005}).

\bibitem[{\citenamefont{Neugebauer et~al.}(2006)\citenamefont{Neugebauer,
  Gritsenko, and Baerends}}]{tddft-ct-7}
\bibinfo{author}{\bibfnamefont{J.}~\bibnamefont{Neugebauer}},
  \bibinfo{author}{\bibfnamefont{O.}~\bibnamefont{Gritsenko}},
  \bibnamefont{and} \bibinfo{author}{\bibfnamefont{E.}~\bibnamefont{Baerends}},
  \bibinfo{journal}{J. Chem. Phys.} \textbf{\bibinfo{volume}{124}},
  \bibinfo{pages}{214102} (\bibinfo{year}{2006}).

\bibitem[{\citenamefont{Erickson and Holmes}(2011)}]{erickson-all}
\bibinfo{author}{\bibfnamefont{N.}~\bibnamefont{Erickson}} \bibnamefont{and}
  \bibinfo{author}{\bibfnamefont{R.}~\bibnamefont{Holmes}},
  \bibinfo{journal}{J. Appl. Phys.} \textbf{\bibinfo{volume}{110}},
  \bibinfo{pages}{084515} (\bibinfo{year}{2011}).

\bibitem[{\citenamefont{Giannozzi et~al.}(2009)\citenamefont{Giannozzi, Baroni,
  Bonini, Calandra, Car, Cavazzoni, Ceresoli, Chiarotti, Cococcioni, Dabo
  et~al.}}]{espresso}
\bibinfo{author}{\bibfnamefont{P.}~\bibnamefont{Giannozzi}},
  \bibinfo{author}{\bibfnamefont{S.}~\bibnamefont{Baroni}},
  \bibinfo{author}{\bibfnamefont{N.}~\bibnamefont{Bonini}},
  \bibinfo{author}{\bibfnamefont{M.}~\bibnamefont{Calandra}},
  \bibinfo{author}{\bibfnamefont{R.}~\bibnamefont{Car}},
  \bibinfo{author}{\bibfnamefont{C.}~\bibnamefont{Cavazzoni}},
  \bibinfo{author}{\bibfnamefont{D.}~\bibnamefont{Ceresoli}},
  \bibinfo{author}{\bibfnamefont{G.}~\bibnamefont{Chiarotti}},
  \bibinfo{author}{\bibfnamefont{M.}~\bibnamefont{Cococcioni}},
  \bibinfo{author}{\bibfnamefont{I.}~\bibnamefont{Dabo}}, \bibnamefont{et~al.},
  \bibinfo{journal}{J. Phys. Condens. Matter} \textbf{\bibinfo{volume}{21}},
  \bibinfo{pages}{395502} (\bibinfo{year}{2009}).

\bibitem[{\citenamefont{Frisch et~al.}()\citenamefont{Frisch, Trucks, Schlegel,
  Scuseria, Robb, Cheeseman, Scalmani, Barone, Mennucci, Petersson
  et~al.}}]{g09}
\bibinfo{author}{\bibfnamefont{M.~J.} \bibnamefont{Frisch}},
  \bibinfo{author}{\bibfnamefont{G.~W.} \bibnamefont{Trucks}},
  \bibinfo{author}{\bibfnamefont{H.~B.} \bibnamefont{Schlegel}},
  \bibinfo{author}{\bibfnamefont{G.~E.} \bibnamefont{Scuseria}},
  \bibinfo{author}{\bibfnamefont{M.~A.} \bibnamefont{Robb}},
  \bibinfo{author}{\bibfnamefont{J.~R.} \bibnamefont{Cheeseman}},
  \bibinfo{author}{\bibfnamefont{G.}~\bibnamefont{Scalmani}},
  \bibinfo{author}{\bibfnamefont{V.}~\bibnamefont{Barone}},
  \bibinfo{author}{\bibfnamefont{B.}~\bibnamefont{Mennucci}},
  \bibinfo{author}{\bibfnamefont{G.~A.} \bibnamefont{Petersson}},
  \bibnamefont{et~al.}, \emph{\bibinfo{title}{Gaussian09 {R}evision {A}.1}},
  \bibinfo{note}{gaussian Inc. Wallingford CT 2009}.

\bibitem[{\citenamefont{Perdew et~al.}(1996)\citenamefont{Perdew, Burke, and
  Ernzerhof}}]{pbe}
\bibinfo{author}{\bibfnamefont{J.}~\bibnamefont{Perdew}},
  \bibinfo{author}{\bibfnamefont{K.}~\bibnamefont{Burke}}, \bibnamefont{and}
  \bibinfo{author}{\bibfnamefont{M.}~\bibnamefont{Ernzerhof}},
  \bibinfo{journal}{Phys. Rev. Lett} \textbf{\bibinfo{volume}{77}},
  \bibinfo{pages}{3865} (\bibinfo{year}{1996}).

\bibitem[{\citenamefont{Vanderbilt}(1990)}]{vanderbilt}
\bibinfo{author}{\bibfnamefont{D.}~\bibnamefont{Vanderbilt}},
  \bibinfo{journal}{Phys. Rev. B} \textbf{\bibinfo{volume}{41}},
  \bibinfo{pages}{7892} (\bibinfo{year}{1990}).

\bibitem[{\citenamefont{Bauernschmitt and Ahlrichs}(1996)}]{g09-impl-1}
\bibinfo{author}{\bibfnamefont{R.}~\bibnamefont{Bauernschmitt}}
  \bibnamefont{and} \bibinfo{author}{\bibfnamefont{R.}~\bibnamefont{Ahlrichs}},
  \bibinfo{journal}{Chem. Phys. Lett} \textbf{\bibinfo{volume}{256}},
  \bibinfo{pages}{454} (\bibinfo{year}{1996}).

\bibitem[{\citenamefont{Casida et~al.}(1998{\natexlab{b}})\citenamefont{Casida,
  Jamorski, Casida, and Salahub}}]{g09-impl-2}
\bibinfo{author}{\bibfnamefont{M.}~\bibnamefont{Casida}},
  \bibinfo{author}{\bibfnamefont{C.}~\bibnamefont{Jamorski}},
  \bibinfo{author}{\bibfnamefont{K.}~\bibnamefont{Casida}}, \bibnamefont{and}
  \bibinfo{author}{\bibfnamefont{D.}~\bibnamefont{Salahub}},
  \bibinfo{journal}{J. Chem. Phys} \textbf{\bibinfo{volume}{108}},
  \bibinfo{pages}{4439} (\bibinfo{year}{1998}{\natexlab{b}}).

\bibitem[{\citenamefont{Stratmann et~al.}(1998)\citenamefont{Stratmann,
  Scuseria, and Frisch}}]{g09-impl-3}
\bibinfo{author}{\bibfnamefont{R.}~\bibnamefont{Stratmann}},
  \bibinfo{author}{\bibfnamefont{G.}~\bibnamefont{Scuseria}}, \bibnamefont{and}
  \bibinfo{author}{\bibfnamefont{M.}~\bibnamefont{Frisch}},
  \bibinfo{journal}{J. Chem. Phys} \textbf{\bibinfo{volume}{109}},
  \bibinfo{pages}{8218} (\bibinfo{year}{1998}).

\bibitem[{\citenamefont{Van~Caillie and Amos}(1999)}]{g09-impl-4}
\bibinfo{author}{\bibfnamefont{C.}~\bibnamefont{Van~Caillie}} \bibnamefont{and}
  \bibinfo{author}{\bibfnamefont{R.}~\bibnamefont{Amos}},
  \bibinfo{journal}{Chem. Phys. Lett} \textbf{\bibinfo{volume}{308}},
  \bibinfo{pages}{249} (\bibinfo{year}{1999}).

\bibitem[{\citenamefont{Van~Caillie and Amos}(2000)}]{g09-impl-5}
\bibinfo{author}{\bibfnamefont{C.}~\bibnamefont{Van~Caillie}} \bibnamefont{and}
  \bibinfo{author}{\bibfnamefont{R.}~\bibnamefont{Amos}},
  \bibinfo{journal}{Chem. Phys. Lett.} \textbf{\bibinfo{volume}{317}},
  \bibinfo{pages}{159} (\bibinfo{year}{2000}).

\bibitem[{\citenamefont{Furche and Ahlrichs}(2002)}]{g09-impl-6}
\bibinfo{author}{\bibfnamefont{F.}~\bibnamefont{Furche}} \bibnamefont{and}
  \bibinfo{author}{\bibfnamefont{R.}~\bibnamefont{Ahlrichs}},
  \bibinfo{journal}{J. Chem. Phys} \textbf{\bibinfo{volume}{117}},
  \bibinfo{pages}{7433} (\bibinfo{year}{2002}).

\bibitem[{\citenamefont{Scalmani et~al.}(2006)\citenamefont{Scalmani, Frisch,
  Mennucci, Tomasi, Cammi, and Barone}}]{g09-impl-7}
\bibinfo{author}{\bibfnamefont{G.}~\bibnamefont{Scalmani}},
  \bibinfo{author}{\bibfnamefont{M.}~\bibnamefont{Frisch}},
  \bibinfo{author}{\bibfnamefont{B.}~\bibnamefont{Mennucci}},
  \bibinfo{author}{\bibfnamefont{J.}~\bibnamefont{Tomasi}},
  \bibinfo{author}{\bibfnamefont{R.}~\bibnamefont{Cammi}}, \bibnamefont{and}
  \bibinfo{author}{\bibfnamefont{V.}~\bibnamefont{Barone}},
  \bibinfo{journal}{J. Chem. Phys} \textbf{\bibinfo{volume}{124}},
  \bibinfo{pages}{094107} (\bibinfo{year}{2006}).

\bibitem[{\citenamefont{Becke}(1993)}]{b3lyp-1}
\bibinfo{author}{\bibfnamefont{A.}~\bibnamefont{Becke}}, \bibinfo{journal}{J.
  Chem. Phys} \textbf{\bibinfo{volume}{98}}, \bibinfo{pages}{1372}
  (\bibinfo{year}{1993}).

\bibitem[{\citenamefont{Lee et~al.}(1988)\citenamefont{Lee, Yang, and
  Parr}}]{b3lyp-2}
\bibinfo{author}{\bibfnamefont{C.}~\bibnamefont{Lee}},
  \bibinfo{author}{\bibfnamefont{W.}~\bibnamefont{Yang}}, \bibnamefont{and}
  \bibinfo{author}{\bibfnamefont{R.}~\bibnamefont{Parr}},
  \bibinfo{journal}{Phys. Rev. B} \textbf{\bibinfo{volume}{37}},
  \bibinfo{pages}{785} (\bibinfo{year}{1988}).

\bibitem[{\citenamefont{Zhao and Truhlar}(2008)}]{m06}
\bibinfo{author}{\bibfnamefont{Y.}~\bibnamefont{Zhao}} \bibnamefont{and}
  \bibinfo{author}{\bibfnamefont{D.}~\bibnamefont{Truhlar}},
  \bibinfo{journal}{Theor. Chem. Acc.} \textbf{\bibinfo{volume}{120}},
  \bibinfo{pages}{215} (\bibinfo{year}{2008}).

\bibitem[{\citenamefont{Anisimov et~al.}(1991)\citenamefont{Anisimov, Zaanen,
  and Andersen}}]{anisimov-1991}
\bibinfo{author}{\bibfnamefont{V.}~\bibnamefont{Anisimov}},
  \bibinfo{author}{\bibfnamefont{J.}~\bibnamefont{Zaanen}}, \bibnamefont{and}
  \bibinfo{author}{\bibfnamefont{O.}~\bibnamefont{Andersen}},
  \bibinfo{journal}{Phys. Rev. B} \textbf{\bibinfo{volume}{44}},
  \bibinfo{pages}{943} (\bibinfo{year}{1991}).

\bibitem[{\citenamefont{Anisimov et~al.}(1993)\citenamefont{Anisimov, Solovyev,
  Korotin, Czy{\.z}yk, and Sawatzky}}]{anisimov-1993}
\bibinfo{author}{\bibfnamefont{V.}~\bibnamefont{Anisimov}},
  \bibinfo{author}{\bibfnamefont{I.}~\bibnamefont{Solovyev}},
  \bibinfo{author}{\bibfnamefont{M.}~\bibnamefont{Korotin}},
  \bibinfo{author}{\bibfnamefont{M.}~\bibnamefont{Czy{\.z}yk}},
  \bibnamefont{and} \bibinfo{author}{\bibfnamefont{G.}~\bibnamefont{Sawatzky}},
  \bibinfo{journal}{Phys. Rev. B} \textbf{\bibinfo{volume}{48}},
  \bibinfo{pages}{16929} (\bibinfo{year}{1993}).

\bibitem[{\citenamefont{Mazin and Anisimov}(1997)}]{mazin-1997}
\bibinfo{author}{\bibfnamefont{I.}~\bibnamefont{Mazin}} \bibnamefont{and}
  \bibinfo{author}{\bibfnamefont{V.}~\bibnamefont{Anisimov}},
  \bibinfo{journal}{Phys. Rev. B} \textbf{\bibinfo{volume}{55}},
  \bibinfo{pages}{12822} (\bibinfo{year}{1997}).

\bibitem[{\citenamefont{Cococcioni and De~Gironcoli}(2005)}]{Ucalc}
\bibinfo{author}{\bibfnamefont{M.}~\bibnamefont{Cococcioni}} \bibnamefont{and}
  \bibinfo{author}{\bibfnamefont{S.}~\bibnamefont{De~Gironcoli}},
  \bibinfo{journal}{Phys. Rev. B} \textbf{\bibinfo{volume}{71}},
  \bibinfo{pages}{35105} (\bibinfo{year}{2005}).

\bibitem[{\citenamefont{Campo~Jr and Cococcioni}(2010)}]{HubV}
\bibinfo{author}{\bibfnamefont{V.}~\bibnamefont{Campo~Jr}} \bibnamefont{and}
  \bibinfo{author}{\bibfnamefont{M.}~\bibnamefont{Cococcioni}},
  \bibinfo{journal}{J. Phys. Condens. Matter} \textbf{\bibinfo{volume}{22}},
  \bibinfo{pages}{055602} (\bibinfo{year}{2010}).

\bibitem[{\citenamefont{Kokalj}(2003)}]{xcrysden}
\bibinfo{author}{\bibfnamefont{A.}~\bibnamefont{Kokalj}},
  \bibinfo{journal}{Comp. Mater. Sci.} \textbf{\bibinfo{volume}{28}},
  \bibinfo{pages}{155} (\bibinfo{year}{2003}).

\bibitem[{\citenamefont{Van~Leeuwen and Baerends}(1994)}]{ct-correct-1}
\bibinfo{author}{\bibfnamefont{R.}~\bibnamefont{Van~Leeuwen}} \bibnamefont{and}
  \bibinfo{author}{\bibfnamefont{E.}~\bibnamefont{Baerends}},
  \bibinfo{journal}{Phys. Rev. A} \textbf{\bibinfo{volume}{49}},
  \bibinfo{pages}{2421} (\bibinfo{year}{1994}).

\bibitem[{\citenamefont{Vasiliev and Martin}(2004)}]{ct-correct-2}
\bibinfo{author}{\bibfnamefont{I.}~\bibnamefont{Vasiliev}} \bibnamefont{and}
  \bibinfo{author}{\bibfnamefont{R.}~\bibnamefont{Martin}},
  \bibinfo{journal}{Phys. Rev. A} \textbf{\bibinfo{volume}{69}},
  \bibinfo{pages}{052508} (\bibinfo{year}{2004}).

\bibitem[{\citenamefont{Gritsenko and Baerends}(2004)}]{ct-correct-3}
\bibinfo{author}{\bibfnamefont{O.}~\bibnamefont{Gritsenko}} \bibnamefont{and}
  \bibinfo{author}{\bibfnamefont{E.}~\bibnamefont{Baerends}},
  \bibinfo{journal}{J. Chem. Phys.} \textbf{\bibinfo{volume}{121}},
  \bibinfo{pages}{655} (\bibinfo{year}{2004}).

\bibitem[{\citenamefont{Tawada et~al.}(2004)\citenamefont{Tawada, Tsuneda,
  Yanagisawa, Yanai, and Hirao}}]{ct-correct-4}
\bibinfo{author}{\bibfnamefont{Y.}~\bibnamefont{Tawada}},
  \bibinfo{author}{\bibfnamefont{T.}~\bibnamefont{Tsuneda}},
  \bibinfo{author}{\bibfnamefont{S.}~\bibnamefont{Yanagisawa}},
  \bibinfo{author}{\bibfnamefont{T.}~\bibnamefont{Yanai}}, \bibnamefont{and}
  \bibinfo{author}{\bibfnamefont{K.}~\bibnamefont{Hirao}}, \bibinfo{journal}{J.
  Chem. Phys.} \textbf{\bibinfo{volume}{120}}, \bibinfo{pages}{8425}
  (\bibinfo{year}{2004}).

\bibitem[{\citenamefont{Rohrdanz et~al.}(2009)\citenamefont{Rohrdanz, Martins,
  and Herbert}}]{ct-correct-5}
\bibinfo{author}{\bibfnamefont{M.}~\bibnamefont{Rohrdanz}},
  \bibinfo{author}{\bibfnamefont{K.}~\bibnamefont{Martins}}, \bibnamefont{and}
  \bibinfo{author}{\bibfnamefont{J.}~\bibnamefont{Herbert}},
  \bibinfo{journal}{J. Chem. Phys.} \textbf{\bibinfo{volume}{130}},
  \bibinfo{pages}{054112} (\bibinfo{year}{2009}).

\bibitem[{\citenamefont{Song et~al.}(2009)\citenamefont{Song, Watson, and
  Hirao}}]{ct-correct-6}
\bibinfo{author}{\bibfnamefont{J.}~\bibnamefont{Song}},
  \bibinfo{author}{\bibfnamefont{M.}~\bibnamefont{Watson}}, \bibnamefont{and}
  \bibinfo{author}{\bibfnamefont{K.}~\bibnamefont{Hirao}}, \bibinfo{journal}{J.
  Chem. Phys.} \textbf{\bibinfo{volume}{131}}, \bibinfo{pages}{144108}
  (\bibinfo{year}{2009}).

\bibitem[{\citenamefont{Della~Sala and G{\"o}rling}(2003)}]{ct-correct-7}
\bibinfo{author}{\bibfnamefont{F.}~\bibnamefont{Della~Sala}} \bibnamefont{and}
  \bibinfo{author}{\bibfnamefont{A.}~\bibnamefont{G{\"o}rling}},
  \bibinfo{journal}{Int. J. Quantum Chem.} \textbf{\bibinfo{volume}{91}},
  \bibinfo{pages}{131} (\bibinfo{year}{2003}).

\bibitem[{\citenamefont{Jacquemin et~al.}(2008)\citenamefont{Jacquemin,
  Perp{\`e}te, Scuseria, Ciofini, and Adamo}}]{ct-correct-8}
\bibinfo{author}{\bibfnamefont{D.}~\bibnamefont{Jacquemin}},
  \bibinfo{author}{\bibfnamefont{E.}~\bibnamefont{Perp{\`e}te}},
  \bibinfo{author}{\bibfnamefont{G.}~\bibnamefont{Scuseria}},
  \bibinfo{author}{\bibfnamefont{I.}~\bibnamefont{Ciofini}}, \bibnamefont{and}
  \bibinfo{author}{\bibfnamefont{C.}~\bibnamefont{Adamo}}, \bibinfo{journal}{J.
  Chem. Theory Comput.} \textbf{\bibinfo{volume}{4}}, \bibinfo{pages}{123}
  (\bibinfo{year}{2008}).

\bibitem[{\citenamefont{Stein et~al.}(2009{\natexlab{a}})\citenamefont{Stein,
  Kronik, and Baer}}]{kronik-1}
\bibinfo{author}{\bibfnamefont{T.}~\bibnamefont{Stein}},
  \bibinfo{author}{\bibfnamefont{L.}~\bibnamefont{Kronik}}, \bibnamefont{and}
  \bibinfo{author}{\bibfnamefont{R.}~\bibnamefont{Baer}}, \bibinfo{journal}{J.
  Am. Chem. Soc.} \textbf{\bibinfo{volume}{131}}, \bibinfo{pages}{2818}
  (\bibinfo{year}{2009}{\natexlab{a}}).

\bibitem[{\citenamefont{Stein et~al.}(2009{\natexlab{b}})\citenamefont{Stein,
  Kronik, and Baer}}]{kronik-2}
\bibinfo{author}{\bibfnamefont{T.}~\bibnamefont{Stein}},
  \bibinfo{author}{\bibfnamefont{L.}~\bibnamefont{Kronik}}, \bibnamefont{and}
  \bibinfo{author}{\bibfnamefont{R.}~\bibnamefont{Baer}}, \bibinfo{journal}{J.
  Chem. Phys.} \textbf{\bibinfo{volume}{131}}, \bibinfo{pages}{244119}
  (\bibinfo{year}{2009}{\natexlab{b}}).

\bibitem[{\citenamefont{Kuritz et~al.}(2011)\citenamefont{Kuritz, Stein, Baer,
  and Kronik}}]{kronik-3}
\bibinfo{author}{\bibfnamefont{N.}~\bibnamefont{Kuritz}},
  \bibinfo{author}{\bibfnamefont{T.}~\bibnamefont{Stein}},
  \bibinfo{author}{\bibfnamefont{R.}~\bibnamefont{Baer}}, \bibnamefont{and}
  \bibinfo{author}{\bibfnamefont{L.}~\bibnamefont{Kronik}},
  \bibinfo{journal}{J. Chem. Theory Comput.}  (\bibinfo{year}{2011}).

\bibitem[{\citenamefont{Hu et~al.}(2006)\citenamefont{Hu, Sugino, and
  Miyamoto}}]{tddft-dscf-mix-1}
\bibinfo{author}{\bibfnamefont{C.}~\bibnamefont{Hu}},
  \bibinfo{author}{\bibfnamefont{O.}~\bibnamefont{Sugino}}, \bibnamefont{and}
  \bibinfo{author}{\bibfnamefont{Y.}~\bibnamefont{Miyamoto}},
  \bibinfo{journal}{Phys. Rev. A} \textbf{\bibinfo{volume}{74}},
  \bibinfo{pages}{032508} (\bibinfo{year}{2006}).

\bibitem[{\citenamefont{Hu and Sugino}(2007)}]{tddft-dscf-mix-2}
\bibinfo{author}{\bibfnamefont{C.}~\bibnamefont{Hu}} \bibnamefont{and}
  \bibinfo{author}{\bibfnamefont{O.}~\bibnamefont{Sugino}},
  \bibinfo{journal}{J. Chem. Phys} \textbf{\bibinfo{volume}{126}},
  \bibinfo{pages}{074112} (\bibinfo{year}{2007}).

\bibitem[{\citenamefont{Ziegler
  et~al.}(1977{\natexlab{b}})\citenamefont{Ziegler, Rauk, and
  Baerends}}]{spin-pur-1}
\bibinfo{author}{\bibfnamefont{T.}~\bibnamefont{Ziegler}},
  \bibinfo{author}{\bibfnamefont{A.}~\bibnamefont{Rauk}}, \bibnamefont{and}
  \bibinfo{author}{\bibfnamefont{E.}~\bibnamefont{Baerends}},
  \bibinfo{journal}{Theor. Chem. Acc.} \textbf{\bibinfo{volume}{43}},
  \bibinfo{pages}{261} (\bibinfo{year}{1977}{\natexlab{b}}).

\bibitem[{\citenamefont{Ziegler}(1991)}]{spin-pur-2}
\bibinfo{author}{\bibfnamefont{T.}~\bibnamefont{Ziegler}},
  \bibinfo{journal}{Chem. Rev.} \textbf{\bibinfo{volume}{91}},
  \bibinfo{pages}{651} (\bibinfo{year}{1991}).

\bibitem[{\citenamefont{Frank et~al.}(1998)\citenamefont{Frank, Hutter, Marx,
  and Parrinello}}]{spin-pur-3}
\bibinfo{author}{\bibfnamefont{I.}~\bibnamefont{Frank}},
  \bibinfo{author}{\bibfnamefont{J.}~\bibnamefont{Hutter}},
  \bibinfo{author}{\bibfnamefont{D.}~\bibnamefont{Marx}}, \bibnamefont{and}
  \bibinfo{author}{\bibfnamefont{M.}~\bibnamefont{Parrinello}},
  \bibinfo{journal}{J. Chem. Phys.} \textbf{\bibinfo{volume}{108}},
  \bibinfo{pages}{4060} (\bibinfo{year}{1998}).

\bibitem[{\citenamefont{Behler et~al.}(2008)\citenamefont{Behler, Reuter, and
  Scheffler}}]{nsp-o2-1}
\bibinfo{author}{\bibfnamefont{J.}~\bibnamefont{Behler}},
  \bibinfo{author}{\bibfnamefont{K.}~\bibnamefont{Reuter}}, \bibnamefont{and}
  \bibinfo{author}{\bibfnamefont{M.}~\bibnamefont{Scheffler}},
  \bibinfo{journal}{Phys. Rev. B} \textbf{\bibinfo{volume}{77}},
  \bibinfo{pages}{115421} (\bibinfo{year}{2008}).

\bibitem[{\citenamefont{Behler et~al.}(2005)\citenamefont{Behler, Delley,
  Lorenz, Reuter, and Scheffler}}]{nsp-o2-2}
\bibinfo{author}{\bibfnamefont{J.}~\bibnamefont{Behler}},
  \bibinfo{author}{\bibfnamefont{B.}~\bibnamefont{Delley}},
  \bibinfo{author}{\bibfnamefont{S.}~\bibnamefont{Lorenz}},
  \bibinfo{author}{\bibfnamefont{K.}~\bibnamefont{Reuter}}, \bibnamefont{and}
  \bibinfo{author}{\bibfnamefont{M.}~\bibnamefont{Scheffler}},
  \bibinfo{journal}{Phys. Rev. Lett.} \textbf{\bibinfo{volume}{94}},
  \bibinfo{pages}{36104} (\bibinfo{year}{2005}).

\bibitem[{\citenamefont{Alducin et~al.}(2008)\citenamefont{Alducin, Busnengo,
  and Mui{\~n}o}}]{nsp-o2-3}
\bibinfo{author}{\bibfnamefont{M.}~\bibnamefont{Alducin}},
  \bibinfo{author}{\bibfnamefont{H.}~\bibnamefont{Busnengo}}, \bibnamefont{and}
  \bibinfo{author}{\bibfnamefont{R.}~\bibnamefont{Mui{\~n}o}},
  \bibinfo{journal}{The Journal of chemical physics}
  \textbf{\bibinfo{volume}{129}}, \bibinfo{pages}{224702}
  (\bibinfo{year}{2008}).

\bibitem[{\citenamefont{Maurer and Reuter}(2011)}]{nsp-new}
\bibinfo{author}{\bibfnamefont{R.}~\bibnamefont{Maurer}} \bibnamefont{and}
  \bibinfo{author}{\bibfnamefont{K.}~\bibnamefont{Reuter}},
  \bibinfo{journal}{J. Chem. Phys.} \textbf{\bibinfo{volume}{135}},
  \bibinfo{pages}{224303} (\bibinfo{year}{2011}).

\bibitem[{\citenamefont{Levy and Nagy}(1999)}]{nagy}
\bibinfo{author}{\bibfnamefont{M.}~\bibnamefont{Levy}} \bibnamefont{and}
  \bibinfo{author}{\bibfnamefont{{\'A}.}~\bibnamefont{Nagy}},
  \bibinfo{journal}{Phys. Rev. Lett} \textbf{\bibinfo{volume}{83}},
  \bibinfo{pages}{4361} (\bibinfo{year}{1999}).

\bibitem[{\citenamefont{Ayers and Levy}(2009)}]{nagy-2}
\bibinfo{author}{\bibfnamefont{P.}~\bibnamefont{Ayers}} \bibnamefont{and}
  \bibinfo{author}{\bibfnamefont{M.}~\bibnamefont{Levy}},
  \bibinfo{journal}{Phys. Rev. A} \textbf{\bibinfo{volume}{80}},
  \bibinfo{pages}{012508} (\bibinfo{year}{2009}).

\bibitem[{\citenamefont{Fritsche}(1986)}]{dft-wvfc-1}
\bibinfo{author}{\bibfnamefont{L.}~\bibnamefont{Fritsche}},
  \bibinfo{journal}{Phys. Rev. B} \textbf{\bibinfo{volume}{33}},
  \bibinfo{pages}{3976} (\bibinfo{year}{1986}).

\bibitem[{\citenamefont{Cordes and Fritsche}(1989)}]{dft-wvfc-2}
\bibinfo{author}{\bibfnamefont{J.}~\bibnamefont{Cordes}} \bibnamefont{and}
  \bibinfo{author}{\bibfnamefont{L.}~\bibnamefont{Fritsche}},
  \bibinfo{journal}{Z. Phys. D} \textbf{\bibinfo{volume}{13}},
  \bibinfo{pages}{345} (\bibinfo{year}{1989}).

\bibitem[{\citenamefont{Gross et~al.}(1988{\natexlab{a}})\citenamefont{Gross,
  Oliveira, and Kohn}}]{edft-1}
\bibinfo{author}{\bibfnamefont{E.}~\bibnamefont{Gross}},
  \bibinfo{author}{\bibfnamefont{L.}~\bibnamefont{Oliveira}}, \bibnamefont{and}
  \bibinfo{author}{\bibfnamefont{W.}~\bibnamefont{Kohn}},
  \bibinfo{journal}{Phys. Rev. A} \textbf{\bibinfo{volume}{37}},
  \bibinfo{pages}{2805} (\bibinfo{year}{1988}{\natexlab{a}}).

\bibitem[{\citenamefont{Gross et~al.}(1988{\natexlab{b}})\citenamefont{Gross,
  Oliveira, and Kohn}}]{edft-2}
\bibinfo{author}{\bibfnamefont{E.}~\bibnamefont{Gross}},
  \bibinfo{author}{\bibfnamefont{L.}~\bibnamefont{Oliveira}}, \bibnamefont{and}
  \bibinfo{author}{\bibfnamefont{W.}~\bibnamefont{Kohn}},
  \bibinfo{journal}{Phys. Rev. A} \textbf{\bibinfo{volume}{37}},
  \bibinfo{pages}{2809} (\bibinfo{year}{1988}{\natexlab{b}}).

\bibitem[{\citenamefont{Oliveira et~al.}(1988)\citenamefont{Oliveira, Gross,
  and Kohn}}]{edft-3}
\bibinfo{author}{\bibfnamefont{L.}~\bibnamefont{Oliveira}},
  \bibinfo{author}{\bibfnamefont{E.}~\bibnamefont{Gross}}, \bibnamefont{and}
  \bibinfo{author}{\bibfnamefont{W.}~\bibnamefont{Kohn}},
  \bibinfo{journal}{Phys. Rev. A} \textbf{\bibinfo{volume}{37}},
  \bibinfo{pages}{2821} (\bibinfo{year}{1988}).

\bibitem[{\citenamefont{Kohn}(1986)}]{kohn-edft}
\bibinfo{author}{\bibfnamefont{W.}~\bibnamefont{Kohn}}, \bibinfo{journal}{Phys.
  Rev. A} \textbf{\bibinfo{volume}{34}}, \bibinfo{pages}{737}
  (\bibinfo{year}{1986}).

\bibitem[{\citenamefont{Kulik and Marzari}(2011)}]{hubv-molecule}
\bibinfo{author}{\bibfnamefont{H.}~\bibnamefont{Kulik}} \bibnamefont{and}
  \bibinfo{author}{\bibfnamefont{N.}~\bibnamefont{Marzari}},
  \bibinfo{journal}{J. Chem. Phys} \textbf{\bibinfo{volume}{134}},
  \bibinfo{pages}{094103} (\bibinfo{year}{2011}).

\bibitem[{\citenamefont{Belozerov et~al.}(2012)\citenamefont{Belozerov,
  Korotin, Anisimov, and Poteryaev}}]{vo2}
\bibinfo{author}{\bibfnamefont{A.}~\bibnamefont{Belozerov}},
  \bibinfo{author}{\bibfnamefont{M.}~\bibnamefont{Korotin}},
  \bibinfo{author}{\bibfnamefont{V.}~\bibnamefont{Anisimov}}, \bibnamefont{and}
  \bibinfo{author}{\bibfnamefont{A.}~\bibnamefont{Poteryaev}},
  \bibinfo{journal}{Phys. Rev. B} \textbf{\bibinfo{volume}{85}},
  \bibinfo{pages}{045109} (\bibinfo{year}{2012}).

\bibitem[{\citenamefont{Baldo et~al.}(2000{\natexlab{a}})\citenamefont{Baldo,
  Adachi, and Forrest}}]{baldo-irppy3}
\bibinfo{author}{\bibfnamefont{M.}~\bibnamefont{Baldo}},
  \bibinfo{author}{\bibfnamefont{C.}~\bibnamefont{Adachi}}, \bibnamefont{and}
  \bibinfo{author}{\bibfnamefont{S.}~\bibnamefont{Forrest}},
  \bibinfo{journal}{Phys. Rev. B} \textbf{\bibinfo{volume}{62}},
  \bibinfo{pages}{10967} (\bibinfo{year}{2000}{\natexlab{a}}).

\bibitem[{\citenamefont{Goushi et~al.}(2004)\citenamefont{Goushi, Kwong, Brown,
  Sasabe, and Adachi}}]{goushi-irppy3}
\bibinfo{author}{\bibfnamefont{K.}~\bibnamefont{Goushi}},
  \bibinfo{author}{\bibfnamefont{R.}~\bibnamefont{Kwong}},
  \bibinfo{author}{\bibfnamefont{J.}~\bibnamefont{Brown}},
  \bibinfo{author}{\bibfnamefont{H.}~\bibnamefont{Sasabe}}, \bibnamefont{and}
  \bibinfo{author}{\bibfnamefont{C.}~\bibnamefont{Adachi}},
  \bibinfo{journal}{J. Appl. Phys} \textbf{\bibinfo{volume}{95}},
  \bibinfo{pages}{7798} (\bibinfo{year}{2004}).

\bibitem[{\citenamefont{Eom et~al.}(2009)\citenamefont{Eom, Zheng,
  Wrzesniewski, Lee, Chopra, So, and Xue}}]{eom-irppy3}
\bibinfo{author}{\bibfnamefont{S.}~\bibnamefont{Eom}},
  \bibinfo{author}{\bibfnamefont{Y.}~\bibnamefont{Zheng}},
  \bibinfo{author}{\bibfnamefont{E.}~\bibnamefont{Wrzesniewski}},
  \bibinfo{author}{\bibfnamefont{J.}~\bibnamefont{Lee}},
  \bibinfo{author}{\bibfnamefont{N.}~\bibnamefont{Chopra}},
  \bibinfo{author}{\bibfnamefont{F.}~\bibnamefont{So}}, \bibnamefont{and}
  \bibinfo{author}{\bibfnamefont{J.}~\bibnamefont{Xue}},
  \bibinfo{journal}{Appl. Phys. Lett} \textbf{\bibinfo{volume}{94}},
  \bibinfo{pages}{153303} (\bibinfo{year}{2009}).

\bibitem[{\citenamefont{Adachi et~al.}(2001{\natexlab{a}})\citenamefont{Adachi,
  Kwong, and Forrest}}]{adachi-irppy3}
\bibinfo{author}{\bibfnamefont{C.}~\bibnamefont{Adachi}},
  \bibinfo{author}{\bibfnamefont{R.}~\bibnamefont{Kwong}}, \bibnamefont{and}
  \bibinfo{author}{\bibfnamefont{S.}~\bibnamefont{Forrest}},
  \bibinfo{journal}{Org. Electron.} \textbf{\bibinfo{volume}{2}},
  \bibinfo{pages}{37} (\bibinfo{year}{2001}{\natexlab{a}}).

\bibitem[{\citenamefont{Tsuboi and Tanigawa}(2003)}]{tsuboi-irppy3}
\bibinfo{author}{\bibfnamefont{T.}~\bibnamefont{Tsuboi}} \bibnamefont{and}
  \bibinfo{author}{\bibfnamefont{M.}~\bibnamefont{Tanigawa}},
  \bibinfo{journal}{Thin Solid Films} \textbf{\bibinfo{volume}{438}},
  \bibinfo{pages}{301} (\bibinfo{year}{2003}).

\bibitem[{\citenamefont{Lamansky et~al.}(2001)\citenamefont{Lamansky,
  Djurovich, Murphy, Abdel-Razzaq, Lee, Adachi, Burrows, Forrest, and
  Thompson}}]{lamansky-irppy3}
\bibinfo{author}{\bibfnamefont{S.}~\bibnamefont{Lamansky}},
  \bibinfo{author}{\bibfnamefont{P.}~\bibnamefont{Djurovich}},
  \bibinfo{author}{\bibfnamefont{D.}~\bibnamefont{Murphy}},
  \bibinfo{author}{\bibfnamefont{F.}~\bibnamefont{Abdel-Razzaq}},
  \bibinfo{author}{\bibfnamefont{H.}~\bibnamefont{Lee}},
  \bibinfo{author}{\bibfnamefont{C.}~\bibnamefont{Adachi}},
  \bibinfo{author}{\bibfnamefont{P.}~\bibnamefont{Burrows}},
  \bibinfo{author}{\bibfnamefont{S.}~\bibnamefont{Forrest}}, \bibnamefont{and}
  \bibinfo{author}{\bibfnamefont{M.}~\bibnamefont{Thompson}},
  \bibinfo{journal}{J. Am. Chem. Soc.} \textbf{\bibinfo{volume}{123}},
  \bibinfo{pages}{4304} (\bibinfo{year}{2001}).

\bibitem[{\citenamefont{Hofbeck and
  Yersin}(2010{\natexlab{a}})}]{hofbeck-irppy3}
\bibinfo{author}{\bibfnamefont{T.}~\bibnamefont{Hofbeck}} \bibnamefont{and}
  \bibinfo{author}{\bibfnamefont{H.}~\bibnamefont{Yersin}},
  \bibinfo{journal}{Inorg. Chem.}  (\bibinfo{year}{2010}{\natexlab{a}}).

\bibitem[{\citenamefont{Colombo et~al.}(1994)\citenamefont{Colombo, Brunold,
  Riedener, Guedel, Fortsch, and Buergi}}]{colombo-irppy3}
\bibinfo{author}{\bibfnamefont{M.}~\bibnamefont{Colombo}},
  \bibinfo{author}{\bibfnamefont{T.}~\bibnamefont{Brunold}},
  \bibinfo{author}{\bibfnamefont{T.}~\bibnamefont{Riedener}},
  \bibinfo{author}{\bibfnamefont{H.}~\bibnamefont{Guedel}},
  \bibinfo{author}{\bibfnamefont{M.}~\bibnamefont{Fortsch}}, \bibnamefont{and}
  \bibinfo{author}{\bibfnamefont{H.}~\bibnamefont{Buergi}},
  \bibinfo{journal}{Inorg. Chem.} \textbf{\bibinfo{volume}{33}},
  \bibinfo{pages}{545} (\bibinfo{year}{1994}).

\bibitem[{\citenamefont{Tsuboyama et~al.}(2003)\citenamefont{Tsuboyama,
  Iwawaki, Furugori, Mukaide, Kamatani, Igawa, Moriyama, Miura, Takiguchi,
  Okada et~al.}}]{tsuboyama-irppy3}
\bibinfo{author}{\bibfnamefont{A.}~\bibnamefont{Tsuboyama}},
  \bibinfo{author}{\bibfnamefont{H.}~\bibnamefont{Iwawaki}},
  \bibinfo{author}{\bibfnamefont{M.}~\bibnamefont{Furugori}},
  \bibinfo{author}{\bibfnamefont{T.}~\bibnamefont{Mukaide}},
  \bibinfo{author}{\bibfnamefont{J.}~\bibnamefont{Kamatani}},
  \bibinfo{author}{\bibfnamefont{S.}~\bibnamefont{Igawa}},
  \bibinfo{author}{\bibfnamefont{T.}~\bibnamefont{Moriyama}},
  \bibinfo{author}{\bibfnamefont{S.}~\bibnamefont{Miura}},
  \bibinfo{author}{\bibfnamefont{T.}~\bibnamefont{Takiguchi}},
  \bibinfo{author}{\bibfnamefont{S.}~\bibnamefont{Okada}},
  \bibnamefont{et~al.}, \bibinfo{journal}{J. Am. Chem. Soc.}
  \textbf{\bibinfo{volume}{125}}, \bibinfo{pages}{12971}
  (\bibinfo{year}{2003}).

\bibitem[{\citenamefont{Holzer et~al.}(2005)\citenamefont{Holzer, Penzkofer,
  and Tsuboi}}]{holzer-irppy3}
\bibinfo{author}{\bibfnamefont{W.}~\bibnamefont{Holzer}},
  \bibinfo{author}{\bibfnamefont{A.}~\bibnamefont{Penzkofer}},
  \bibnamefont{and} \bibinfo{author}{\bibfnamefont{T.}~\bibnamefont{Tsuboi}},
  \bibinfo{journal}{Chem. Phys.} \textbf{\bibinfo{volume}{308}},
  \bibinfo{pages}{93} (\bibinfo{year}{2005}).

\bibitem[{\citenamefont{Ichimura et~al.}(1987)\citenamefont{Ichimura,
  Kobayashi, King, and Watts}}]{ichimura-irppy3}
\bibinfo{author}{\bibfnamefont{K.}~\bibnamefont{Ichimura}},
  \bibinfo{author}{\bibfnamefont{T.}~\bibnamefont{Kobayashi}},
  \bibinfo{author}{\bibfnamefont{K.}~\bibnamefont{King}}, \bibnamefont{and}
  \bibinfo{author}{\bibfnamefont{R.}~\bibnamefont{Watts}}, \bibinfo{journal}{J.
  Phys. Chem.} \textbf{\bibinfo{volume}{91}}, \bibinfo{pages}{6104}
  (\bibinfo{year}{1987}).

\bibitem[{\citenamefont{Baldo et~al.}(2000{\natexlab{b}})\citenamefont{Baldo,
  Adachi, and Forrest}}]{baldo-optical-lumo}
\bibinfo{author}{\bibfnamefont{M.}~\bibnamefont{Baldo}},
  \bibinfo{author}{\bibfnamefont{C.}~\bibnamefont{Adachi}}, \bibnamefont{and}
  \bibinfo{author}{\bibfnamefont{S.}~\bibnamefont{Forrest}},
  \bibinfo{journal}{Phys. Rev. B} \textbf{\bibinfo{volume}{62}},
  \bibinfo{pages}{10967} (\bibinfo{year}{2000}{\natexlab{b}}).

\bibitem[{\citenamefont{Djurovich et~al.}(2009)\citenamefont{Djurovich, Mayo,
  Forrest, and Thompson}}]{djurovich-optical-lumo}
\bibinfo{author}{\bibfnamefont{P.}~\bibnamefont{Djurovich}},
  \bibinfo{author}{\bibfnamefont{E.}~\bibnamefont{Mayo}},
  \bibinfo{author}{\bibfnamefont{S.}~\bibnamefont{Forrest}}, \bibnamefont{and}
  \bibinfo{author}{\bibfnamefont{M.}~\bibnamefont{Thompson}},
  \bibinfo{journal}{Org. Electron.} \textbf{\bibinfo{volume}{10}},
  \bibinfo{pages}{515} (\bibinfo{year}{2009}).

\bibitem[{\citenamefont{Seki and Kanai}(2006)}]{ipes-1}
\bibinfo{author}{\bibfnamefont{K.}~\bibnamefont{Seki}} \bibnamefont{and}
  \bibinfo{author}{\bibfnamefont{K.}~\bibnamefont{Kanai}},
  \bibinfo{journal}{Mol. Cryst. Liq. Cryst} \textbf{\bibinfo{volume}{455}},
  \bibinfo{pages}{145} (\bibinfo{year}{2006}).

\bibitem[{\citenamefont{Cahen and Kahn}(2003)}]{ipes-2}
\bibinfo{author}{\bibfnamefont{D.}~\bibnamefont{Cahen}} \bibnamefont{and}
  \bibinfo{author}{\bibfnamefont{A.}~\bibnamefont{Kahn}},
  \bibinfo{journal}{Adv. Mater.} \textbf{\bibinfo{volume}{15}},
  \bibinfo{pages}{271} (\bibinfo{year}{2003}).

\bibitem[{\citenamefont{Smith et~al.}(2011)\citenamefont{Smith, Riley, Lo,
  Burn, Gentle, and Powell}}]{ircomplex-relativistic}
\bibinfo{author}{\bibfnamefont{A.}~\bibnamefont{Smith}},
  \bibinfo{author}{\bibfnamefont{M.}~\bibnamefont{Riley}},
  \bibinfo{author}{\bibfnamefont{S.}~\bibnamefont{Lo}},
  \bibinfo{author}{\bibfnamefont{P.}~\bibnamefont{Burn}},
  \bibinfo{author}{\bibfnamefont{I.}~\bibnamefont{Gentle}}, \bibnamefont{and}
  \bibinfo{author}{\bibfnamefont{B.}~\bibnamefont{Powell}},
  \bibinfo{journal}{Phys. Rev. B} \textbf{\bibinfo{volume}{83}},
  \bibinfo{pages}{041105} (\bibinfo{year}{2011}).

\bibitem[{\citenamefont{D’Andrade et~al.}(2005)\citenamefont{D’Andrade,
  Datta, Forrest, Djurovich, Polikarpov, and Thompson}}]{solution}
\bibinfo{author}{\bibfnamefont{B.}~\bibnamefont{D’Andrade}},
  \bibinfo{author}{\bibfnamefont{S.}~\bibnamefont{Datta}},
  \bibinfo{author}{\bibfnamefont{S.}~\bibnamefont{Forrest}},
  \bibinfo{author}{\bibfnamefont{P.}~\bibnamefont{Djurovich}},
  \bibinfo{author}{\bibfnamefont{E.}~\bibnamefont{Polikarpov}},
  \bibnamefont{and} \bibinfo{author}{\bibfnamefont{M.}~\bibnamefont{Thompson}},
  \bibinfo{journal}{Org. Electron.} \textbf{\bibinfo{volume}{6}},
  \bibinfo{pages}{11} (\bibinfo{year}{2005}).

\bibitem[{\citenamefont{Hay}(2002)}]{hay-irppy3}
\bibinfo{author}{\bibfnamefont{P.}~\bibnamefont{Hay}}, \bibinfo{journal}{J.
  Phys. Chem. A} \textbf{\bibinfo{volume}{106}}, \bibinfo{pages}{1634}
  (\bibinfo{year}{2002}).

\bibitem[{\citenamefont{{\'S}widerek and Paneth}(2009)}]{ir-complex-calc-1}
\bibinfo{author}{\bibfnamefont{K.}~\bibnamefont{{\'S}widerek}}
  \bibnamefont{and} \bibinfo{author}{\bibfnamefont{P.}~\bibnamefont{Paneth}},
  \bibinfo{journal}{J. Phys. Org. Chem.} \textbf{\bibinfo{volume}{22}},
  \bibinfo{pages}{845} (\bibinfo{year}{2009}).

\bibitem[{\citenamefont{Jansson et~al.}(2007)\citenamefont{Jansson, Minaev,
  Schrader, and Agren}}]{ir-complex-calc-2}
\bibinfo{author}{\bibfnamefont{E.}~\bibnamefont{Jansson}},
  \bibinfo{author}{\bibfnamefont{B.}~\bibnamefont{Minaev}},
  \bibinfo{author}{\bibfnamefont{S.}~\bibnamefont{Schrader}}, \bibnamefont{and}
  \bibinfo{author}{\bibfnamefont{H.}~\bibnamefont{Agren}},
  \bibinfo{journal}{Chem. Phys.} \textbf{\bibinfo{volume}{333}},
  \bibinfo{pages}{157} (\bibinfo{year}{2007}).

\bibitem[{\citenamefont{Kukhta et~al.}(2007)\citenamefont{Kukhta, Kukhta,
  Bagnich, Kazakov, Andreev, Neyra, and Meza}}]{ir-complex-calc-3}
\bibinfo{author}{\bibfnamefont{A.}~\bibnamefont{Kukhta}},
  \bibinfo{author}{\bibfnamefont{I.}~\bibnamefont{Kukhta}},
  \bibinfo{author}{\bibfnamefont{S.}~\bibnamefont{Bagnich}},
  \bibinfo{author}{\bibfnamefont{S.}~\bibnamefont{Kazakov}},
  \bibinfo{author}{\bibfnamefont{V.}~\bibnamefont{Andreev}},
  \bibinfo{author}{\bibfnamefont{O.}~\bibnamefont{Neyra}}, \bibnamefont{and}
  \bibinfo{author}{\bibfnamefont{E.}~\bibnamefont{Meza}},
  \bibinfo{journal}{Chem. Phys. Lett.} \textbf{\bibinfo{volume}{434}},
  \bibinfo{pages}{11} (\bibinfo{year}{2007}).

\bibitem[{\citenamefont{Hofbeck and
  Yersin}(2010{\natexlab{b}})}]{mlct-irppy3-1}
\bibinfo{author}{\bibfnamefont{T.}~\bibnamefont{Hofbeck}} \bibnamefont{and}
  \bibinfo{author}{\bibfnamefont{H.}~\bibnamefont{Yersin}},
  \bibinfo{journal}{Inorg. Chem.} \textbf{\bibinfo{volume}{49}},
  \bibinfo{pages}{9290} (\bibinfo{year}{2010}{\natexlab{b}}).

\bibitem[{\citenamefont{Tang et~al.}(2004)\citenamefont{Tang, Liu, Chen
  et~al.}}]{mlct-irppy3-2}
\bibinfo{author}{\bibfnamefont{K.}~\bibnamefont{Tang}},
  \bibinfo{author}{\bibfnamefont{K.}~\bibnamefont{Liu}},
  \bibinfo{author}{\bibfnamefont{I.}~\bibnamefont{Chen}}, \bibnamefont{et~al.},
  \bibinfo{journal}{Chem. Phys. Lett.} \textbf{\bibinfo{volume}{386}},
  \bibinfo{pages}{437} (\bibinfo{year}{2004}).

\bibitem[{\citenamefont{Himmetoglu et~al.}(2011)\citenamefont{Himmetoglu,
  Wentzcovitch, and Cococcioni}}]{hwc}
\bibinfo{author}{\bibfnamefont{B.}~\bibnamefont{Himmetoglu}},
  \bibinfo{author}{\bibfnamefont{R.~M.} \bibnamefont{Wentzcovitch}},
  \bibnamefont{and}
  \bibinfo{author}{\bibfnamefont{M.}~\bibnamefont{Cococcioni}},
  \bibinfo{journal}{Phys. Rev. B} \textbf{\bibinfo{volume}{84}},
  \bibinfo{pages}{115108} (\bibinfo{year}{2011}).

\bibitem[{\citenamefont{Tsuboi et~al.}(2008)\citenamefont{Tsuboi, Murayama,
  Yeh, Wu, and Chen}}]{tsuboi-firpic}
\bibinfo{author}{\bibfnamefont{T.}~\bibnamefont{Tsuboi}},
  \bibinfo{author}{\bibfnamefont{H.}~\bibnamefont{Murayama}},
  \bibinfo{author}{\bibfnamefont{S.}~\bibnamefont{Yeh}},
  \bibinfo{author}{\bibfnamefont{M.}~\bibnamefont{Wu}}, \bibnamefont{and}
  \bibinfo{author}{\bibfnamefont{C.}~\bibnamefont{Chen}},
  \bibinfo{journal}{Opt. Mater.} \textbf{\bibinfo{volume}{31}},
  \bibinfo{pages}{366} (\bibinfo{year}{2008}).

\bibitem[{\citenamefont{Adachi et~al.}(2001{\natexlab{b}})\citenamefont{Adachi,
  Kwong, Djurovich, Adamovich, Baldo, Thompson, and Forrest}}]{adachi-firpic}
\bibinfo{author}{\bibfnamefont{C.}~\bibnamefont{Adachi}},
  \bibinfo{author}{\bibfnamefont{R.}~\bibnamefont{Kwong}},
  \bibinfo{author}{\bibfnamefont{P.}~\bibnamefont{Djurovich}},
  \bibinfo{author}{\bibfnamefont{V.}~\bibnamefont{Adamovich}},
  \bibinfo{author}{\bibfnamefont{M.}~\bibnamefont{Baldo}},
  \bibinfo{author}{\bibfnamefont{M.}~\bibnamefont{Thompson}}, \bibnamefont{and}
  \bibinfo{author}{\bibfnamefont{S.}~\bibnamefont{Forrest}},
  \bibinfo{journal}{Appl. Phys. Lett.} \textbf{\bibinfo{volume}{79}},
  \bibinfo{pages}{2082} (\bibinfo{year}{2001}{\natexlab{b}}).

\bibitem[{\citenamefont{Holmes et~al.}(2003)\citenamefont{Holmes, Forrest,
  Tung, Kwong, Brown, Garon, and Thompson}}]{holmes-firpic}
\bibinfo{author}{\bibfnamefont{R.}~\bibnamefont{Holmes}},
  \bibinfo{author}{\bibfnamefont{S.}~\bibnamefont{Forrest}},
  \bibinfo{author}{\bibfnamefont{Y.}~\bibnamefont{Tung}},
  \bibinfo{author}{\bibfnamefont{R.}~\bibnamefont{Kwong}},
  \bibinfo{author}{\bibfnamefont{J.}~\bibnamefont{Brown}},
  \bibinfo{author}{\bibfnamefont{S.}~\bibnamefont{Garon}}, \bibnamefont{and}
  \bibinfo{author}{\bibfnamefont{M.}~\bibnamefont{Thompson}},
  \bibinfo{journal}{Appl. Phys. Lett.} \textbf{\bibinfo{volume}{82}},
  \bibinfo{pages}{2422} (\bibinfo{year}{2003}).

\bibitem[{\citenamefont{Tokito et~al.}(2003)\citenamefont{Tokito, Iijima,
  Suzuri, Kita, Tsuzuki, and Sato}}]{tokito-firpic}
\bibinfo{author}{\bibfnamefont{S.}~\bibnamefont{Tokito}},
  \bibinfo{author}{\bibfnamefont{T.}~\bibnamefont{Iijima}},
  \bibinfo{author}{\bibfnamefont{Y.}~\bibnamefont{Suzuri}},
  \bibinfo{author}{\bibfnamefont{H.}~\bibnamefont{Kita}},
  \bibinfo{author}{\bibfnamefont{T.}~\bibnamefont{Tsuzuki}}, \bibnamefont{and}
  \bibinfo{author}{\bibfnamefont{F.}~\bibnamefont{Sato}},
  \bibinfo{journal}{Applied physics letters} \textbf{\bibinfo{volume}{83}},
  \bibinfo{pages}{569} (\bibinfo{year}{2003}).

\bibitem[{\citenamefont{You and Park}(2005)}]{you-firpic}
\bibinfo{author}{\bibfnamefont{Y.}~\bibnamefont{You}} \bibnamefont{and}
  \bibinfo{author}{\bibfnamefont{S.}~\bibnamefont{Park}}, \bibinfo{journal}{J.
  Am. Chem. Soc.} \textbf{\bibinfo{volume}{127}}, \bibinfo{pages}{12438}
  (\bibinfo{year}{2005}).

\bibitem[{\citenamefont{Lee et~al.}(2009)\citenamefont{Lee, Hsu, Chen, Chen,
  Chen, and Wang}}]{lee-firpic}
\bibinfo{author}{\bibfnamefont{H.}~\bibnamefont{Lee}},
  \bibinfo{author}{\bibfnamefont{Y.}~\bibnamefont{Hsu}},
  \bibinfo{author}{\bibfnamefont{T.}~\bibnamefont{Chen}},
  \bibinfo{author}{\bibfnamefont{J.}~\bibnamefont{Chen}},
  \bibinfo{author}{\bibfnamefont{K.}~\bibnamefont{Chen}}, \bibnamefont{and}
  \bibinfo{author}{\bibfnamefont{J.}~\bibnamefont{Wang}},
  \bibinfo{journal}{Inorg. Chem.} \textbf{\bibinfo{volume}{48}},
  \bibinfo{pages}{1263} (\bibinfo{year}{2009}).

\bibitem[{\citenamefont{De~Angelis et~al.}(2007)\citenamefont{De~Angelis,
  Fantacci, Evans, Klein, Zakeeruddin, Moser, Kalyanasundaram, Bolink,
  Gr{\"a}tzel, and Nazeeruddin}}]{func-1}
\bibinfo{author}{\bibfnamefont{F.}~\bibnamefont{De~Angelis}},
  \bibinfo{author}{\bibfnamefont{S.}~\bibnamefont{Fantacci}},
  \bibinfo{author}{\bibfnamefont{N.}~\bibnamefont{Evans}},
  \bibinfo{author}{\bibfnamefont{C.}~\bibnamefont{Klein}},
  \bibinfo{author}{\bibfnamefont{S.}~\bibnamefont{Zakeeruddin}},
  \bibinfo{author}{\bibfnamefont{J.}~\bibnamefont{Moser}},
  \bibinfo{author}{\bibfnamefont{K.}~\bibnamefont{Kalyanasundaram}},
  \bibinfo{author}{\bibfnamefont{H.}~\bibnamefont{Bolink}},
  \bibinfo{author}{\bibfnamefont{M.}~\bibnamefont{Gr{\"a}tzel}},
  \bibnamefont{and}
  \bibinfo{author}{\bibfnamefont{M.}~\bibnamefont{Nazeeruddin}},
  \bibinfo{journal}{Inorg. Chem.} \textbf{\bibinfo{volume}{46}},
  \bibinfo{pages}{5989} (\bibinfo{year}{2007}).

\bibitem[{\citenamefont{Lo et~al.}(2006)\citenamefont{Lo, Shipley, Bera,
  Harding, Cowley, Burn, and Samuel}}]{func-2}
\bibinfo{author}{\bibfnamefont{S.}~\bibnamefont{Lo}},
  \bibinfo{author}{\bibfnamefont{C.}~\bibnamefont{Shipley}},
  \bibinfo{author}{\bibfnamefont{R.}~\bibnamefont{Bera}},
  \bibinfo{author}{\bibfnamefont{R.}~\bibnamefont{Harding}},
  \bibinfo{author}{\bibfnamefont{A.}~\bibnamefont{Cowley}},
  \bibinfo{author}{\bibfnamefont{P.}~\bibnamefont{Burn}}, \bibnamefont{and}
  \bibinfo{author}{\bibfnamefont{I.}~\bibnamefont{Samuel}},
  \bibinfo{journal}{Chem. Mater.} \textbf{\bibinfo{volume}{18}},
  \bibinfo{pages}{5119} (\bibinfo{year}{2006}).

\bibitem[{\citenamefont{Su et~al.}(2008)\citenamefont{Su, Gonmori, Sasabe, and
  Kido}}]{su-pqir}
\bibinfo{author}{\bibfnamefont{S.}~\bibnamefont{Su}},
  \bibinfo{author}{\bibfnamefont{E.}~\bibnamefont{Gonmori}},
  \bibinfo{author}{\bibfnamefont{H.}~\bibnamefont{Sasabe}}, \bibnamefont{and}
  \bibinfo{author}{\bibfnamefont{J.}~\bibnamefont{Kido}},
  \bibinfo{journal}{Adv. Mater.} \textbf{\bibinfo{volume}{20}},
  \bibinfo{pages}{4189} (\bibinfo{year}{2008}).

\bibitem[{\citenamefont{D'Andrade and Forrest}(2004)}]{d2004-pqir}
\bibinfo{author}{\bibfnamefont{B.}~\bibnamefont{D'Andrade}} \bibnamefont{and}
  \bibinfo{author}{\bibfnamefont{S.}~\bibnamefont{Forrest}},
  \bibinfo{journal}{Adv. Mater.} \textbf{\bibinfo{volume}{16}},
  \bibinfo{pages}{1585} (\bibinfo{year}{2004}).

\bibitem[{\citenamefont{D'Andrade et~al.}(2004)\citenamefont{D'Andrade, Holmes,
  and Forrest}}]{d2004-pqir2}
\bibinfo{author}{\bibfnamefont{B.}~\bibnamefont{D'Andrade}},
  \bibinfo{author}{\bibfnamefont{R.}~\bibnamefont{Holmes}}, \bibnamefont{and}
  \bibinfo{author}{\bibfnamefont{S.}~\bibnamefont{Forrest}},
  \bibinfo{journal}{Adv. Mater.} \textbf{\bibinfo{volume}{16}},
  \bibinfo{pages}{624} (\bibinfo{year}{2004}).

\bibitem[{\citenamefont{Yersin}(2008)}]{oled-book}
\bibinfo{author}{\bibfnamefont{H.}~\bibnamefont{Yersin}},
  \emph{\bibinfo{title}{Highly efficient OLEDs with phosphorescent materials}}
  (\bibinfo{publisher}{Wiley Online Library}, \bibinfo{year}{2008}).

\bibitem[{\citenamefont{Jacko et~al.}(2010)\citenamefont{Jacko, Powell, and
  McKenzie}}]{model-1}
\bibinfo{author}{\bibfnamefont{A.}~\bibnamefont{Jacko}},
  \bibinfo{author}{\bibfnamefont{B.}~\bibnamefont{Powell}}, \bibnamefont{and}
  \bibinfo{author}{\bibfnamefont{R.}~\bibnamefont{McKenzie}},
  \bibinfo{journal}{J. Chem. Phys.} \textbf{\bibinfo{volume}{133}},
  \bibinfo{pages}{124314} (\bibinfo{year}{2010}).

\bibitem[{\citenamefont{Jacko and Powell}(2011)}]{model-2}
\bibinfo{author}{\bibfnamefont{A.}~\bibnamefont{Jacko}} \bibnamefont{and}
  \bibinfo{author}{\bibfnamefont{B.}~\bibnamefont{Powell}},
  \bibinfo{journal}{Chem. Phys. Lett.} \textbf{\bibinfo{volume}{508}},
  \bibinfo{pages}{22} (\bibinfo{year}{2011}).

\bibitem[{\citenamefont{Slater}(1974)}]{slater}
\bibinfo{author}{\bibfnamefont{J.}~\bibnamefont{Slater}},
  \emph{\bibinfo{title}{The self-consistent field for molecular and solids,
  quantum theory of molecular and solids, vol. 4}} (\bibinfo{year}{1974}).

\bibitem[{\citenamefont{Gross and Dreizler}(1995)}]{gross-book}
\bibinfo{author}{\bibfnamefont{E.}~\bibnamefont{Gross}} \bibnamefont{and}
  \bibinfo{author}{\bibfnamefont{R.}~\bibnamefont{Dreizler}},
  \emph{\bibinfo{title}{Density functional theory}}, vol. \bibinfo{volume}{337}
  (\bibinfo{publisher}{Springer}, \bibinfo{year}{1995}).

\end{thebibliography}
\end{document}